\documentclass[honours,titlesmallcaps, copyrightpage]{mqthesis}

\usepackage{graphicx, subfigure, amsmath}

\newtheorem{theorem}{Theorem}[chapter]
\newtheorem{conjecture}[theorem]{Conjecture}

\newcommand{\ket}[1]{\left| #1\right\rangle}		
\newcommand{\bra}[1]{\left\langle #1\right|}		
\newcommand{\op}[2]{\ket{#1}\!\bra{#2}}			
\newcommand{\ip}[2]{\left\langle #1|#2\right\rangle}			
\newcommand{\tr}{\mbox{Tr}}					

\newcommand{\Acal}{\mathcal{A}}

\newcommand{\Dcal}{\mathcal{D}}
\newcommand{\Ecal}{\mathcal{E}}
\newcommand{\Fcal}{\mathcal{F}}

\newcommand{\Ical}{\mathcal{I}}
\newcommand{\Jcal}{\mathcal{J}}

\newcommand{\Mcal}{\mathcal{M}}
\newcommand{\Ncal}{\mathcal{N}}

\newcommand{\Pcal}{\mathcal{P}}

\newcommand{\Scal}{\mathcal{S}}
\newcommand{\Tcal}{\mathcal{T}}
\newcommand{\Vcal}{\mathcal{V}}

\newcommand{\qed}{\hfill \ensuremath{\Box}} 
\newcommand{\id}{\mathbb{I}}					

\begin{document}

\frontmatter


\title{Non-Completely Positive Maps: Properties and Applications}
\author{Christopher Wood}
\department{Physics}  


\titlepage

\chapter{Acknowledgements}

I begin with the obvious and offer tremendous thanks to my supervisors Alexei Gilchrist and Daniel Terno. They managed to kept me on track throughout the year, whether with the stick or the carrot will not be mentioned, and without their support and guidance this thesis would not exist.

I would also like to thank Ian Benn and John Holdsworth from the University of Newcastle. The former who's often shambolic but always interesting undergraduate courses cultivated my interest in theoretical physics, and the latter who told me to follow my heart and suggested the idea of completing my Honours year at a different university.

Special mention also goes to Mark Butler who's excellent teaching in high school physics derailed me from a planned career in graphic design and set me on the path of science. Through happenstance he also introduced me to members of the quantum information group at Macquarie University where I ended up enrolling for my Honours year.

I also thank my family whose constant love and support is greatly appreciated. Also free food and accommodation is always welcome to a financially challenged university student.

Finally, acknowledgments go to my fellow Honours students, who were both a blessing and a curse. They provided ample distractions from physics with extended lunch breaks, which were good for sanity, but not so good for getting work done. They also provided the comfort of shared misery when deadlines loomed and sleep deprivation took its toll.

\chapter{Abstract}

In this thesis we investigate the evolution of open quantum systems in the presence of initial correlations with an environment. In the presence of such initial correlations the standard formalism of describing evolution by completely positive trace preserving (CPTP) quantum operations can fail and non-completely positive (non-CP) maps may be observed.

We investigate a new method of classification of correlations between a system and environment using the so-called quantum discord. We found an issue with this classification as quantum discord is not a symmetric quantity between exchange of systems. This leads to ambiguity in classifications --- states which are both quantum and classically correlated depending on the order of the two systems.

Quantum process tomography is investigated with regard to non-CP maps. We examine two methods of performing tomography, standard quantum process tomography (SQPT) and ancilla assisted process tomography (AAPT). We pay particular attention to the effect of state preparation in the presence of initial correlations between the system and environment. We find that in the case of SQPT the preparation procedure can influence the complete-positivity of the reconstructed quantum operation. We examine a recently proposed method of state preparation by using projective measurements, and propose our own protocol that uses a single measurement that is followed by unitary rotations. In the case of the former the evolution can be non-CP while the later will always give rise to a CP map. State preparation in AAPT was found always to give rise to CP evolution. In addition we briefly investigate a proposed method of performing bilinear process tomography.

We investigate the effect of statistical noise in process tomography tomography, and how it can result in the identification of a map as non-CP when the evolution should be CP. We study a probability distribution for quantum operations reconstructed by process tomography in the presence of statistical noise. We found that the variance of the distribution for reconstructed processes is inversely proportional to the number of copies of a state used to perform tomography. As a result, by increasing this number one can distinguish between the distributions of CP and true non-CP processes with a high degree of accuracy.  Finally, we detail an experiment using currently available linear optics quantum computation devices to demonstrate non-CP maps arising in SQPT.

\tableofcontents

\listoffigures
\listoftables

\mainmatter

\chapter{Introduction}

One hundred years after its discovery and preliminary development, quantum physics continues to offer both new insight into the natural world, and exciting technological developments. Techniques exploiting quantum mechanical effects to store, manipulate, transmit and process information have been developed in unison with extraordinary scientific breakthroughs in our understanding of quantum physics. This knowledge and its technological applications are encapsulated in \emph{Quantum information theory}. Quantum information is a new branch of research which seeks to develop information processing technologies based on quantum mechanical effects. Not only has it found application in information centric areas such as totally-secure communication and computation, but it has also given remarkable insights into other branches of physics such as relativistic field theory and condensed matter physics~\cite{Nielsen:2000}.

In this project we shall investigate the theoretical framework which underpins our understanding of the processes which occur in quantum information. We will begin by briefly introducing the concept of an \emph{open} quantum system, and how its evolution is described. Following this we shall discuss in some detail the fundamental mathematics and physics used in quantum information theory to describe the dynamics of quantum systems. In particular, we deal with the concept of describing the state of a quantum system by its \emph{density matrix}, and its evolution by \emph{quantum operations} using the formalism of \emph{completely positive} maps. We shall also describe the important technique of \emph{quantum tomography} which is used to experimentally measure and determine the state and dynamics of a quantum system. Finally we shall discuss the limitations of completely positive maps in describing quantum operations and recent developments which attempt to address these problems.

Quantum systems fall into two categories, they are either considered to be \emph{open}, or \emph{closed}. Closed quantum systems are completely isolated from their environment and hence have no interactions with their surroundings. The dynamics of these systems are directly described by the postulates of quantum mechanics~\cite{Dirac:1982}. More interesting however, are \emph{open} quantum systems. These are systems which exhibit some degree of interaction with their environment --- all real world systems are of this type to some degree. Understanding how these systems evolve is of considerable interest as it is imperative for the construction and operation of actual quantum devices to implement quantum information processing.

In many cases we are not interested in the detailed time-evolution of a process, just in the end result. In this case quantum dynamics are described by \emph{quantum operations}, which are maps from a valid initial quantum state to a valid final quantum state for our system of interest. According to the standard formalism, quantum operations satisfy two requirements: that they are \emph{Completely Positive} (CP), and \emph{Trace Preserving} (TP)~\cite{Kraus:1983}. This is discussed in detail later. Despite being indispensable for much of the theory of quantum information science, this formalism has significant limitations. The approach based on CP maps assumes that there are no initial correlations between the system and its environment. If initial correlations are present the output state of the system after a valid physical process may be predicted to have a negative probability~\cite{Pechukas:1994}. In this case the quantum operation is a \emph{Non-Completely Positive} (non-CP) map. Since a valid physical process requires non-negative probabilities, the CP map formalism has failed to model the physics involved.

The problem is that correlations with an environment arise naturally and may be present in any experiment. Recent research has focused on trying to classify different classes of correlations, however the r\^ole of these correlations and a general formalism for non-CP maps are not well understood.  Unlike the situation with CP maps, only partial results are known~\cite{Terno:2008,Sudarshan:2008}.

\section{Overview of the Thesis}
In this thesis we investigate certain situations where non-CP maps arise from initial correlations between an open system and the environment. The layout of material covered is as follows.

In Chapters~\ref{chap:qinfo} and \ref{chap:quantumops} we introduce the essential mathematical background required for the remaining chapters of this thesis. We introduce the concept of quantum information and encoding information into the state of a quantum system. We describe the representation of the state of a quantum systems by its density matrix, and some useful decompositions of this representation. We introduce the standard formalism for describing the evolution of open quantum systems by quantum operations, represented by completely positive trace preserving (CPTP) maps. In particular we describe several different mathematical representations for CPTP that were encountered in the literature. We provide clear summary of the relationships between these representations as this was not found in the literature.

In Chapter~\ref{chap:correlations} we introduce non-CP maps and discuss how they can be described using \emph{assignment maps}~\cite{Terno:2008}. We also investigate the classification of initial correlations between a system and environment. We follow a recently proposed method of classifying correlations based on a quantity called \emph{quantum discord}~\cite{Ollivier:2002,Henderson:2001,Sudarshan:2008}. However, we found problems with this approach as quantum discord is not a symmetric quantity. We show this by counter-example. The asymmetry prompted us to conjecture that a theorem~\cite{Sudarshan:2008} concerning the relationship between CP evolution and quantum discord does not hold when we exchange the roles of two systems.

In Chapter~\ref{chap:tomography} we investigate the quantum process tomography, which is the process of characterizing an unknown quantum operation. We introduce two methods, standard quantum process tomography (SQPT) and ancilla assisted process tomography (AAPT). As a preliminary step we describe the characterization of unknown quantum states by state tomography. Here we interpret these schemes in a notation consistent with that used in Chapter~\ref{chap:quantumops}. We also discuss sources of statistical noise in process tomography, how this can give non-CP results, and current schemes for dealing with the the noise by maximum likelihood process tomography. We raise the issue of this scheme not being able to distinguish between true non-CP results arising from initial correlations, and non-CP results due to statistical noise.

In Chapter~\ref{chap:stateprep} we investigate the effect of initial correlations on several state preparation techniques used in process tomography. We begin by investigating preparation schemes proposed in the literature, and then we propose a new scheme. We extend the previous work by relating state preparation to non-CP evolution by contrasting the effect of the different schemes when we take into account initial correlations. At the end of the chapter we introduce a bilinear process tomography scheme proposed in Ref~\cite{Kuah:2007}. However, we find one of the asserted properties of the matrix describing bilinear evolution incorrect. We provide the corrected result.

In Chapters~\ref{chap:statsim} we investigate the statistical noise, introduced in Section~\ref{sec:ncptomo}, arising in optical implementations of process tomography. We propose original ideas for distinguishing between non-CP results arising from from noise, and those that legitimately arise from initial correlations.

In Chapter~\ref{chap:linoptics} we bring together our results from the preceding chapters to outline an original experiment which could be performed with currently available linear optical techniques to demonstrate how the different state preparation procedures, introduced in Sections~\ref{ssec:qptrot} and \ref{sec:sqptproj}, can result in non-CP evolution.

Finally, in Chapter~\ref{chap:conclusion} we review the main results from this thesis and discuss directions for future investigations which have arisen from our work. 


\chapter{Quantum Information}        \label{chap:qinfo}

We begin with some essential background in the field of quantum information science. Before we can describe the evolution of open quantum systems in the presence of initial correlations, we must first introduce some fundamental tools used in quantum information theory. In this chapter we will briefly introduce the idea of storing information in the state of a quantum system, and the basic mathematics required for the description of these quantum systems and their states.

\section{Qubits}

In classical information theory all information is described by strings of bits, each of which can have a value of 0 or 1. Computation is then achieved by performing logical operations on these strings. In \emph{quantum} information theory one represents information as \emph{quantum} bits (qubits). A qubit represents a two-level quantum system whose state space is  spanned by the state vectors $\ket{0}$ and $\ket{1}$\footnote{Strictly speaking, a qubit is represented mathematically by a two-dimensional complex Hilbert Space with orthonormal basis $\{\ket{0},\ket{1}\}$. This choice of basis vectors is known as the \emph{computational basis} , and the vectors are labeled to correspond to their classical bit counterparts.}.
Actual two-level physical systems which make good candidates for a qubit include the spin of a particle (\emph{up} or \emph{down}), the polarization of a photon (\emph{vertical} or \emph{horizontal}), or the state of a two level atom (\emph{ground} or \emph{excited}).

An important distinction between a qubit and a classical bit is that unlike its classical counterpart, the state of a qubit can be any normalized complex linear combination of its basis states:
\begin{equation}
\ket{\psi}=\alpha \ket{0} + \beta \ket{1}, \hspace{2em} |\alpha|^2 + |\beta|^2 = 1.
\label{eqn:qstate}
\end{equation}

Physically, if we were to measure the qubit with a device capable of detecting either $\ket{0}$ or $\ket{1}$, then the result would be $\ket{0}$ with probability $|\alpha|^2$, or $\ket{1}$ with probability $|\beta|^2$. This is known as \emph{Born's Rule} and it connects the complex coefficients $\alpha$ and $\beta$, known as amplitudes, with the probability of measuring results.

To model larger systems consisting of multiple qubits we introduce further notation. If we have two qubits labeled $A$ and $B$ with basis states $\ket{0}_A, \ket{1}_A$ and $\ket{0}_B, \ket{1}_B$  respectively, the basis states for the composite system are given by the tensor product of those for the individual systems. Hence, for example, in a 2-qubit system the basis states are
\[
\{ \ket{00}, \ket{01}, \ket{10}, \ket{11} \},  \hspace{2em}\mbox{where }
\ket{00} \equiv \ket{0}_A \otimes \ket{0}_B,
\]
this will be discussed in more detail in Section~\ref{ssec:tensorproduct}.
We will now introduce several important mathematical tools needed for the description of the states of a quantum system.

\section{The Density Matrix} \label{sec:densitymatrix}

To this point we have described our quantum systems in the language of state vectors. However, there is a more general approach using what is known as the \emph{density operator} or \emph{density matrix}\footnote{When dealing with finite-dimensional quantum systems the terms operator and matrix can be used interchangeably. This is because given a basis for the state space of the system, all linear operators acting on the space have a matrix representation with respect to the given basis.}.
The density matrix approach is mathematically richer than the state vector approach, and provides a more convenient framework for many scenarios encountered in quantum information science.

To be precise, if a quantum system is in one of several states $\ket{\psi_i}$ with corresponding probability $p_i$, the density matrix, $\rho$, for the system is defined by:
\begin{equation}
\rho = \sum_i p_i \op{\psi_i}{\psi_i}.
\label{eqn:densityop}
\end{equation}

If a quantum system is known to be in a state $\ket{\psi}$ with unit probability, its density matrix is given by $\rho=\op{\psi}{\psi}$ and it is said to be in a \emph{pure state}. This case is equivalent to the state vector description in Eqn.~(\ref{eqn:qstate}). A density matrix defined as a mixture of different pure states, as in Eqn.~(\ref{eqn:densityop}), is said to be in a \emph{mixed state}. Note that if the density matrix for a qubit is given by $\rho = \id / 2$, where $\id$ is the identity matrix, $\rho$ is said to be in a \emph{maximally mixed state}.

For example, if we had a single qubit in the state
\[
\rho= p_0 \op{0}{0} + p_1 \op{1}{1}, \hspace{2em} \mbox{where }\  p_0+p_1=1,
\]
then the probability of measuring a $\ket{0}$ or $\ket{1}$ is $p_0$ or $p_1$ respectively. If $p_0=p_1= 1/2$, this is a maximally mixed state, while if either $p_0=0$ or $p_1=0$ this will be a pure state. A measure of the \emph{purity} of a state $\rho$ is given by $\tr(\rho^2)$. Pure states satisfy $\tr(\rho^2)=1$ while for mixed states $\tr(\rho^2)<1$

For $\rho$ to be a valid density matrix for our system it is necessary and sufficient that it satisfies two conditions.
\begin{enumerate}
\item \textbf{$\rho$ is a positive-semidefinite matrix:}
A matrix $\rho$ is positive-semidefinite if it is \emph{positive} and \emph{hermitian}.
\begin{itemize}
 \item A matrix $\rho$ is hermitian if $\rho^\dagger=\rho$, where $^\dagger$ denotes conjugate-transposition, $(\rho^\dagger)_{mn}=\rho_{nm}^*$. This is equivalent to requiring that all the eigenvalues of $\rho$ are real.
 \item A positive matrix $\rho$ satisfies $\bra{\psi} \rho\ket{\psi}\ge0$ for any vector $\ket{\psi}$. This is equivalent to requiring that all the eigenvalues of $\rho$ are non-negative.
 \end{itemize}
\item \textbf{$\rho$ has unit trace:}
 That is to say $\tr(\rho)=\sum_m \rho_{mm}=1$. Along with the requirement of positive-semidefinite, this is equivalent to the eigenvalues of $\rho$ being real, non-negative and summing to 1.
\end{enumerate}

The requirement of positivity is related to measurement probabilities. If $\ket{\psi}$ is a pure state, then $\bra{\psi} \rho\ket{\psi}$ is interpreted as the probability of measuring $\ket{\psi}$ given $\rho$. This is why we require it to be non-negative. Requiring $\tr(\rho)=1$ guarantees that the probabilities add up to 1. In general we will refer to a positive-semidefinite matrix simply as a positive matrix, denoted by $\rho\ge0$. We also note that positivity implies hermiticity in a complex Hilbert space.

\section{Bloch Sphere}          \label{ssec:bloch}

\begin{figure}[ht]
\begin{center}
\includegraphics[width=0.46\textwidth]{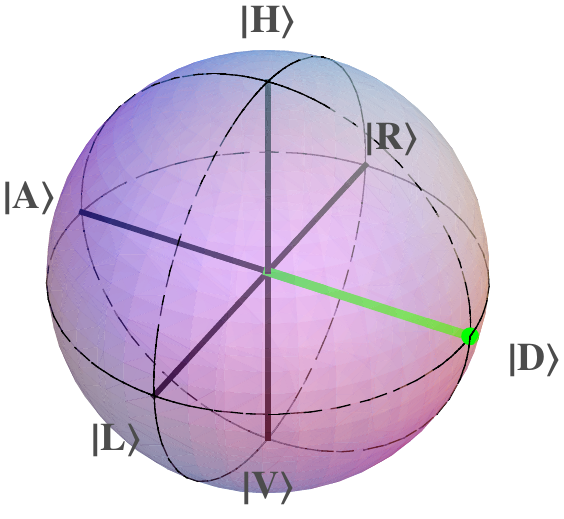}
\end{center}
\caption{The pure state $\ket{D}=(\ket{0}+\ket{1})/\sqrt{2}$ represented on the Bloch sphere}
\label{fig:blochsphere}
\end{figure}

We can visually represent the state of a qubit as a point on the \emph{Bloch sphere}. An example is shown in Fig.~(\ref{fig:blochsphere}). For convenience, the axes of the sphere are labeled to correspond to polarization states of a photon. The correspondence between the polarization states and the computation basis $\{\ket{0},\ket{1}\}$ is shown in Table~(\ref{tab:polarization}).

\begin{table}[htbp]
\begin{center}
\begin{tabular}{|lll|}
  \hline
  Photon polarization &     & State vector in \\
  state vector        &     & computation basis\\
  \hline
  Horizeontal        & $\ket{H}\,\equiv$  &   $\ket{0}$ \\
  Vertical          & $\ket{V}\,\equiv$  &   $\ket{1}$ \\
  Diagonal          & $\ket{D}\,\equiv$  &   $\ket{+}=\frac{1}{\sqrt{2}}\left(\ket{0}+\ket{1}\right)$ \\
  Anti-Diagonal     & $\ket{A}\,\equiv$  &   $\ket{-}=\frac{1}{\sqrt{2}}\left(\ket{0}-\ket{1}\right)$ \\
  Right Circular    & $\ket{R}\,\equiv$  &   $\ket{+i}=\frac{1}{\sqrt{2}}\left(\ket{0}+i\ket{1}\right)$ \\
  Left Circular     & $\ket{L}\,\equiv$  &   $\ket{-i}=\frac{1}{\sqrt{2}}\left(\ket{0}-i\ket{1}\right)$ \\
  \hline
\end{tabular}
\caption{Correspondence between polarization state of a photon and the computational basis for a two-level quantum system}
\label{tab:polarization}
\end{center}
\end{table}

Any pure state vector $\ket{\psi}$ of a qubit can be parameterized by two angles $\theta, \phi$ by
\[
\ket{\psi}=\cos\theta\ket{0}+e^{i\phi}\sin\theta\ket{1}.
\]
This corresponds to a point on the surface of the Bloch sphere specified by the vector $\vec{\alpha}\in\mathbb{R}^3$, where $\vec{\alpha}=(\alpha_1,\alpha_2,\alpha_3)=(\sin2\theta\cos\phi,\sin2\theta\sin\phi,\cos2\theta)$. The vector $\vec{\alpha}$ is called the \emph{Bloch vector} for the state $\ket{\phi}$.

A density matrix can also be represented in terms of its \emph{Bloch vector}, $\vec{\alpha}$, except now $\vec{\alpha}$ need not be of unit length. Pure states are represented by points on the surface of the sphere and have $|\vec{\alpha}|=1$. Mixed states are represented by points inside the Bloch sphere and have $|\vec{\alpha}|<1$. The maximally mixed state corresponds to the center of the Bloch sphere and has $\vec{\alpha}=0$.
For an arbitrary density matrix $\rho$, the \emph{Bloch representation} is given by
\[
\rho=\frac{1}{2}(\id+\sum_{i=1}^3 \alpha_i\sigma_i),
\]
where $\sigma_i$ are the Pauli matrices
\begin{equation}
\sigma_1\equiv X=\left[
	\begin{array}{ c c }
     	0 & 1 \\
   	    1 & 0
  \end{array} \right],\quad
\sigma_2\equiv Y=\left[
	\begin{array}{ c c }
     	0 & -i \\
        i & 0
  \end{array} \right],\quad
\sigma_3\equiv Z=\left[
	\begin{array}{ c c }
     	1 & 0 \\
    	0 & -1
  \end{array} \right],\label{eqn:pauli}
\end{equation}

The set $\{\id,\sigma_1,\sigma_2,\sigma_3\}$ is an orthogonal basis for all $2\times 2$ complex matrices with respect to the \emph{Hilbert-Schmidt inner product} $(A,B)\equiv\tr(A^\dagger B)$. We can recover the elements of the Bloch vector for a state $\rho$ by $\alpha_i=\tr(\rho\sigma_i)$.

\section{Spectral Decomposition}          \label{ssec:spectralthm}

A particularly useful result from linear algebra is the \emph{spectral theorem}~\cite{Horn:1985}. This theorem applies to \emph{normal matrices}. A matrix $M$ is normal if and only if $M^\dagger M=MM^\dagger$. Hermitian and unitary matrices are both subsets of normal matrices.

The spectral theorem states that for any normal matrix $M$ on a vector space $\Vcal$ there exists an orthonormal basis for $\Vcal$ which diagonalizes $M$. The basis is given by the normalized eigenvectors, $\ket{e_i}$, of $M$, and the entries of the resulting diagonal matrix $D_M$ are the eigenvalues, $\lambda_i$, of $M$. Hence we can form the \emph{spectral decomposition} of $M$,

\[
M=E_M D_m E_M^\dagger = \sum_i \lambda_i \op{e_i}{e_i},
\]
where $\ket{e_i}$ are the columns of $E_M$.

In the case of density matrices, the eigenvalues are real, non-negative and sum to 1. This is an important result as it means that \emph{every} density matrix can be expressed as a \emph{convex-linear} sum of its eigenstates $\op{e_i}{e_i}$. A sum $\rho=\sum_j p_j\rho_j$ is convex-linear if the coefficients $p_j$ are all non-negative and sum to 1. Conversely, a density matrix is a pure state if it only has one non-zero eigenvalue.

\section{Postulates Of Quantum Mechanics}        \label{ssec:postulates}

Quantum mechanics can be formulated using density matrices and the following four postulates~\cite{Nielsen:2000}:
\begin{itemize}
\item \textbf{Postulate 1}:
The state of any \emph{closed} physical system is completely described by its \emph{density matrix}. The density matrix, $\rho$, is a positive matrix which acts on the state space of the system and has trace one. Here a closed system refers to a system which does not interact with its environment. If the system is in a state $\rho_i$ with probability $p_i$, the systems density matrix is given by $\sum_i p_i \rho_i$.

\item \textbf{Postulate 2}:
Closed quantum systems evolve under \emph{unitary transformation}. That is, if a system evolves from $\rho$ to $\rho^\prime$ at times $t_1$ and $t_2$ respectively, the states are related by a unitary operator $U\equiv U(t_2,t_1)$ by the equation
\begin{equation}
\rho^\prime = U\rho U^\dagger,
\label{eqn:uevo}
\end{equation}
where $U^\dagger$ is the adjoint of the operator $U$ (an operator $U$ is unitary if and only if $U^\dagger U=\id$). This is equivalent to saying that closed quantum systems evolve according to Schr\"odinger's equation.

\item \textbf{Postulate 3}:
Quantum measurements are described by a collection of positive operators, $\{M_m\}$, acting on the state space of the system being measured. These operators satisfy the \emph{completeness relation} $\sum_m M^{\dagger}_m M_m = \id$. If, prior to measurement the system is in state $\rho$, the probability that result $m$ occurs is given by
\[
p(m)=\tr(M^{\dagger}_m M_m \rho),
\]
and the post measurement state of the system is
\[
\rho_m=\frac{M_m \rho M^{\dagger}_m}{\tr(M^{\dagger}_m M_m \rho)}.
\]

\item \textbf{Postulate 4}:
For a composite physical system, the state space is given by the \emph{tensor product} of state spaces of its component systems. If we have $n$ constituent systems, with the $i$ system being prepared in state $\rho_i$, the joint state of the composite system is $\rho_1\otimes\rho_2\otimes\ldots\otimes\rho_n$.

\end{itemize}

\section{The Tensor Product}             \label{ssec:tensorproduct}

The \emph{tensor product} mentioned in Postulate 4 is a mathematical operation for combining two or more matrices of arbitrary dimension into a larger \emph{block matrix}~\cite{Horn:1991}. If we consider two matrices $A$ and $B$, with matrix elements $a_{mn}$, and $b_{kl}$ respectively, then their tensor product is given by the block matrix
\[
A\otimes B = \left(
               \begin{array}{cccc}
                 a_{11} B & a_{12} B & \ldots & a_{1n} B \\
                 a_{21} B & \ddots   &       & \vdots    \\
                 \vdots   &          &       &           \\
                 a_{m1} B & \ldots   &       & a_{mn} B  \\
               \end{array}
             \right).
\]
This operation can be applied to both density matrices and state vectors, we treat the latter as $d\times 1$ matrices.

For matrices (or vectors) $A$, $B$, $C$, $D$ and scalar $\alpha$ the tensor product satisfies the following relations:
\begin{eqnarray*}
\mbox{Left and right distributivity: }&& (A+B)\otimes(C+D)= A\otimes B + C\otimes D\\
\mbox{Bilinearity: }&&    (\alpha A)\otimes B=A\otimes (\alpha B) =\alpha (A\otimes B)\\
\mbox{Associativity: }&&  (A\otimes B)\otimes C=A\otimes (B\otimes C)\\
\mbox{Multiplication: }&& (A\otimes B)(C\otimes D)= (AC)\otimes(BD)\\
\mbox{Trace distributivity: }&& \tr(A\otimes B)=\tr(A)\tr(B)\\
\mbox{Adjoint, Transpose, Inverse, Complex-conjugate: }&& (A\otimes B)^\circ = A^\circ\otimes B^\circ
\,\,\mbox{where}\,\, ^\circ=^\dagger, ^T,^{-1},^*.
\end{eqnarray*}

\section{Composite Quantum Systems}      \label{sec:compsystems}

We shall now introduce notation for describing composite systems of two qubits. However, by repeated application one can extend this to systems of many qubits. The Bloch representation of a single qubit can be generalized to a system of two qubits, $A$ and $B$, by taking the tensor products of the basis elements $\{\id,\sigma_1,\sigma_2,\sigma_3\}$. In this notation we can express an arbitrary two qubit state as
\begin{equation}
\rho_{AB}=\frac{1}{4}\sum_{ij}\left( \id\otimes\id
+ \alpha_i\sigma_i\otimes\id + \beta_j\id\otimes\sigma_j
+ \gamma_{ij}\sigma_i\otimes\sigma_j\right).\label{eqn:fano}
\end{equation}
where $\vec{\alpha}=(\alpha_1,\alpha_2,\alpha_3)$ and $\vec{\beta}=(\beta_1,\beta_2,\beta_3)$ are the Bloch vectors for the reduced states of system $A$ and $B$ respectively, and $\gamma_{ij}$ are real parameters describing \emph{correlations} between the systems.

If the state of the combined system can be written as $\rho_{AB}=\rho_A\otimes\rho_B$, then $\rho_{AB}$ is said to be \emph{simply separable}, or a \emph{product state}. If $\rho_{AB}$ can be expressed as a convex-linear sum of product states, $\rho_{AB}=\sum_i p_i\rho^A_i\otimes\rho^B_i$, then $\rho_{AB}$ is said to be \emph{separable}.

An important feature distinguishing quantum systems from classical ones is that when we combine two systems $A$ and $B$, arbitrary states of the joint system $AB$ may have correlations present which cannot be described classically. In the cases where $\rho_{AB}$ is not separable, the state is said to be \emph{entangled}. Entanglement is a feature unique to quantum physics and is the reason for many of the intriguing results of quantum information, such as quantum teleportation and totally secure communication~\cite{Nielsen:2000}. We consider correlations in more detail in Chapter~\ref{chap:correlations}.

\section{Partial Trace}                  \label{ssec:partialtrace}

If we know the state of a composite system, the state of one of the component systems can be described using the \emph{partial trace}. Consider two systems, labeled $A$ and $B$, with the composite system described by the density matrix $\rho_{AB}$. The density matrix for the system $A$ is then given by the reduced density matrix
\begin{equation}
\rho_A \equiv \tr_B (\rho_{AB}).
\label{eqn:ptrace}
\end{equation}
Here $\tr_B$ is called the \emph{partial trace} over system $B$.

If we have a system of two arbitrary matrices $A\otimes B$, then $\tr_B$ is defined as:
\begin{equation}
 \tr_B(A\otimes B) \equiv A\, \tr(B).
 \label{eqn:ptracedef}
\end{equation}

By linearity the above expression can be extended to a more general correlated states $\rho_{AB}$, which we can express in the form of Eqn.~(\ref{eqn:fano}) or its higher dimension generalization. From here we can directly apply the partial trace as defined by Eqn.~(\ref{eqn:ptracedef}). Since the Pauli matrices are traceless the reduced states are
\[
\rho_A=\frac{1}{2}\left( \id+\sum_{i=1}^3 \alpha_i\sigma_i\right),
\quad
\rho_B=\frac{1}{2}\left( \id+\sum_{j=1}^3 \beta_j\sigma_i\right),
\]
which is the Bloch sphere representation from Section~\ref{ssec:bloch}.

A second method of calculating the partial trace of a density matrix $\rho_{AB}$ is in terms of any orthonormal basis $\{\ket{i}\}$ for the system we wish to trace over. In this case we have
\begin{eqnarray}
\rho_A          &=&\tr_B{\rho_{AB}}\\
                &=&\sum_i \bra{i}^B\rho_{AB}\ket{i}^B\\
                &=&\sum_i \tr(\rho_{AB}\op{i}{i}^B)\\
(\rho_A)_{mn}   &=&\sum_{i} (\rho_{AB})_{mi,ni} \, ,
\end{eqnarray}
where $\{\ket{i}\}$ is an orthonormal basis for $B$ and the superscript $B$ denotes that the operator acts only on system $B$, i.e. if $B$ is a qubit
\[
\ket{i}^B=\id\otimes\ket{i}=
\left[
  \begin{array}{cc}
    u_i & 0 \\
    v_i & 0 \\
    0 & u_i \\
    0 & v_i \\
  \end{array}
\right],\quad
\mbox{where }
\ket{i}=
\left[
  \begin{array}{c}
    u_i \\
    v_i \\
  \end{array}
\right]
.
\]
Similarly we can define operators $\ket{i}^A=\ket{i}\otimes\id$ to trace out system $A$.

\section{Schmidt Decomposition}          \label{ssec:schmidt}

Another important result which we will briefly mention is the \emph{Schmidt decomposition}~\cite{Nielsen:2000}. This theorem states that for any pure state $\ket{\psi}$ of a composite system $AB$, there exist orthonormal bases $\{\ket{i_A}\}$ and $\{\ket{i_B}\}$ for systems $A$ and $B$ respectively such that
\[
\ket{\psi}=\sum_i \lambda_i\ket{i_A}\otimes\ket{i_B}.
\]
The coefficients $\lambda_i$, called the \emph{Schmidt coefficients}, are real, non-negative numbers satisfying $\sum_i \lambda_i^2=1$. The number of non-zero Schmidt coefficients is called the \emph{Schmidt number} for $\ket{\psi}$ and represented by $Sch(\ket{\psi})$.

The Schmidt number can be used as a measure of entanglement. For any state $\ket{\psi}$, $Sch(\ket{\psi})=1$ if and only if $\ket{\psi}$ is a product state, while $Sch(\ket{\psi})>1$ if and only if $\ket{\psi}$ is entangled.

Generalizing this to density matrices gives
\[
\rho=\sum_{i,j} \lambda_i\lambda_j \op{i_A}{j_A}\otimes\op{i_B}{j_B}.
\]
The density matrices for the reduced states are given by $\rho_A=\tr_B(\rho)= \sum_i \lambda_i^2 \op{i_A}{i_A}$ and $\rho_B=\tr_A(\rho)= \sum_i \lambda_i^2 \op{i_B}{i_B}$. Hence the reduced states $\rho_A$ and $\rho_B$ have the same eigenvalues. For a density matrix a Schmidt number of 1 corresponds to a \emph{simply separable} state.

\section{Vectorization of Matrices}          \label{ssec:vecmat}

Since the space of matrices is a linear space, we can represent density matrices as vectors on a higher dimensional Hilbert space know as a \emph{Liouville space}~\cite{Horn:1991}. This is useful for describing certain topics encountered later on. This process is called \emph{vectorization}, and transforms an $n\times m$ matrix $M$ into a $(mn)\times 1$ column vector denoted by $\mbox{vec}(M)$ or $\ket{M}$. This is done by stacking the columns of $M$ to form a vector, so if $M_{ij}=m_{ij}$ then
$\mbox{vec}(M)\equiv\ket{M}=[m_{11},...,m_{m1},m_{12},...,m_{m2} ... m_{1n},...m_{mn}]^T$.

In the case of a single qubit we have
\[
\rho=\left(
	\begin{array}{ c c }
     	p_{11} & p_{12} \\
   	p_{21} & p_{22}
  \end{array} \right)
\quad\mbox{ then }
\ket{\rho}=\left[
\begin{array}{ c }
     	p_{11} \\ p_{21} \\ p_{12} \\ p_{22}
\end{array}\right].
\]
The inverse of vectorization is the $mat$ function, so $mat\ket{\rho}=\rho$.
We will now list some useful properties of vectorized matrices~\cite{Horn:1991}. Let $A, B$ and $C$ be complex square matrices. Then,
\begin{enumerate}
\item The vector inner product is the Hilbert-Schmidt inner product: $\ip{A}{B}= \tr(A^\dagger B)$.
\item The Schmidt number of $\ket{A}$ is equal to the rank of $A$.
\item Vectorization is linear: For matrices $A_i$ and scalars $a_i$,
        $\ket{\sum_i a_i A_i}=\sum_i a_i\ket{A_i}$
\item The action of a matrix $A\otimes\id$ on a vectorized matrix $\ket{B}$ is $A\otimes \id\ket{B}=\ket{BA^T}$
\item More generally, we have the identity $\ket{ABC}=(C^T\otimes A)\ket{B}$.
\item For computational basis $\{\ket{i}\}$, if $A=\sum_{ij}a_{ij}\op{i}{j}$, then $\ket{A}=\sum_{ij} a_{ij}\ket{j}\otimes\ket{i}$.
\item Finally, for any matrix $A$, $\ket{A}=A^T\otimes\id\ket{\varphi}=\id\otimes A\ket{\varphi}$ where $\ket{\varphi}=\sum_i\ket{i}\otimes\ket{i}$.
\end{enumerate}


\chapter{Quantum Operations}        \label{chap:quantumops}

Now that we have introduced the essential mathematical tools for describing the states of quantum systems, we move onto the description of their evolution. Postulate 2 is only sufficient for describing the dynamics of \emph{closed} quantum systems. However, in the real world there will always be some degree of interaction between a system and the environment. We require a more robust framework to describe the dynamics of these open quantum systems. The standard mathematical formalism for the evolution of open systems is quantum operations, though this formalism does have its limitations which we will discuss.

A \emph{quantum operation} is a map, $\Ecal$, acting on the state space of a quantum system. The map describes how quantum states, represented by density matrices $\rho$, transform. A formal definition will be given in Section~\ref{ssec:cpaxioms}. Two elementary examples of quantum operations are unitary transformation $\Lambda(\rho)=U\rho U^\dagger$,  and quantum measurement $\Lambda_m(\rho)=M_m\rho M^{\dagger}_m/\tr[M^\dagger M\rho]$, which were mentioned in postulates 2 and 3, respectively, of Section~\ref{ssec:postulates}.

There are several different approaches for describing general quantum operations which we will introduce in the following sections. The formal definition of a completely-positive trace-preserving map (CPTP) is based on a set of physically motivated axioms such an operation should satisfy. There are then several useful mathematical representations for describing CPTP maps, these include the Kraus representation, process matrix, and the superoperator. All these methods are equivalent under the condition of complete positivity, however there are certain advantages to each approach~\cite{Choi:1975,Kraus:1983,Nielsen:2000,Bengtsson:2006}. Finally there is a physical interpretation where we consider the system and environment together to be a closed system.

\section{Completely Positive Maps}     \label{ssec:cpaxioms}
First we shall introduce the formal definition for a quantum operation. This definition is based on a set of axioms encompassing the physical constraints the evolution of a quantum system should satisfy. It mostly follows from the properties of density matrices in Section~\ref{sec:densitymatrix}, which a map describing quantum evolution \emph{should} preserve. If the output state is to be a valid density matrix we would expect the map to be trace preserving, convex-linear, and positive. It turns out that requiring the map to be positive is not strong enough, and instead complete positivity is required. The reason for this requirement is explained below.

A quantum operation for a system $S$ is a map,
\begin{equation}
\Ecal: \rho \mapsto \Ecal(\rho),
\end{equation}
acting on density matrices for a system $S$, which satisfies the following axioms:
\begin{enumerate}
\item $\Ecal$ is \emph{trace preserving}. That is to say  $\tr[\Ecal(\rho)]=\tr(\rho)$.
\item $\Ecal$ is a \emph{convex-linear map}. That is
	\begin{equation}
		\Ecal\left( \sum_i p_i \rho_i \right) = \sum_i p_i \Ecal(\rho_i)
		\quad \mbox{where }  p_i\ge 0 \,\forall i, \,\mbox{ and }\, \sum_i p_i=1
		\label{eqn:convexlin}
	\end{equation}
\item $\Ecal$ is a \emph{completely-positive} map. This means that $\Ecal$ is positive (maps positive matrices to positive matrices), and also if we introduce an auxiliary system of arbitrary dimension then the map $\Ical\otimes\Ecal$ on the joint system is positive, where $\Ical$ is the identity map on the auxiliary system.
\end{enumerate}
The trace preserving property of $\Ecal$ ensures that the output density matrix will satisfy the trace condition. Convex-linearity ensures that the transformation of a mixed state is equivalent to the probabilistic sum of the transformations of constituent pure states. Finally, complete positivity of the map ensures that the positivity condition is satisfied by the output state of \emph{any} combined system. Requiring that $\Ecal$ be only a positive map instead of completely positive is not sufficient as it does not guarantee that the transformation of a positive state on a composite system is itself positive. We will show this by example.

Consider the transposition map $\Tcal(\rho)=\rho^T$. Since for any density matrix $\rho^T\ge0$, $\Tcal$ is a positive map. However if we consider the action of $\Ical\otimes\Tcal$ on the maximally entangled $2$-qubit state $\frac{1}{\sqrt{2}}\left(\ket{00}+\ket{11}\right)$, the density matrix for the final state of the joint system is
\[
\frac{1}{2}\left(
  \begin{array}{cccc}
    1 & 0 & 0 & 0 \\
    0 & 0 & 1 & 0 \\
    0 & 1 & 0 & 0 \\
    0 & 0 & 0 & 1 \\
  \end{array}
\right),
\]
which has a negative eigenvalue of $-1/2$ and so is not a physical density matrix for the joint system.

A map satisfying these three axioms is referred to as a completely positive, trace preserving (CPTP) map. It is possible to relax the trace condition to $\tr[\Ecal(\rho)]\le\tr[\rho]$, allowing for completely positive trace decreasing maps. We will only be concerned with CPTP maps and for the remainder of this thesis, except when explicitly specified, we shall refer to maps satisfying all three of these axioms simply as completely-positive maps (CP maps).

\section{Mathematical Representations of Completely Positive Maps} \label{sec:cpmaps}

Now that we have defined quantum operations to be completely positive maps, we need a mathematical representation for them. There are several different representations used in literature, and we will now introduce three of them. These are the Kraus representation~\cite{Kraus:1983}, the process matrix~\cite{Choi:1975}, and the superoperator~\cite{Bengtsson:2006}. We will also reveal the relationships between these representations, and how each can be converted into the others. Following this, we will show how these representations can arise from considering unitary evolution of the system and environment when taken from the point of view of the system alone.

\subsection{Kraus Representation}       \label{ssec:krausrep}

Our first representation, and the most widely used, is the \emph{Kraus representation}, and is the result of the following theorem~\cite{Choi:1975}.

\begin{theorem}[Kraus Representation]\label{thm:kraus}
A map $\Ecal$ acting on density matrices of a system $S$ is CPTP if and only if there exists a set of operators $\{K_n\}$ acting on the state space of system $S$ such that
\begin{equation}
\Ecal(\rho) = \sum_n K_n \rho K_n^\dagger
\quad \mbox{ where }\quad \sum_n K_n^\dagger K_n = \id.
\label{eqn:krausform}
\end{equation}
\end{theorem}

This form of expressing a quantum operation is also refereed to as the \emph{operator-sum formalism}, and proof of this theorem can be found in most textbooks concerning quantum information~\cite{Nielsen:2000,Choi:1975,Bengtsson:2006}. The operators $K_n$ are called \emph{Kraus matrices} and they satisfy $\sum_n K_n^\dagger K_n = \id$, which is known as the  \emph{completeness relation}.

An elementary example is unitary evolution, where we only have one Kraus matrix. In general the evolution described here need not be unitary. The Kraus representation allows us to completely characterize the dynamics of an open system $S$ by a map acting only on the state space of $S$. In other words we do not have to explicitly consider properties of environment, they are all accounted for in the Kraus operators $\{K_n\}$. Another feature of the Kraus representation is that it is not unique. This can be useful as different system-environment interactions may still give rise to the same reduced dynamics on the system.

\subsection{Process Matrix}    \label{ssec:processmap}

Our second method describing a completely positive map $\Ecal$ acting on a system $S$ is with the \emph{process matrix}, $\Lambda_\Ecal$. We define the unique process matrix for a CPTP map $\Ecal$ as a matrix on a fictitious space $A\otimes S$, where the auxiliary system (\emph{ancilla}) $A$ is a copy of $S$. It is given by
\begin{equation}
\Lambda_\Ecal=\sum_{i,j}E_{ij} \otimes  \Ecal(E_{ij}), \label{eqn:processmap}\\
\end{equation}
where $E_{ij}=\op{i}{j}$ is the matrix with a 1 in the $(ij)^{th}$ entry and zeros elsewhere.

The evolution of a state $\rho$ by a quantum operation $\Ecal$ can then expressed in terms of the process matrix by
\begin{equation}
\Ecal(\rho)=\tr_A\left[\Lambda_\Ecal(\rho^T\otimes\id)\right]=\tr_A\left[(\rho^T\otimes\id)\Lambda_\Ecal\right]. \label{eqn:pmapevo}
\end{equation}
The proof of this is given in Appendix~\ref{app:processmatrix}~\footnote{One could also define the process matrix by
$\Lambda_\Ecal=\sum_{i,j} \Ecal(E_{ij}) \otimes E_{ij}$. Our evolution would then be given by $\Ecal(\rho)=\tr_B[\Lambda_\Ecal(\id\otimes\rho^T)]$. Since we can treat $\Lambda_\Ecal$ as a bipartite matrix, this change in definition is just a swap of the two components $A\leftrightarrow B$. The relationship between $\Lambda_\Ecal$ and the other representations would also change with respect to this swap.}.

There is an equivalence between the completely positivity of a quantum operation $\Ecal$ and properties of its process matrix. This equivalence is given by the following theorem.
\begin{theorem}[Process Matrix]\label{thm:process}
A quantum operation $\Ecal$ on a $d$-dimensional system $S$ is CPTP if and only if its process matrix $\Lambda_\Ecal$, defined by Eqn.~(\ref{eqn:processmap}), satisfies
\begin{enumerate}
\item $\Lambda_\Ecal$ is positive-semidefinite $(\Lambda_\Ecal\ge0)$.
\item $\tr_B(\Lambda_\Ecal)=\id$.
\end{enumerate}
\end{theorem}
For a proof we refer the reader to~\cite{Choi:1975} or \cite{Bengtsson:2006}.

From Eqn.~(\ref{eqn:processmap}) it is easy to see how to construct a process matrix from a Kraus representation. If we wish to switch representations in the other direction we note that for a completely positive map $\Ecal$ the process matrix $\Lambda_\Ecal$ is positive and hermitian. Hence $\Lambda_\Ecal$ has non-negative eigenvalues and orthonormal eigenvectors, $\lambda_i$ and $\ket{e_i}$ respectively. We construct the Kaus matrices by
\[
K_n=\sqrt{\lambda_n}\,mat\ket{e_i}.
\]
The proof of this can be found in Appendix~\ref{app:pmaptokraus}


\subsubsection{Process Matrices as States}

$\Lambda_\Ecal$ has all the properties of a density matrix except the normalization~\footnote{The normalization is an aesthetic choice in order to simplify the related equations.}. It is a positive hermitian matrix, however it is $d^2\times d^2$ instead of $d\times d$, and has trace $d$ instead of 1. This similarity is due to a relationship known as the \emph{Jamiolkowski isomorphism}~\cite{Choi:1975}, which gives a direct correspondence between process matrices for a  $d$-dimensional system $S$, and density matrices for a $d^2$-dimensional system $A\otimes S$, where we have introduced an ancilla $A$ of the same dimension as $S$.

To see this we note that
\[
\sum_{i,j} \frac{1}{d} E_{ij}\otimes E_{ij}
=\sum_{i,j}  \frac{1}{d}\op{i}{j}\otimes\op{i}{j}
=\op{\phi}{\phi},
\]
where $\ket{\phi}=\sum_i(\ket{i}\otimes\ket{i})/\sqrt{d}$ is a maximally entangled state of the joint system $A\otimes S$.

\begin{figure}[htbp]
\begin{center}
\includegraphics[width=0.45\textwidth]{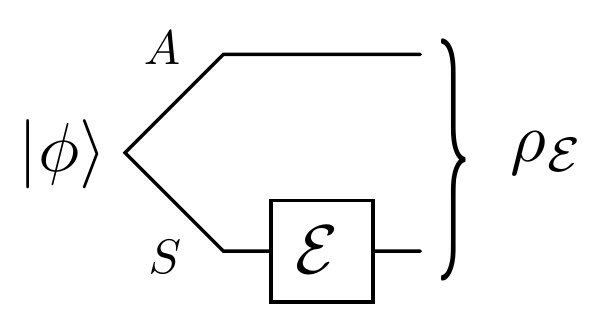}
\end{center}
\caption{The relationship between a process matrix and a density matrix by the Jamiolkowski isomorphism}
\label{fig:jamisom}
\end{figure}

Thus the correspondence is given by $\rho_\Ecal=(\Ical\otimes\Ecal)(\op{\phi}{\phi})=\Lambda/d$. This can be thought of as introducing an ancilla system $A$ as illustrated in Fig.~(\ref{fig:jamisom}). One important application of this approach is in process tomography which we investigate in Chapter~\ref{chap:tomography}.

\subsection{Linear Superoperator}     \label{ssec:superop}

Our final mathematical representation of a quantum operator $\Ecal$ is as a \emph{linear superoperator} $\Phi_\Ecal$, which acts on vectorized density matrices. This evolution is given by
\begin{equation}
\ket{\Ecal(\rho)}=\Phi_\Ecal\ket{\rho}.
\label{eqn:sopevo}
\end{equation}

We give a novel definition for the superoperator, which is not found in the literature,
\begin{equation}
\Phi_\Ecal=\sum_{ij}\op{\Ecal(E_{ij})}{E_{ij}},
\label{eqn:sodef}
\end{equation}
where $E_{ij}=\op{i}{j}$ is the matrix with a 1 in the $(ij)^{th}$ entry and zeros elsewhere. Proof of this can be found in Appendix~\ref{app:superop}

While a linear superoperator is a convenient representation of a CPTP map $\Ecal$, proving an arbitrary map is CP directly from $\Phi_\Ecal$ is not very illuminating. What is usually done~\cite{Bengtsson:2006} is to transform $\Phi_\Ecal$ into the process matrix $\Lambda_\Ecal$. This is done by a process called \emph{reshuffling}. Reshuffling is a rearrangement of the matrix elements of the superoperator. It is given by
$(\Lambda_\Ecal)_{mn,\mu\nu}=(\Phi_\Ecal)_{\mu m,\nu n}$. Full details of this procedure can be found in~\cite{Bengtsson:2006}.

However, we noticed a more intuitively picture of reshuffling by using our definition for the superoperator. If we compare the definitions of $\Phi_\Ecal$ in Eqn.~(\ref{eqn:sodef}) and $\Lambda_\Ecal$ in Eqn~({\ref{eqn:processmap}) we can see that this process swaps
\[
\op{\Ecal(E_{ij})}{E_{ij}} \leftrightarrow E_{ij}\otimes \Ecal(E_{ij}).
\]

The superoperator can easily be calculated from the Kraus representation by vectorizing and using the identities in Section~\ref{ssec:vecmat}.
\begin{eqnarray}
\ket{\Ecal(\rho)}
&=& \mbox{vec}\left(\sum_n K_n\rho K_n^\dagger\right)\\
&=& \sum_n K_n^*\otimes K_n \ket{\rho}
\end{eqnarray}
so $\Phi=\sum_n K_n^*\otimes K_n$. The reverse process is not so simple. One approach is to reshuffle $\Phi_\Ecal$ to form $\Lambda_\Ecal$, and the apply the procedure for constructing a set of Kraus operators from a process matrix.

\subsection{Summary of Relationships Between Mathematical Representations}

We have now introduced several different mathematical descriptions for CPTP quantum processes. In this section we will briefly summarize the relationships between them. To our knowledge such a compact summary is not presented anywhere in the literature though it is a useful computational aid.

For a CPTP quantum operation $\Ecal$ acting on system $A$ there exists a process matrix $\Lambda_\Ecal$, a superoperator $\Phi_\Ecal$, and Kraus operators $\{K_n\}$ as defined in Sections~\ref{ssec:processmap}, \ref{ssec:processmap}, and \ref{ssec:superop} respectively.
Evolution of density matrix $\rho$ of system $A$ is then given by
\[
\Ecal(\rho)=\left\{
  \begin{array}{l}
    \tr_A\left[\Lambda_\Ecal(\rho^T\otimes\id)\right], \\
    mat(\Phi_\Ecal\ket{\rho}),\\
    \sum_n K_n\rho K_n^\dagger. \\
  \end{array}
\right.
\]

To switch between representations we can use the relationships summarized in Table~(\ref{tab:cpreps})

\begin{table}[htbp]
\begin{center}
\begin{tabular}{|c||c|c|c|}
  \hline
  To $\backslash$ From &Process Matrix $(\Lambda_\Ecal)$& Superoperator $(\Phi_\Ecal)$ &Kraus Representation $(K_n)$\\
  \hline\hline
  $\Lambda_\Ecal =$ & $\Lambda_\Ecal$ & Reshuffle($\Phi_\Ecal$) &  $\sum_{ij} E_{ij}\otimes\left(\sum_n K_n E_{ij} K_n^\dagger \right)$   \\
    & &$(\Lambda_\Ecal)_{mn,\mu\nu}=(\Phi_\Ecal)_{\mu m,\nu n}$ & \\ \hline
  $\Phi_\Ecal =$    & Reshuffle($\Lambda_\Ecal$)    & $\Phi_\Ecal$       &   $\sum_n K_n^*\otimes K_n$ \\
                    & $(\Phi_\Ecal)_{\mu m,\nu n}=(\Lambda_\Ecal)_{mn,\mu\nu}$& & \\ \hline
  $K_n =$           & $\sqrt{\lambda_n}\, mat\ket{e_i}$ & Reshuffle($\Phi_\Ecal$) & $K_n$ \\
                    &    & $\sqrt{\lambda_n}\, mat\ket{e_i}$ &\\
  \hline
\end{tabular}
\end{center}
\caption{Relationships between mathematical representations of CP maps.}
\label{tab:cpreps}
\end{table}

\section{System-Environment Model}       \label{ssec:sys-env}

Our final approach for modeling open quantum systems is built on an intuitive picture of the joint system comprised of our \emph{system} of interest, and the \emph{environment} it is interacting with. We will denote these by quantum systems $S$ and $E$, respectively. We also assume that the state space of environment is of the same dimension as the system~\cite{Nielsen:2000}. The composite system $SE$ is assumed to be a closed system and thus its dynamics are described by the postulates of quantum mechanics mentioned in Section~\ref{ssec:postulates}.

Suppose our system is in a state $\rho$. If we send this state into a black box~\footnote{A black box is a device which does \emph{something} to our system, and the specifics are irrelevant. All we are concerned with is is what we put in one side, and what comes out the other.} where it undergoes evolution while interacting with the environment, the final state $\Ecal(\rho)$ will in general not be related to the initial state $\rho$ by a unitary transformation. To describe the evolution of $\rho$ we must consider the unitary transformation of the whole system. We can then recover the transformed state of the system by performing the partial trace over the environment.  This is illustrated in Fig.~(\ref{fig:seevo}).

\begin{figure}[htbp]
\begin{center}
\includegraphics[width=0.45\textwidth]{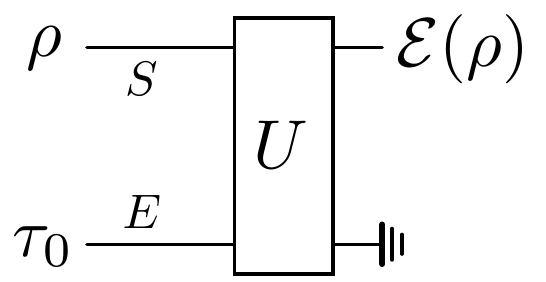}
\end{center}
\caption{Model of an open quantum system as a system interacting with its environment~\cite{Nielsen:2000}}
\label{fig:seevo}
\end{figure}

To do this we first suppose the environment is initially in a state $\tau_0$. We also assume that the there are no correlations between the system and the environment, and hence the state of the combined system is given by the $\rho\otimes\tau_0$. This is a crucial assumption. The final state of the system is then given by:
\begin{equation}
\Ecal(\rho)=\tr_E \left[ U (\rho\otimes\tau_0)U^\dagger \right].
\label{eqn:seevo}
\end{equation}
This leads us to the following theorem
\begin{theorem}\label{thm:productcp}
For an open quantum system $S$ with environment $E$, if the initial state on the joint system $SE$ is a product state, then the quantum operation $\Ecal$, as described in Eqn.~(\ref{eqn:seevo}), will always \emph{always} be CPTP.
\end{theorem}
The proof of this is a shown in Appendix~\ref{app:pstatecp}, and is a matter of constructing a Kraus representation for the evolution.

While this method provides a nice physical picture for the evolution of an open system it is necessary to know details of the environment and the systems interaction with it. Another limitation of this approach is that we have assumed that the composite system is in a product state. This generally requires the experimenter to perform a preparation procedure to remove any correlations between the system and environment. The implications of this are discussed in  Chapter~\ref{chap:stateprep}.

If we now consider a more general situation where there are initial correlations between the system and environment, which is to say the initial state of the joint system may be entangled, or the state of the environment somehow depends on the state of the system, then the formalism of quantum operations as CP maps breaks down~\cite{Pechukas:1994, Pechukas:1995, Sudarshan:2008, Terno:2008, Shaji:2005, Jordan:2004, Salgado:2004, Hayashi:2003, Stelmachovic:2001}. In these cases one cannot construct a Kraus representation, and the process matrix will have negative eigenvalues. There are no general mathematical representations of non-completely positive maps~\cite{Terno:2008}.

This completes our review of the mathematics used in the standard formalism for describing the evolution of open quantum systems. In the next chapter we consider the nature of initial correlations between a system and environment, which can cause the presented formalism to break down. 


\chapter{Non-Completely Positive Maps}       \label{chap:correlations}

In the previous two chapters we introduced several mathematical tools needed to describe the state and evolution of open quantum systems. Open quantum systems were assumed to interact with an environment, and the system together with its environment was treated as a closed system. The most general description of the state of a system was given by its density matrix, and evolution was described by a completely positive trace preserving map which acts on the density matrices. We mentioned that the CP formalism can break down in the presence of initial correlations between an open system and its environment. In this chapter we investigate such correlations between two quantum systems, and their effect on quantum evolution.

\section{Separable and Entangled States} \label{sec:entangled}

As was introduced in Section~\ref{ssec:sys-env}, we will be considering  $S$ which is interacting with an environment system $E$ of equal dimension. We will specifically deal with the case of qubits ($d=2$), though these results can be generalized to higher dimensional systems. Recall from Section~\ref{sec:compsystems} that any state of the joint system $SE$ can be written in the form
\[
\rho_{SE}=\frac{1}{4}\sum_{i=1}^3\sum_{j=1}^3\left( \id\otimes\id
+ \alpha_i\sigma_i\otimes\id + \beta_j\id\otimes\sigma_j
+ \gamma_{ij}\sigma_i\otimes\sigma_j\right),
\]
where $\sigma_i$ are the Pauli matrices. The states of the reduced systems are given by
\[
\rho_S=\frac{1}{2}\left( \id+\sum_{i=1}^3\alpha_i\sigma_i\right),
\quad\mbox{and}\quad
\rho_E=\frac{1}{2}\left( \id+\sum_{j=1}^3\beta_j\sigma_j\right).
\]

We have mentioned three classes for categorizing the correlations of joint state $\rho_{SE}$. These are

\begin{enumerate}
\item \textbf{Simply separable:} $\rho_{AB}=\rho_{A}\otimes\rho_{B}$.
\item \textbf{Separable:} $\rho_{AB}=\sum_i p_i\rho^A_i\otimes\rho^B_i$. \quad Here $\rho_{AB}$ is a convex-linear sum of simply separable states.
\item \textbf{Entangled:} $\rho_{AB}$ is not separable.
\end{enumerate}

Traditionally, it was thought that \emph{classical} correlations (where the system and environment are separable) give rise to CP dynamics, while \emph{quantum} correlations (where system and environment are entangled) could give rise to non-CP dynamics~\cite{Werner:1989}. However, while simply separable systems always have CP dynamics, it has been shown that non-entangled systems can still give rise to non-CP dynamics~\cite{Sudarshan:2008,Terno:2008}. We will illustrate this with an example.

\subsection{Example of a Non-Completely Positive Map} \label{ssec:ncpex}

We will use an example from~\cite{Terno:2008}, where the state of a joint system $SE$ is dependent \emph{only} on the reduced state of system $S$. The initial state of a two qubit system is
\begin{equation}
\rho_{SE}=\frac{1}{4}\left(\id_{SE}
+\sum_{i=1}^3 \alpha_i \sigma_i\otimes\id
+ a\sum_{i=1}^3 \sigma_i\otimes\sigma_i \right), \label{eqn:ternoex}
\end{equation}
where $a$ is a fixed parameter.
In this case $\rho_{SE}\ge0$ if $0\le |a| \le \frac{1}{3}\left(\sqrt{4-3|\alpha|^2}-1\right)$, so if we fix the value of $a$, only certain reduced states of $S$ will give a physical state of the joint system. In addition the state $\rho_{SE}$ is always separable for $a>0$. This is a case of an \emph{assignment map} which we will formally introduce momentarily.

If we evolve this state by a two-qubit unitary rotation
\[
U=\left(
    \begin{array}{cccc}
      1 & 0 & 0 & 0 \\
      0 & \cos\theta & \sin\theta & 0 \\
      0 & -\sin\theta & \cos\theta & 0 \\
      0 & 0 & 0 & 1 \\
    \end{array}
  \right).
\]
This state and unitary evolution lead to non-CP evolution on the first system~\cite{Terno:2008}. This can be seen by the eigenvalues of the process matrix taking negative values for certain values of $\theta$. The eigenvalues for a fixed value of $a=0.2$ is shown in Fig.~(\ref{fig:ternoex}).

\begin{figure}[ht]
\begin{center}
\includegraphics[width=0.50\textwidth]{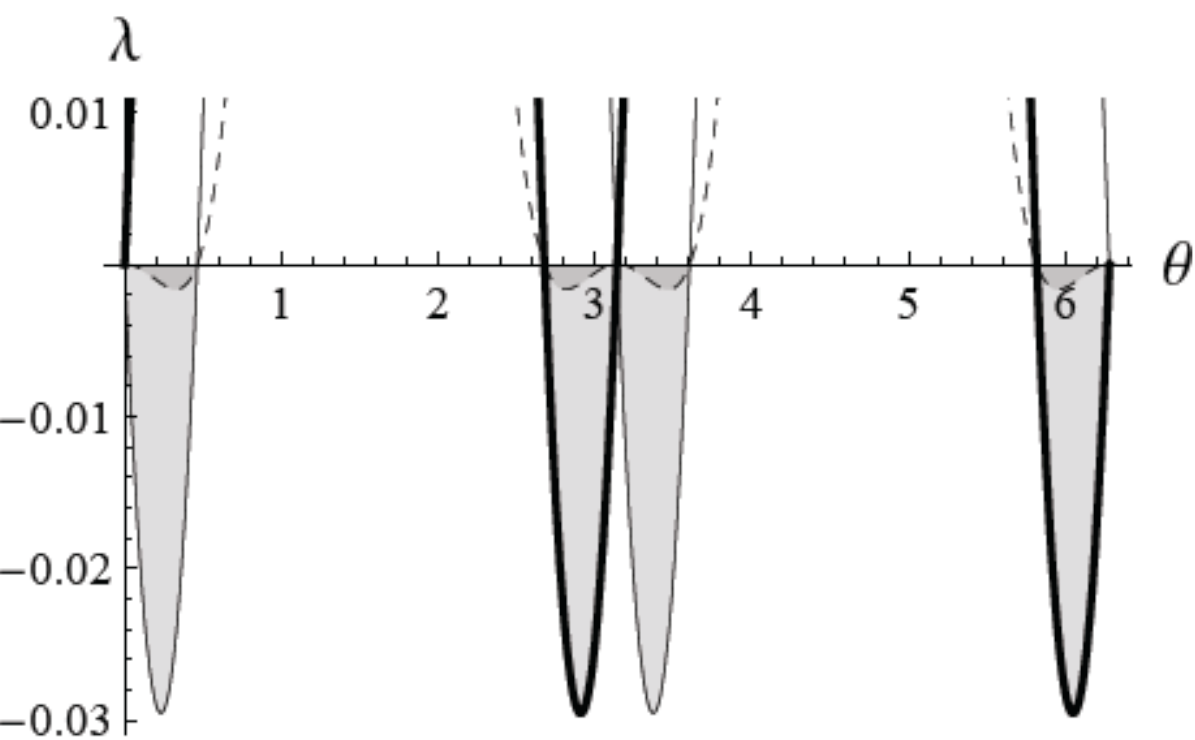}
\caption{Three of the eigenvalues of the process matrix for evolution of $\rho_{AB}$ by $U(\theta)$. Reproduced from \cite{Terno:2008} with permission.}
\label{fig:ternoex}
\end{center}
\end{figure}

This indicates that using entanglement as the definition of quantum correlations is too restrictive. A more recent formalism for describing quantum correlations in terms of what is known as \emph{quantum discord} will be discussed in Section~\ref{sec:discord}. However, first we must describe \emph{assignment maps} and how they can give a physical meaning to non-CP dynamics.

\section{Describing Non-CP Dynamics by Assignment Maps} \label{sec:assmap}

One approach to characterize quantum operations, CP and non-CP, is with the notion of an \emph{assignment}~\cite{Alicki:1995} or \emph{extension}~\cite{Kuah:2005} map. Let us consider an open quantum system $S$ interacting with an environment system $E$. In the approach used in Section~\ref{ssec:sys-env} we started with a state of the joint system $\rho_{SE}$ with the property $\rho_S=\tr_E(\rho_{SE})$, and considered its unitary evolution. Any correlations between the system and environment were encompassed in the initial state $\rho_{SE}$. We now take the opposite approach and now start with a state $\rho_S$ of system $S$ only. We then wish to \emph{embed} this state into a state on the combined system. This is achieved by an \emph{assignment} map: a map $\Acal$ from a density matrix of system $S$ to a density matrix of the combined system $SE$ given by
\begin{equation}
\Acal: \rho_S \mapsto \Acal(\rho_S)=\rho_{SE},\label{eqn:assignmentmap}
\end{equation}
where we still require that $\tr_E(\rho_{SE})=\rho_A$. Any correlations between the system and environment are now a accounted for in the assignment map $\Acal$.

When using assignment maps in our description, any trace preserving quantum operation $\Ecal$ on system $A$ can be expressed as the composition of an assignment map $\Acal$ and a CP map $\Fcal$ by
\[
\Ecal(\rho) = \tr_E\left(\Fcal\circ \Acal(\rho)\right).
\]
$\Fcal$ is simply the unitary evolution of the composite system $\Fcal\left(\Acal(\rho)\right)= U \left(\Acal(\rho)\right) U^\dagger$. This is almost the same as Eqn.~(\ref{eqn:seevo}) used to describe the evolution in our system-environment model for CP quantum operations, however the assignment map means the state of the composite system is not restricted to being a product state. In this case $\Ecal$ is a CP map if and only if the assignment map $\Acal$ is CP~\cite{Kuah:2005}.

The problem with assignment maps is that their definition is ambiguous, there is no restriction on how we embed our system into the combined system. This has been a cause of controversy in relation to using non-CP maps to describe physical processes~\cite{Pechukas:1995,Alicki:1995}. We need a method for characterizing which non-CP maps are physically relevant.

A definition for a non-CP map $\Ecal$ being physically \emph{accessible}~\cite{Terno:2008} is that there must exist an assignment map $\Acal$ such that $\Ecal(\rho)=\tr_E[U\Acal(\rho)U^\dagger]$, where $\Acal$ need only act on a finite-volume subset of the state space of $S$. In addition a non-CP map only has a physical meaning when acting on its \emph{domain of positivity}, where $\Ecal(\rho)\ge0$. Physically relevant maps should also be able to be identified by \emph{quantum process tomography}, which is a technique for characterizing quantum operations that we introduce in Chapter~\ref{chap:tomography}.

Terno \emph{et al.}~\cite{Terno:2008} define two main classes of assignment maps which are \emph{physically accessible}: Linear, and non-linear. These give rise to linear and non-linear system-environment correlations respectively. The linear case is the simplest scenario where the state of the environment is independent of the state of the system. For example $\Acal(\rho)=\gamma_{(\rho)}$ where $\tr_E[\gamma_{(\rho))}]=\rho$ but $\tr_S[\gamma_{(\rho)}]=\tau_0\, \forall \rho$ This type of assignment map will always lead to CP evolution as the state of the environment is independent of the state of the system~\cite{Terno:2008}.

In the non-linear case the state of the environment may be a function of the input state $\rho$. Not much is known about the non-linear case other than such assignment maps often lead to non-CP dynamics~\cite{Terno:2008}. We will show this with example in Section~\ref{sec:tomoex}, when what is effectively a bilinear assignment results in a non-CP quantum operation.

\section{Quantum Discord} \label{sec:discord}

Our example in Section~\ref{ssec:ncpex} showed that even separable, or classically correlated, initial states can give rise to non-CP evolution. This suggests that taking classical and quantum correlations to be synonymous with separable and entangled states respectively is inadequate. A new method for defining quantum and classical correlations was proposed by Ollivier and Zurek~\cite{Ollivier:2002}, and independently by Henderson and Vedal~\cite{Henderson:2001}. This method, which we will now introduce, is known as \emph{quantum discord} and uses two inequivalent quantum versions of the classically equivalent formulas for the \emph{mutual information} that quantifies the correlations between two systems. We begin by briefly reviewing some classical information theory.

\subsection{Classical Information Theory}       \label{ssec:cinfo}

In classical information theory~\cite{Cover:1991} the Shannon entropy is a measure of the ignorance, or missing information, about a random variable $A$. It is given by
\[
H(A)=-\sum_a p(A=a)\log p(A=a),
\]
where $p(A=a)$ is the probability of $A$ taking the value $a$. If we now consider two random variables $A$ and $B$, the conditional entropy of $A$ given $B$ is
\begin{eqnarray*}
H(A|B)&=&\sum_b P(B=b)H(A|B=b) \\
&=& \sum_b P(B=b) \,\sum_a P(A=a|B=b)\log P(A=a|B=b),
\end{eqnarray*}
where $P(A=a|B=b)$ is the probability of $A$ having value $a$ given that we know $B=b$.

The correlations between $A$ and $B$ are then measured by the \emph{mutual information}
\begin{equation}
\Jcal(A:B)=H(A)-H(A|B).     \label{eqn:Jclass}
\end{equation}
In all expressions the probabilities are derived from the joint probability distribution $P(A,B)$.

It is possible to formulate an equivalent expression for Eqn.~(\ref{eqn:Jclass}). By using Bayes theorem one can rewrite the conditional entropy as $H(A|B)=H(A,B)-H(B)$~\cite{Cover:1991}. This gives us a second expression for the classical mutual information
\begin{equation}
\Ical(A:B)=H(A)+H(B)-H(A,B).         \label{eqn:Iclass}
\end{equation}

Our next step is to generalize the expressions $\Jcal(A:B)$ and $\Ical(A:B)$ to quantum systems.

\subsection{Quantum Expressions for Mutual Information}     \label{ssec:mutualinfo}

To generalize mutual information to quantum systems~\cite{Nielsen:2000}, the random variables now represent the states of quantum systems $A$ and $B$. The equivalent of the joint probability distribution is the density matrix of the state of the combined system, $\rho_{AB}$. The reduced density matrices $\rho_A=\tr_B(\rho_{AB})$ and $\rho_B=\tr_A(\rho_{AB})$ for systems $A$ and $B$ respectively are the equivalent expressions for the probability distributions of the individual systems.

Our measure of ignorance about the state $\rho$ of a system is then given by the von Neumann entropy
\begin{eqnarray}
S(\rho)&=&-\tr(\rho\log\rho)\\
        &=&\sum_i -\lambda_i\log\lambda_i
\end{eqnarray}
where $\{\lambda_i\}$ are the eigenvalues of $\rho$, and we define $0\log0\equiv0$ if zero is an eigenvalue of $\rho$. The logarithms are taken to base 2.

A quantum expression for Eqn.~(\ref{eqn:Iclass}) is then obtained by replacing the Shannon entropies with their equivalent von-Neumann expressions. So the quantum mutual information is
\begin{equation}
\Ical_{A:B}(\rho_{AB})=S(\rho_A)+S(\rho_B)-S(\rho_{AB}).      \label{eqn:Iquant}
\end{equation}

In the case of Eqn.~(\ref{eqn:Jclass}) however, the generalization is not trivial. In the case of conditional entropy, to express the state of system $A$ given that we know the state of system $B$ requires us to perform a set of measurements on system $B$. This is done by a complete set of 1-dimensional orthogonal projectors $\{\Pi_i^B\}$ acting on the system $B$. If the outcome corresponding to a measurement of $\Pi_i^A$ is detected, the state of the joint system will be given by
\begin{equation}
\rho_{A|\Pi_i^B}=\frac{1}{p_i}\Pi^B_i\rho_{AB}\Pi^B_i,       \label{eqn:conddensity}
\end{equation}
where $p_i=\tr(\Pi^B_i\rho_{AB})$ is the probability of the $i^{th}$ outcome.

The generalized expression for the conditional entropy is then obtained by averaging the entropy over the post-measurement states. That is
\begin{equation}
S(\rho_{AB}|\{\Pi^B_i\})=\sum_i p_i \,S(\rho_{A|\Pi_i^B}).      \label{eqn:condentropy}
\end{equation}

Our second quantum expression for mutual information gives a measure of the information about system $A$ which can be obtained by measuring system $B$. It is given by
\begin{equation}
\Jcal_{A:B}(\rho_{AB})_{\{\Pi_i^B\}}=S(\rho_A)-S(\rho_{AB}|\{\Pi^B_i\}),
\end{equation}
and depends not only on the joint system state $\rho_{AB}$, but also on the measurement set $\{\Pi_i^B\}$.

The difference between the quantum generalizations of mutual information gives us a a measure of the non-classical correlations of a joint state. This difference depends on the choice of projectors used in $\Jcal$. In general we are interested in the minimum value of such a quantity, so we define the quantum discord to be minimum difference over all possible choices of projectors,
\begin{eqnarray}
\Dcal_{A:B}(\rho_{AB})&\equiv&\min_{\{\Pi_i^B\}}
\left[\Ical_{A:B}(\rho_{AB})-\Jcal_{A:B}(\rho_{AB})\right].     \label{eqn:discordab} \\
&=& \min_{\{\Pi_i^B\}}
\left[ S(\rho_B) - S(\rho_{AB})+\sum_i p_i S(\rho_{A|\Pi_{i}^B})\right] \nonumber
\end{eqnarray}

Some properties of quantum discord are
\begin{enumerate}
\item Quantum discord is a non-negative quantity, i.e. $\Dcal_{A:B}(\rho_{AB})\ge 0$.
\item For a state $\rho_{AB}$, $\Dcal_{A:B}(\rho_{AB})=0$ if and only if there exists a set of orthogonal projectors $\{\Pi_i^B\}$ acting on system $B$ such that
    \begin{equation}
    \rho_{AB}=\sum_i \Pi_i^B\rho_{AB}\Pi_i^B=\sum_i \tau_i\otimes\Pi_i,   \label{eqn:qddecomp}
    \end{equation}
    where $\tau_i=\tr_B[\Pi_i^B\rho_{AB}\Pi_i^B]$.
\end{enumerate}
Proofs of these results can be found in~\cite{Ollivier:2002}.

Now that we are done with the mathematical derivation of quantum discord we move on to discuss its physical significance.

\subsection{Quantum Discord in Open Quantum Systems}       \label{ssec:discord}

Quantum discord provides us with a quantitative measure of the non-classical correlations in a bipartite system. In a bipartite system, if we can obtain information from one system by performing measurements on the other then the systems are correlated. The extreme case of this is quantum entanglement. However, we can have non-entangled states which still exhibit this property characterized by a discord greater than zero.

For example, consider a bipartite system of two qubits in the state
\[
\rho=\frac{1}{4}\left( \rho_H\otimes\rho_D + \rho_V\otimes\rho_A
+\rho_D\otimes\rho_V+\rho_A\otimes\rho_H \right),
\]
where $\rho_H=\op{H}{H}, \rho_V=\op{V}{V} \rho_D=\op{D}{D}$ and $\rho_A=\op{A}{A}.$
This state is separable, and the reduced density matrices of both systems are maximally mixed. We consider orthogonal projectors of the form $\Pi^B_i=\id\otimes\op{\psi_i}{\psi_i}$ where
\begin{eqnarray*}
\ket{\psi_0}    &=& \cos(\theta)\ket{0}+e^{i\varphi}\sin(\theta)\ket{1},\\
\ket{\psi_1}    &=& \sin(\theta)\ket{0}-e^{i\varphi}\cos(\theta)\ket{1}.
\end{eqnarray*}

A plot of $\Ical_{A:B}-\Jcal_{A:B}$ of $\rho$ as a function of $\theta$ and $\varphi$ is shown in Fig.~(\ref{fig:discordex}). One choice of projectors for which the quantity is minimum is $\theta=\varphi=0$. In this case we can calculate the discord to be
\[
\Dcal_{A:B}(\rho)=\frac{3}{4}\log\left(\frac{3}{4}\right)\approx 0.33128.
\]
%
%
\begin{figure}[htbp]
\begin{center}
\subfigure[Separable state $\rho$]
    {\label{fig:discordex}
       \includegraphics[width=0.45\textwidth]{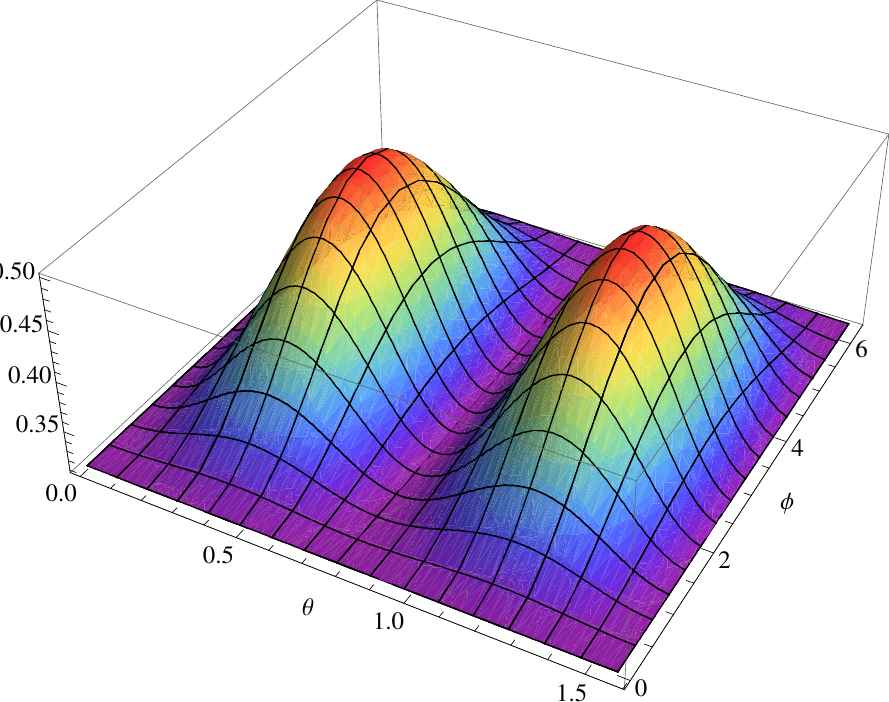}}
\subfigure[Maximally entangled state $\op{\phi}{\phi}$]
    {\label{fig:belldiscord}
        \includegraphics[width=0.45\textwidth]{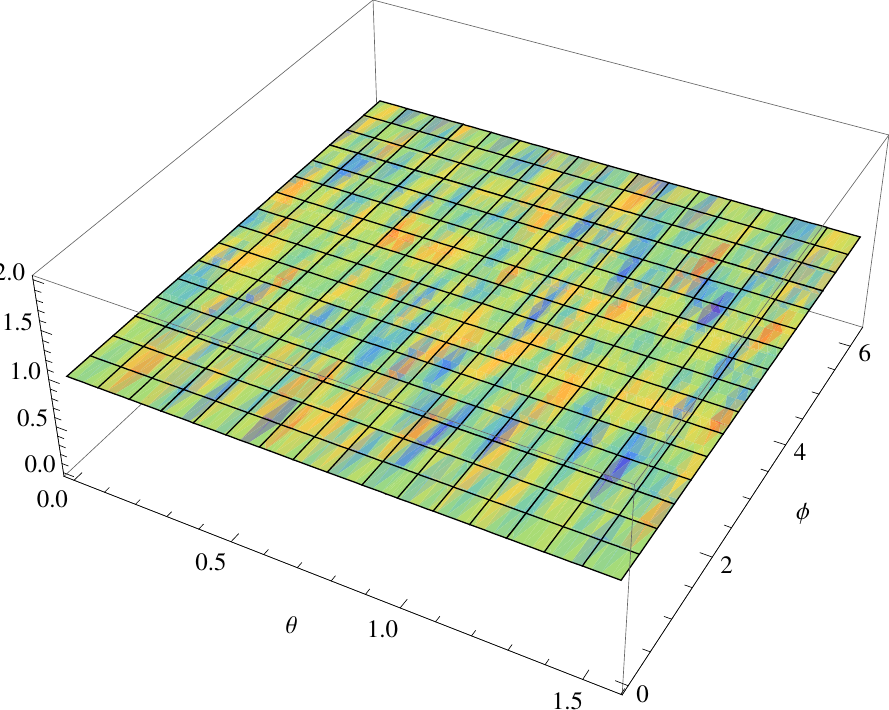}}
\caption{$\Ical_{A:B}-\Jcal_{A:B}$ a function of projectors for the separable state
$\rho= \frac{1}{4}\left( \rho_H\otimes\rho_D + \rho_V\otimes\rho_A
+ \rho_D\otimes\rho_V + \rho_A\otimes\rho_H \right)$ and a maximally entangled Bell state $\ket{\phi}=\frac{1}{\sqrt{d}}(\ket{00}+\ket{11})$.}
\label{fig:discord2ex}
\end{center}
\end{figure}

In the extreme case of a maximally entangled state, for example $\rho_{\phi}=\op{\phi}{\phi}$ where $\ket{\phi}=\frac{1}{\sqrt{2}}(\ket{00}+\ket{11})$, we have that $\Dcal_{A:B}(\rho_{\phi})=1$ for all values of $\theta$ and $\varphi$. This is shown in Fig.~(\ref{fig:belldiscord}).
%

Ollivier \& Zurek~\cite{Ollivier:2002} argue that quantum discord should be used to define what we mean by classical and quantum correlations. The term \emph{classical correlations} should apply to systems with zero quantum discord, and \emph{quantum correlations} to systems with $QD > 0$. In their paper they reverse the order of the systems in their definition of quantum discord when consider a bipartite open quantum system $AB$ consisting of a principle system $A$ interacting with an environment $B$. This means the discord is given by
\begin{equation}
\Dcal_{B:A}(\rho_{AB})=\min_{\{\Pi_i^A\}}
\left(\Ical_{B:A}(\rho_{AB})-\Jcal_{B:A}(\rho_{AB})\right), \label{eqn:discordba}
\end{equation}
where the set of projectors $\{\Pi^A_i\}$ being minimized over act on system $A$. In addition we now have $\Dcal_{B:A}(\rho_{AB})=0$ if and only if there exists a set of orthogonal projectors $\{\Pi_i^A\}$ acting on system $A$ such that
\begin{equation}
\rho_{AB}=\sum_i \Pi_i^A\rho_{AB}\Pi_i^A=\sum_i \Pi_i\otimes\tau_i,   \label{eqn:qddecompb}
\end{equation}

We shall differentiate between the two expressions with the subscripts $A:B$ and $B:A$. Their argument for this is that in general one does not have access to the environment, and thus measurements should be performed over the system $A$.

The main result of~\cite{Sudarshan:2008} is the following theorem.
\begin{theorem}\label{thm:sudarshan}
The reduced evolution of any classically correlated bipartite system \newline($\Dcal_{B:A}(\rho_{AB})=0$) is always completely positive.
\end{theorem}
The original proof for this is given in~\cite{Sudarshan:2008}, however the notation used is not particularly clear. We produce our own proof for this theorem in Appendix~\ref{app:discordbacp} in what we believe to be clearer notation.

The presence of quantum correlations allows for the possibility of non-CP quantum operations. However, while the presence of quantum correlations is found to be \emph{necessary} for non-CP dynamics, it is not \emph{sufficient}. While this provides a convenient method for determining situations where CP evolution is guaranteed, and extends the known class of systems which will always have CP evolution from simply separable to those with classical correlations, it does not help us deal with systems which may potential have non-CP evolution. We have found some interesting issues which arise from the fact that quantum discord is not symmetric.

\subsection{Asymmetry of Quantum Discord}       \label{ssec:aysmdiscord}

The definition of quantum discord in the literature is ambiguous. In Ollivier and Zurek's original derivation~\cite{Ollivier:2002} we are simply considering two systems $A$ and $B$, the choice of orthogonal projectors used in generalizing the condition entropy to quantum systems has the measurements carried out over system $B$, which gives us the definition of Eqn.~(\ref{eqn:discordab}). This approach has been used in several other following studies~\cite{Datta:2008,Luo:2008}. Alternatively we could take the approach used by Sudarshan \emph{et al.}~\cite{Sudarshan:2008} reversing the order of the systems in the definition giving us an expression for discord as defined by Eqn.~(\ref{eqn:discordba})

The issue here is that quantum discord is not a symmetric quantity, in general $\Dcal_{A:B}(\rho)\ne\Dcal_{B:A}(\rho)$. More importantly, a discord of zero in one direction does not imply zero discord in the reverse. We will show this by a counter example.

Suppose $\Dcal_{A:B}(\rho^{AB})=0$. This implies that there exists a set of orthogonal projectors $\{\Pi^B_i\}$ on $B$ satisfying Eqn.~(\ref{eqn:qddecomp}). If this also implied that $\Dcal_{B:A}(\rho^{AB})=0$, then there must also exist a set of orthogonal projects $\{\Pi_i^A\}$ acting on system $A$ which satisfy Eqn~(\ref{eqn:qddecompb}). Consider the case where $\rho^{AB}=\frac{1}{2}\left(\rho_H\otimes\rho_H+\rho_D\otimes\rho_V\right)$.
It is easy to see the orthogonal projectors $\Pi_0^B=\rho_H, \Pi_1^B=\rho_V$ satisfy Eqn.~(\ref{eqn:qddecomp}) and hence $\Dcal_{A:B}(\rho^{AB})=0$.

Now suppose there exists $\Pi_0^A, \Pi_1^A$ satisfying Eqn~(\ref{eqn:qddecompb}). For this to be true we require that
\begin{eqnarray}
\Pi_0^A\rho_H\Pi_0^A+\Pi_1^A\rho_H\Pi_1^A&=&\rho_H,\\
\Pi_0^A\rho_D\Pi_0^A+\Pi_1^A\rho_D\Pi_1^A&=&\rho_D.
\end{eqnarray}
However $\rho_H$ and $\rho_D$ are not orthogonal, so such a set of projectors cannot exist, hence $\Dcal_{B:A}(\rho^{AB})\ne0$. In fact a numerical computation gives the value to be $\Dcal_{B:A}(\rho^{AB})=0.2018$. Fig.~(\ref{fig:discordABBA}) shows how $\Dcal_{A:B}$ and $\Dcal_{B:A}$ vary as function of the projectors used. Here we can also see that the distributions are different.

\begin{figure}[htbp]
\begin{center}
\subfigure[$\Dcal_{A:B}(\rho)$]
    {\label{fig:discordAB}
       \includegraphics[width=0.45\textwidth]{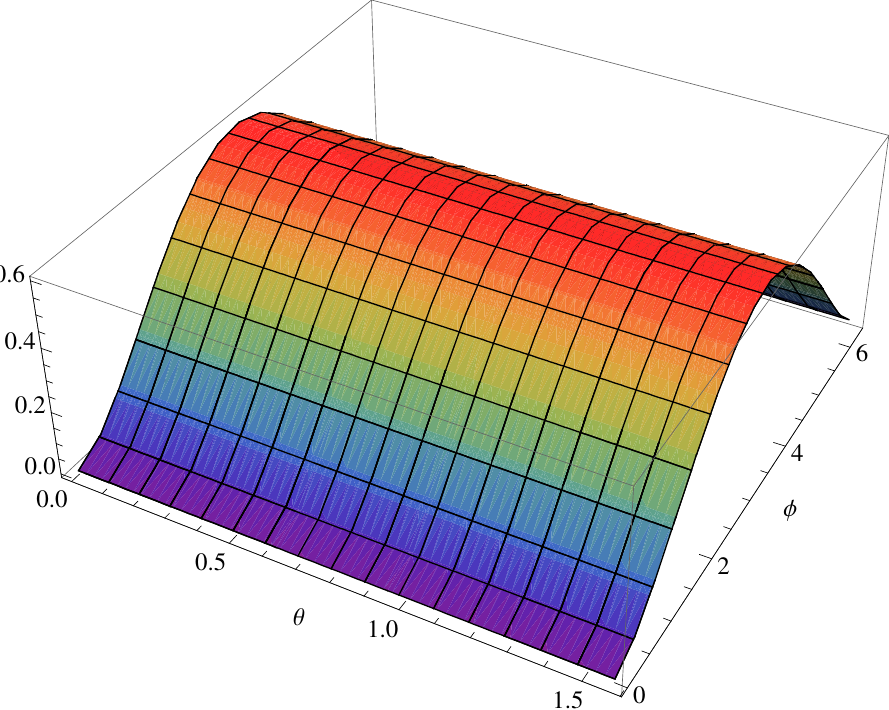}}
\subfigure[$\Dcal_{B:A}(\rho)$]
    {\label{fig:discordBA}
        \includegraphics[width=0.45\textwidth]{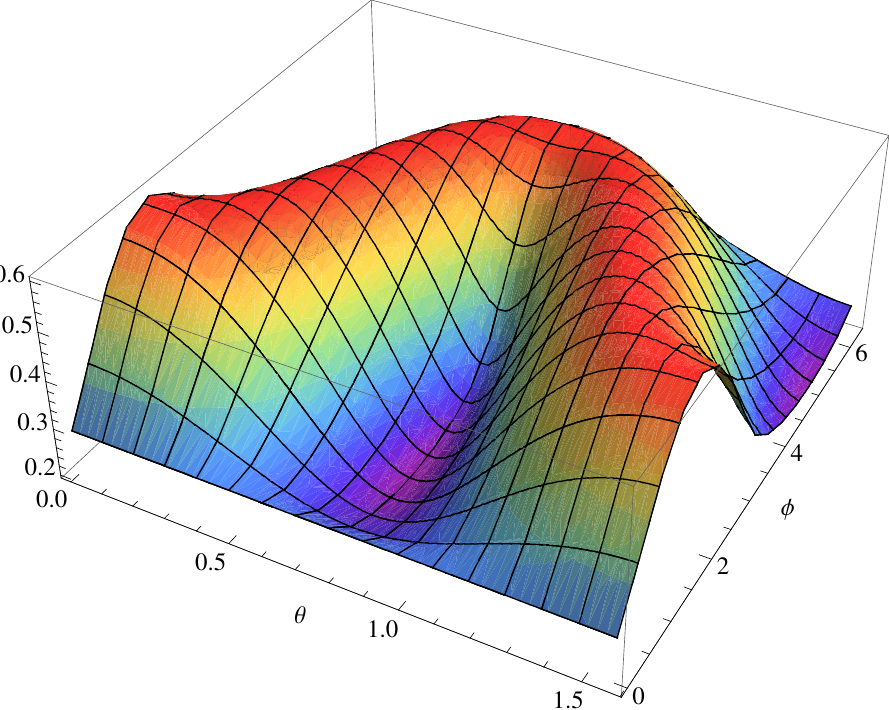}}
\caption{$\Ical-\Jcal$ as a function of projectors for the state
$\rho=\frac{1}{2}\left( \rho_H\otimes\rho_H + \rho_D\otimes\rho_V\right)$.}
\label{fig:discordABBA}
\end{center}
\end{figure}

This presents us with a particular problem in the physical interpretation of quantum discord. If we are to call two systems classically correlated when there is zero discord between them, which definition do we use? One motivation, which is implicit in the interpretation of Sudarshan \emph{et al.}~\cite{Sudarshan:2008}, is that in the case of an open system $A$ interacting with an environment $B$, we can not in general make any measurements of the environment. Thus the definition with projective measurements acting on the system gives us a measure of how much information we can deduce about the environment by only measuring the system.

An alternative to this approach is to ask the question how much information can the environment know about the system? In this case it is the reverse $\Dcal_{A:B}$ which is important, as it reveals how much information about $A$ can be obtained by measurement on its environment.

Each approach is supported by a valid argument and thus we are left with a unique problem. In the case of our example, according to $\Dcal_{A:B}=0$ we have a classically correlated state, yet $\Dcal_{B:A}\ne0$ implies that the state in fact has quantum correlations. Our state \emph{should} be either one or the other, unless \emph{correlations in quantum mechanics are directional}. An important implication of this is in generalizing Thm.~(\ref{thm:sudarshan}).

We propose the following conjecture:

\begin{conjecture} \label{conj:discord}
If an initial state $\rho$ of a bipartite system $AB$ is classically correlated, as defined by $\Dcal_{A:B}(\rho)=0$, yet also has $\Dcal_{B:A}(\rho)\ne0$ (so the correlations are not symmetric), its evolution need not be completely positive.
\end{conjecture}

Our basis for this conjecture is as follows: We try and follow a similar argument to the proof for Thm.~(\ref{thm:sudarshan}) from Appendix~\ref{app:discordbacp}, but using the condition from Eqn.~\ref{eqn:qddecomp} for $\Dcal_{A:B}(\rho)=0$,
\[
\Dcal_{A:B}(\rho)=0 \Longrightarrow \rho=\sum_i \Pi_i^B \rho \Pi_i^B
= \sum_i p_i \tau_i\otimes\Pi_i.
\]
Since the projectors $\Pi_i$ are orthogonal pure states we can write them as $\Pi_i=\op{i}{i}$. Evolution of the joint system by an arbitrary unitary operator $U$ then leads us to
\[
\Ecal(\rho)=U\rho U^\dagger =
\sum_{i,j,n} p_i
\left(\sum_m \bra{n}^B U\ket{m}^B\delta_{mj}\right)\tau_i\left(\sum_m \delta_{mj}\bra{m}^B U^\dagger\ket{n}^B\right)
\]
Now, for each $\tau_i$ there exists a set of projectors $X_{i,j}$ such that
$\tau_i=\sum_j X_{i,j}\tau_i X_{i,j}=\sum_{j} q_{i,j}X_{i,j}$. In general each $X_{i,j}$ are not orthogonal and do not commute for different $i$. This means we use the properties of these orthogonal projectors to remove the dependence of the Kraus operators on $i$ as we did in the previous case. Hence it appears that in general the Kraus operators will be different for different states of the reduced system, suggesting that the evolution in general is not completely positive. Thus Thm.~(\ref{thm:sudarshan}) is only true for $\Dcal_{B:A}(\rho)=0$, it does not appear to generalize to the case of $\Dcal_{A:B}(\rho)=0$.

This indicates that the definition of classical and quantum correlations based on quantum discord is insufficient, as it is not symmetric with respect to the two systems $A$ and $B$. A possible avenue for future research into this result could be to consider what would happen to the definition of quantum discord if we allowed for more arbitrary measurement sets, such as projective operator valued measurements (POVM), or to consider a symmetrized version of the quantum discord.


\chapter{Quantum Process Tomography}    \label{chap:tomography}

We have now introduced the mathematics needed for describing evolution of open quantum systems by completely-positive trace-preserving maps, and also some methods of characterizing correlations of a joint system state. It was shown how certain initial correlations of an open system with its environment could lead to non-completely positive evolution.
In this chapter we will now introduce a technique for characterizing unknown quantum operations on open systems.  This technique is called \emph{quantum process tomography} and allows for the characterization of an unknown quantum process by measuring the output states for a relatively small known set of input states and using this data to reconstruct a process matrix for the evolution.

Several different schemes have been proposed to perform this task. We will investigate two of them in this chapter. The techniques are \emph{standard quantum process tomography}, and \emph{ancilla assisted process tomography}. But first, before we can properly describe the reconstruction of an unknown quantum process, we must introduce a method for reconstructing unknown quantum states.

\section{Quantum State Tomography} \label{sec:statetomo}
If we have a quantum system in an unknown state we require a technique to completely determine the density matrix which represents that state. Such as technique is \emph{quantum state tomography}.

In state tomography an experimenter reconstructs the density matrix by \emph{inverting} probability data obtained from measuring many identical copies of the unknown state with an appropriate set of measurements. Such a set of measurement operators, $\{M_i\}$, must have the property that \emph{any} density matrix of the system can be expressed as a linear combination of these operators, i.e. $\rho=\sum_i a_i M_i$. Such a set is called \emph{tomographically complete}, and forms a basis for the density matrices of the system.
%

In the case of a single qubit represented by the polarization state of a photon, a commonly used tomographically complete measurement set is $\{M_i=\op{i}{i}\}$, where $i=H,V,D,R$. These are the density matrices corresponding to the horizontal, vertical, diagonal, and right-circular polarization states of a photon respectively. Even though $M_i\equiv\rho_i$, as pure state density matrices are projective measurement operators, we will use $M$ to distinguish between states and measurements.

Recall from Section~\ref{ssec:postulates} that for an unknown state $\rho$, the probability of detecting the outcome corresponding to measurement $M_i$ is given by $p_i=\tr(\rho M_i)$. With an appropriate basis $\{D_i\}$ we can reconstruct the density matrix $\rho$ from these probabilities by~\cite{DAriano:2000}
\begin{equation}
\rho=\sum_i p_i D_i. \label{eqn:staterecon}
\end{equation}

The basis $\{D_i\}$ used for reconstruction is called the \emph{dual basis} for $\{ M_i \}$. It is defined by the orthogonality relationship
\begin{equation}
\tr\left(D_i^\dagger M_j\right)=\delta_{ij}. \label{eqn:dualdef}
\end{equation}
Calculations for the dual basis for input basis $\{M_H, M_V, M_D, M_R\}$ can be found in Appendix~\ref{app:duals}. For systems of multiple qubits we can form the tensor products of the measurement set for a single qubit and do joint measurements. For example, if $\rho$ is a 2-qubit state a tomographically complete measurement set will be $\{M_{ij}\}$ where $M_{ij}=M_i\otimes M_j$.

Since measurement in quantum mechanics affects the state of the system, we require many identical copies of an unknown state in order to accurately determine the probabilities $p_i$. If, for each measurement operation $M_i$, we prepare $N$ copies of a state $\rho$, the number of detections recorded, $N_i$, allows us to compute the probability coefficients in the standard way.

Since we are determining the relative probabilities by count statistics, statistical noise becomes a factor in reconstruction. This can lead to the reconstruction of unphysical states. In these cases optimization techniques are employed to optimize the reconstructed state to the closest physical one. This process is discussed in more detail in Section~\ref{ssec:mltomo}.

\section{Quantum Process Tomography} \label{sec:processtomo}
Now that we know how to characterize unknown quantum states we can move on to the characterization of quantum processes. First we shall describe the general motivation common to the various techniques of tomography.

Recall from Section~\ref{ssec:cpaxioms} that the evolution of an open quantum system is commonly described by a quantum operation: a linear, trace-preserving, completely positive map $\Ecal$ acting on the state space of the system. There were several mathematical representations presented in Section~\ref{sec:cpmaps} which allowed us to completely describe the evolution of an arbitrary state $\rho$ by such a CPTP map.

Now we consider the reverse situation. If we are presented with an unknown operation, but believe that it is CP, what technique might we employ to characterize it? One option is to send in known quantum states and measure their outputs. However, it is a practical impossibility to do this for all possible input states. A characterization technique which requires only a finite set of inputs is \emph{process tomography}.

In general we wish to find a mathematical representation for $\Ecal$ through experimental observations. The methods we will describe in the following sections are formulated in terms of the process matrix $\Lambda_\Ecal$ which uniquely describes $\Ecal$. The two methods we will consider are called \emph{standard quantum process tomography} (SQPT), and \emph{ancilla assisted process tomography} (AAPT). We will begin with SQPT.


\subsection{Standard Quantum Process Tomography} \label{sec:sqpt}
One of the earliest proposed methods for performing process tomography is what is now referred to as \emph{standard quantum process tomography} (SQPT)~\cite{Chuang:1997}. The idea of this scheme is shown in Fig.~(\ref{fig:sqpt}). An experimenter prepares a basis of input states, subjects them to an unknown operation $\Ecal$ and determines the output states by state tomography. This information is then used to reconstruct a process matrix for $\Ecal$.

\begin{figure}[htb]
\begin{center}
\includegraphics[width=0.7\textwidth]{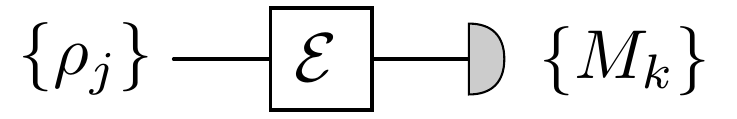}
\end{center}
\caption{Schematic of standard quantum process tomography}
\label{fig:sqpt}
\end{figure}

We start by choosing a basis of density matrices of the principal system $\{\rho_j\}$. This means that any density matrix $\rho$ can be uniquely expressed as a linear combination $\rho=\sum_{j}p_{j}\rho_{j}$. For a $d$-dimensional system, we require a set of $d^2$ linearly independent density matrices.

The choice of an input basis is not unique, and is often determined by experimental convenience. For single qubits in linear quantum optics experiments it is common to choose the states corresponding to horizontal, vertical, diagonal, and right-circular polarized photons. I.e. the basis $\{\ket{H},\ket{V},\ket{D},\ket{R}\}$. The corresponding density matrices for these states are:
\begin{eqnarray*}
\rho_{H} = \frac{1}{2}(\id+\sigma_{3}), &\quad&
\rho_{V}= \frac{1}{2}(\id-\sigma_{3})   \\
\rho_{D}=\frac{1}{2}(\id+\sigma_{1}), &\quad&
\rho_{R}=\frac{1}{2}(\id+\sigma_{2}).
\end{eqnarray*}
Notice that we chose the same basis for input states as for the state tomography measurement operators.

To construct the process matrix we use the approach of Kuah \emph{et al.}~\cite{Kuah:2007}. However, their notation is confusing and hard to generalize. We will formulate an equivalent description inline with the mathematics introduced in Chapter~\ref{chap:quantumops}.

Let $\Ecal$ be an unknown quantum operation acting on an open quantum system. We will denote our basis of input states for the system by $\{ \rho_{j}\}$. The output states will be denoted by $\{ \Ecal(\rho_{j})\}$.

We can then define a process matrix $\Lambda_{\Ecal}$ for our quantum operation by
\begin{equation}
\Lambda_{\Ecal}= \sum_{j} D_j^*\otimes\Ecal(\rho_{j}) \label{eqn:sqptprocess}
\end{equation}
where $D_j$ is the element of the dual basis corresponding to $\rho_j$, and $D_j^*$ designates the complex conjugate of the elements of $D_j$. See Appendix~\ref{app:sqptpmap} for the proof that this expression is equivalent to our original definition of the process matrix.

The output states are determined by performing state tomography as outlined in Section~\ref{sec:statetomo}. If we use the same basis for our input states and measurement set we can combine Eqns.~(\ref{eqn:staterecon}) and (\ref{eqn:sqptprocess}) to give

\begin{equation}
\Lambda_{\Ecal}= \sum_{ij} \tr(\rho_i M_j)\,(D_i^*\otimes D_j), \label{eqn:sqptfullprocess}
\end{equation}
and if our input and measurement sets differ each will just have a different dual basis. So the total number of different input state and measurement combinations is $(d^2)^2$.

From here we can use the equivalences between different representations of CP maps described in Section~\ref{sec:cpmaps} to convert to a Kraus representation, or superoperator if such a representation is preferred.

\subsection{Ancilla Assisted Process Tomography} \label{sec:aapt}
Another method of process tomography is \emph{ancilla assisted process tomography} (AAPT)~\cite{DAriano:2001,White:2003}. This method is mathematically equivalent to SQPT, and exploits the equivalence between quantum operations and quantum states of a larger system we introduced in Section~\ref{ssec:processmap}.

Recall from the Jamiolkowski isomorphism that for a quantum operation $\Ecal$ acting on a $d$-dimensional system $S$, the process matrix could be defined as
\begin{equation}
\frac{\Lambda_\Ecal}{d}=(\Ical\otimes\Ecal)(\op{\phi}{\phi}) =\rho_\Ecal, \label{eqn:aaptprocess}
\end{equation}
where $\ket{\phi}=\sum_j (\ket{j}\otimes\ket{j})/\sqrt{d}$ is a maximally entangled state of the joint system $AS$ where $A$ is an ancilla of the same dimension as $S$.

Hence if we introduce an ancilla $A$ and prepare the maximally entangled state on $AS$, we subject only half of this joint state to the operation $\Ecal$ and determine the output state $\rho_\Ecal$ by state tomography. This is shown in Fig~(\ref{fig:aapt}). We them obtain the process matrix simply by $\Lambda_\Ecal = d \rho_\Ecal$.

\begin{figure}[htb]
\begin{center}
\includegraphics[width=0.6\textwidth]{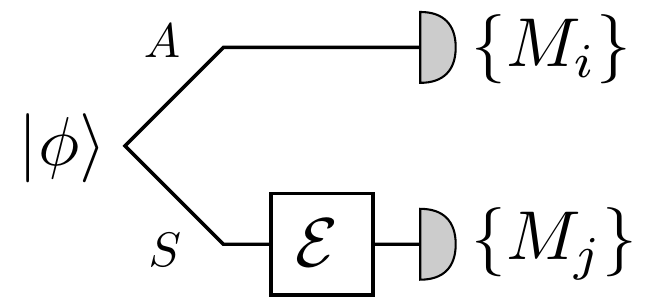}
\end{center}
\caption{Schematic of ancilla assisted process tomography}
\label{fig:aapt}
\end{figure}

AAPT has the advantage of only requiring the preparation of a single input state instead of the $d^2$ required for an $d$-dimension system in SQPT. However, since the input state must be a joint state of our principal system and an ancilla of equal dimension, the required number of measurements to perform state tomography on the output is increased. At the end the total number of measurements is the same as in the case of SQPT, $(d^2)^2$, as we must measure on the larger joint state space.

\subsubsection{AAPT With Alternative Input States}
Eqn.~(\ref{eqn:aaptprocess}) is only valid for the maximally entangled input state $\op{\phi}{\phi}$. Suppose we want to use another pure input state described by a state vector $\ket{M}=\sum_{i,j}m_{ij}\ket{i}\otimes\ket{j}$. Then using our relationship between vectors and matrices from Section~\ref{ssec:vecmat}, we have
\[
\ket{M}=\sqrt{d}\, M^T\otimes\id\ket{\phi} = \sqrt{d}\, \id\otimes M \ket{\phi},
\]
where $M=mat\ket{M}=\sum_{ij}m_{ij}\op{j}{i}$, and $\ket{\phi}\ket{\id}/\sqrt{d}$.

Thus the output state generated by $\Ical\otimes\Ecal$ and the pure input state $\rho_M=\op{M}{M}$ is
\begin{eqnarray*}
(\Ical\otimes\Ecal)(\rho_M)
    &=& (\Ical\otimes\Ecal)(\op{M}{M}) \\
    &=& (\sqrt{d}\,M^T\otimes\id)\left[(\Ical\otimes\Ecal)\op{\phi}{\phi}\right] (\sqrt{d}\,M^*\otimes\id)\\
    &=& (M^T\otimes\id)d\,\rho_\Ecal (M^*\otimes\id)\\
    &=& (M^T\otimes\id)\Lambda_\Ecal (M^*\otimes\id).
\end{eqnarray*}

Hence we can recover the process matrix by
\begin{equation}
\Lambda_\Ecal=(M^T)^{-1}\otimes\id \left[ (\Ecal\otimes\Ical)\rho_M\right](M^*)^{-1}\otimes\id.
\end{equation}
This is possible if and only if $M$ is invertible, ie. if and only if $\mbox{rank}(M)=d$. In terms of the vectorized matrix this is equivalent to requiring $Sch\ket{M}=d$, which means $\ket{M}$ must be entangled.

\subsubsection{AAPT With Separable Input States}
The method of AAPT just described is also referred to as \emph{entanglement assisted process tomography} due to the input being an entangled state. What happens if we wish to use a separable input state $\rho$? By the spectral decomposition, we can express this as the sum of pure states $\rho=\sum_k \op{M_k}{M_k}$ where $\ket{M_k}=\sum_{i,j} (M_k)_{ij}\ket{j}\otimes\ket{i}$. The output state of $\rho$ generated by $\Ecal$ is then
\begin{eqnarray}
(\Ical\otimes\Ecal)(\rho)
&=& \sum_k (M^T_k\otimes \id)\Lambda_\Ecal (M_k^*\otimes \id)\nonumber\\
&=& (\Mcal\otimes\Ical)(\Lambda_\Ecal),     \label{eqn:aaptmixed}
\end{eqnarray}
where $\Mcal(\rho)=\sum_k M^T_k \rho M^*_k$, so the process matrix is recoverable through Eqn.~(\ref{eqn:aaptmixed}) if and only if the map $\Mcal$ is invertible. An input state satisfying this is called \emph{faithful}.

While we need not use an entangled input state for AAPT it has been suggested in literature that these states are the most \emph{efficient}~\cite{DAriano:2001}. Here efficiency refers to requiring a smaller ensemble of copies of our input state to accurately determine the output though state tomography. This is due to statistical errors which result in quantum measurement, the nature of which are discussed in Section~\ref{sec:statetomo}.

A measure of the faithfulness of an input state $\rho_{AB}$ with maximal Schmidt number is given by~\cite{Mohseni:2008}
\[
F(\rho_{AB})=\tr(\rho_A^2)=\sum_{i} \lambda_i^2,
\]
where $\rho_a=\tr_B[\rho_{AB}]$ and $\lambda_i$ are the eigenvalues of $\rho_A$. This coincides with the definition of \emph{purity} in Section~\ref{sec:densitymatrix}. The only pure states with maximal Schmidt number are entangled.

\section{Non-Completely Positive Maps in Process Tomography} \label{sec:ncptomo}
We mentioned in Section~\ref{sec:statetomo} that non-physical output states often occur as a result of statistical noise in process state tomography. The effect of these reconstruction errors can lead to non-CP maps being observed in process tomography. The standard formalism for process tomography is to \emph{assume} that evolution must be completely positive. Then any deviations from complete positivity in our output state \emph{must} be the result of statistical and/or experimental errors. In this case a numerical optimization of the output state is done to enforce the requirement of a CP process map~\cite{White:2001,White:2004,White:2007}. A common implementation of this idea is \emph{maximum likelihood process tomography}, which we will discuss below.

However, there is a problem with this approach. As we have previously mentioned, and indeed showed by example in Section~\ref{ssec:ncpex}, non-CP maps can arise from the presence of initial correlations between a system and its environment. If we can have situations where a non-CP process map \emph{should} be observed through tomography, employing optimization techniques based on the assumption of complete positivity will give us an incorrect process matrix. This leaves us with a situation where we must distinguish between a non-CP process map arising due to statistical noise, and one arising due to initial correlations. This question will be the focus of Chapters~\ref{chap:stateprep} and \ref{chap:statsim}.

\subsection{Maximum Likelihood Process Tomography} \label{ssec:mltomo}

Before we move onto a discussion of state preparation in process tomography we will briefly describe the basic theory of \emph{maximum likelihood process tomography}. This will be important when we investigate the statistical distribution of reconstructed process matrices in Chapter~\ref{chap:statsim}.

In SQPT, for a given process map $\Lambda_\Ecal$ and input states $\{\rho_j\}$, the expected number of counts for $\Ncal$ measurements of an operator $M_m$ is
\begin{equation}
n_{jm}= \Ncal \tr_A\left[(\rho_j^T\otimes M_m)\Lambda_\Ecal \right].     \label{eqn:tomocounts}
\end{equation}

For optical experiments such as we will consider in Chapter~\ref{chap:linoptics}, the photon sources are typically spontaneous. This means they will produce photonic qubits at random intervals, though at an overall average rate. The count statistics in this case will be Poissonian~\cite{Kok:2007}. If we assume that the length of the experiment is sufficiently long, we can approximate the Poissonian distribution by a Gaussian distribution, with mean and variance both equal to $n_{jm}$~\cite{Feller:1966}. Under these assumptions, the probability of detecting a sequence of counts $\{n^e_{jm}\}$ is given by
\begin{equation}
P(\{n_{jm}^e\}|\Lambda_\Ecal)\propto \prod_{jm}\exp\left[-\frac{(n_{jm}^e-n_{jm}(\Lambda_\Ecal))^2}{2n_{jm}(\Lambda_\Ecal)} \right],
\label{eqn:likefunc}
\end{equation}
where $n_{jm}$ is the expected number of counts given by Eqn.~(\ref{eqn:tomocounts}), and $P$ and $n_{jk}$ are functions of $\Lambda_\Ecal$.

Eqn.~(\ref{eqn:likefunc}) is called the \emph{likelihood function}, and we wish to maximize this function over the variable $\Lambda_\Ecal$ to find the process matrix \emph{most likely} to give the detected sequence of counts. This is maximum likelihood estimation. However, it is generally easier to minimize the negative of the log-likelihood function, so the actual optimization task is
\begin{eqnarray*}
\mbox{minimize} && \sum_{jm} \frac{[n_{jm}^e-n_{jm}(\Lambda_\Ecal)]^2}{n_{jm}(\Lambda_\Ecal)}\\
\mbox{with constraints} &&  \Lambda_\Ecal\ge0 \quad\mbox{and}\quad \tr_B(\Lambda_\Ecal)=\id.
\end{eqnarray*}

It should be noted that optimizing this log-likelihood function is equivalent to a weighted least squares problem, however the weights depend on the process matrix. In addition the requirement that the process matrix correspond to a physical state give us our constraints.

It is possible to write this seemingly nonlinear optimization routine in the form of convex optimization called a \emph{semi-definite program}~\cite{Doherty:2006}, for which there are efficient numerical tools for solving~\cite{Boyd:2004}.

\chapter{State Preparation In Process Tomography}           \label{chap:stateprep}
To perform quantum process tomography, as outlined in Chapter~\ref{chap:tomography}, the use of specific input states was required. However, we made no assumptions as to how these states were prepared. In this chapter we investigate methods for preparing an arbitrary state of an open quantum system into the required input states for a tomography experiment. It is a particularly important issue if a system and environment are initially correlated since the evolution of the system may not always be completely positive. We investigate how state preparation can influence the complete positivity of system's evolution. This approach is motivated by the work of Kuah \emph{et al.}~\cite{Kuah:2007} which investigated two types of preparation procedures in SQPT, but both within the context of CP dynamics. We use their paper as a starting point to investigate the situations when non-CP maps can arise.

\section{State Preparation in SQPT} \label{sec:sqptprep}

As mentioned in Chapter~\ref{chap:tomography}, an unknown quantum operation $\Ecal$ can be completely characterized by the use of SQPT. To perform SQPT we require the preparation of a select set of input states. The idea of this procedure is shown in Fig.~(\ref{fig:stateprepsqpt}). If we have an open system $S$ with environment $E$, the initial state of the joint system, $\gamma_0$, may be correlated. State preparation is a procedure applied only to system $S$ to prepare $\gamma_0$ into the required tomographic inputs $\{\rho_j\}$. Following preparation, SQPT is performed as previously described. We will now review methods of state preparation proposed in the literature, and also propose our own scheme, and investigate how they fare in the presence of initial correlations.

\begin{figure}[htb]
\begin{center}
\includegraphics[width=0.8\textwidth]{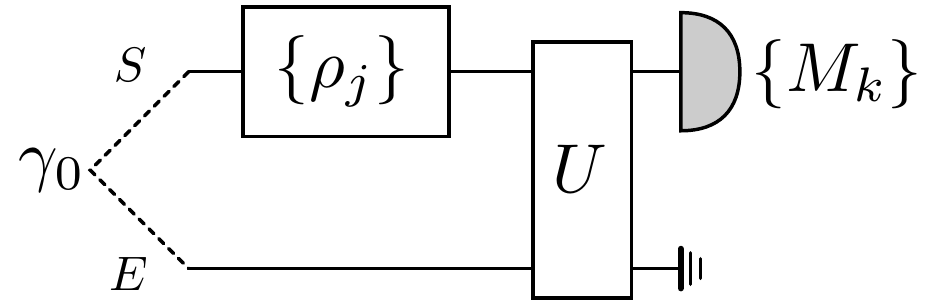}
\end{center}
\caption{State preparation in SQPT}
\label{fig:stateprepsqpt}
\end{figure}

\subsection{Stochastic State Preparation} \label{sec:qptsto}
First we consider the so called \emph{stochastic state preparation}~\cite{Kuah:2007}. The general idea of this procedure is to remove correlations by a process which sends the state of the principal system $S$ to a fixed state regardless of the initial state of the joint system $SE$. For example, cooling a quantum system will bring it into its ground state. From here stochastic unitary operations are used to transform the produced state into the set of input states needed for SQPT.

Mathematically we describe this procedure in terms of a \emph{preparation map}. The first step of this procedure is represented by a map $\Xi$ acting on density matrices of system $S$ which sends \emph{all} states to a fixed pure state $\op{\psi}{\psi}$. So
\begin{eqnarray}
\Xi(\rho)=\op{\psi}{\psi} \quad\forall\, \rho\, \mbox{ of system $S$}.
\end{eqnarray}

Applying this map to states of the joint system gives
\[
(\Xi\otimes\Ical)(\gamma)=\op{\psi}{\psi}\otimes\tau
\]
where $\tau=\tr_{A}\left[ (\Xi\otimes\Ical)(\gamma)\right]$ is the post-preparation state of the environment.

With joint system in the product state $\op{\psi}{\psi}\otimes\tau$, the next step is to prepare the basis of input states $\rho_{j}$ needed to for SQPT. This is done with a set $\{\Scal_j\}$ of so called \emph{stochastic maps} which satisfy $\Scal_{j}(\op{\psi}{\psi})=\rho_{j}$. In fact these are simply a set of unitary operators chosen to give the required states $\rho_j$ when applied to the fixed output state of $\Xi(\rho)$.

Hence the stochastic preparation procedure is given by the collection of maps $\{\Pcal_{j}^{sto}\}$, where
\begin{eqnarray}
\Pcal_{j}^{sto}&\equiv& (\Scal_{j}\circ\Xi)\otimes\Ical \\
\Pcal_{j}^{sto}(\gamma)&=& \rho_{j}\otimes\tau \quad\forall\, \rho\, \mbox{ of the joint system $SE$}.
\end{eqnarray}

If our system is initially in the (possibly correlated) state $\gamma_{0}$, and undergoes unitary evolution via a quantum gate $U$, SQPT performed on the output states
\begin{eqnarray*}
\Ecal(\rho_{j}) &=& \tr_{B}\left[ U\Pcal_{j}^{sto}(\gamma_{0})U^{\dagger}\right]\\
&=& \tr_{B}\left[ U (\rho_{j}\otimes\tau) U^{\dagger}\right].
\end{eqnarray*}

We note that the authors~\cite{Kuah:2007} claim the state $\tau$ is a function of the preparation map $\Xi$ alone, hence it will be fixed for any input $\gamma$. However, if the map $\Xi$ is itself trace preserving this is incorrect. In this case $\tau$ is independent of the map $\Xi$ and \emph{only} depends on the initial state $\gamma$. We prove this in Appendix~\ref{app:ximap}. Since the state of the environment $\tau$ is fixed for all $\rho_j$, by Thm.~(\ref{thm:productcp}) the resulting process matrix from SQPT will always be CP.

\subsection{State Preparation by Measurement and Rotations} \label{ssec:qptrot}

We now propose a preparation procedure similar to the stochastic case which could be implemented in a linear optics SQPT experiment. In \cite{Kuah:2007} the authors consider using a set of projective measurements as a preparation procedure. We will discuss the procedure using only measurements momentarily, but first we will introduce our own procedure based on this idea. We consider replacing the map $\Xi$ from the stochastic procedure with a single projective measurement $\Pi$. Recall from Section~\ref{ssec:postulates} that in quantum mechanics measurement is described by a set of measurement operations $M_j$. If we measure a state $\rho$ and detect outcome $j$, then it is assumed the post-measurement state of our system is $\rho_j=\frac{1}{p_j}M_j\rho M_j^\dagger$, where $p=\tr(\rho\Pi)$ is the probability of detection.

We consider the case of measuring with only a single projective measurement operator $\Pi$. For our example we will use $Pi_H=\op{H}{H}$ which projects onto the horizontal polarization state $\rho_H$. This could be implemented in linear optics for example by a polarization filter set to only let through horizontally polarized light. If a beam of randomly polarized light is incident on the filter, only photons in a polarization state corresponding to the filter will pass through. The sacrifice in this is that we are performing a \emph{trace decreasing} operation. We will \emph{loose} qubits orthogonal to the projection. This means we may have to judiciously select our initial projector based on the initial state of the joint system so that we are not loosing a large portion of our qubits.

In any case, once we have performed the projective measurement our stochastic maps are then the set of unitary rotation matrices $\{ R_{j}\}$ such that $R_{j} \rho_H R_{j}^{\dagger}=\rho_{j}$ where $\rho_{j}$ is one of our SQPT input states $(H,V,D,R)$. In optics such rotations to the polarization state of a photon can be easily implemented by wave plates.

Our preparation maps are then given by
\begin{eqnarray}
\Pcal^{sto}_j(\gamma_0)
&=& \frac{1}{\Gamma_H}(R_j\Pi_H)^A \gamma_0 \,(\Pi_H R^\dagger_j)^A
        \label{eqn:preprotations}\\
&=& \rho_j\otimes\tau_H
\end{eqnarray}
where $(R_j\Pi_H)^A \equiv(R_j\Pi_H)\otimes\id$ is an operator acting only on the first system, $\tau_H=\tr_A\left[(R_j \Pi_H)^A\gamma_0 \,(\Pi_H R^\dagger_j)^A\right]/\Gamma_H$ is the post-preparation state of the environment, and $\Gamma_H=\tr\left[\Pi_H^A\gamma_0\right]$ is the probability of detecting the outcome corresponding $\Pi_H$. Here we see that in general the post-preparation state of the environment is dependent on both the initial state $\gamma$ \emph{and} the measurement operator $\Pi_H$.

\subsection{State Preparation by Projective Measurements} \label{sec:sqptproj}
In our example of a stochastic preparation procedure we introduced the idea of using a projective measurement followed by unitary rotations to prepare the state of our system. We now propose the idea of using \emph{only} measurements to prepare our input states for SQPT. This method of preparation would be convenient in linear optics where such measurements can be implemented with polarizing elements.

If $\{\rho_j\}$ is our required input set for SQPT, we chose a set of projective measurements $\{\Pi_j\}$ where $\Pi_j\equiv\rho_j$. Hence the preparation procedure for projective measurements is given by the collection of maps $\{ \Pcal_{j}^{meas} \}$ where
\begin{eqnarray}
\Pcal_{j}^{meas}(\gamma)
&=& \frac{1}{\Gamma_{j}} \Pi_{j}^A \gamma \Pi_{j}^A    \label{eqn:prepmeas}\\
&&= \rho_{j}\otimes\tau_{j}.
\end{eqnarray}
Here $\tau_j=\tr_A\left[\Pi^A_j\gamma \Pi_j^A\right]/\Gamma_j$ is the post-preparation state of the environment, and $\Gamma_j=\tr\left[\Pi_j^A\gamma\right]$ is the probability of detecting the outcome corresponding $\Pi_j$.

Since we are using measurements alone for preparation we have the possibility of failure in our preparation. If the initial state is orthogonal to one of our projectors $\Pi_j$, then we can never prepare the required state $\rho_j$. If the initial state is instead only close to orthogonal, we will have a very small probability of producing the required state. Since we must prepare many copies of the state for performing state tomography this greatly reduces the efficiency of the experiment.

If our system is initially in a state $\gamma_{0}$, and undergoes unitary evolution via a quantum gate $U$, the output states for our set of tomography input states are given by
\begin{eqnarray}
\Ecal(\rho_j) &=& \tr_{B}\left( U\,\Pcal_{j}^{meas}(\gamma_{0})\,U^{\dagger}\right)\\
&=& \tr_{B}\left( U\,\rho_{j}\otimes\tau_{j}\,U^{\dagger}\right).
\end{eqnarray}

Here we can see that in general the state of the environment, $\tau_j$, depends on the state of the system, $\rho_j$. Since the state of the environment is not fixed, the process matrix determined by SQPT will not necessarily be CP. This situation can occur when we use a correlated initial state $\gamma_0$. We will show this with a simple example in Section~\ref{sec:tomoex}.

\subsection{Non-CP Maps due to Preparation by Measurement} \label{ssec:projncp}
We now ask whether we can find some bounds on the type of correlations which may give rise to non-CP process matrix. Our initial idea was to use the concept of \emph{quantum discord} which we introduced in Section~\ref{sec:discord}.  Recall from Thm.~(\ref{thm:sudarshan}) that the evolution of any initial state $\gamma$ satisfying $\Dcal_{E:S}(\gamma)=0$ will always be completely positive. We find that if one uses projective measurements alone to prepare the input states for SQPT, then this theorem no longer holds. An initial state $\gamma$ satisfying $\Dcal_{E:S}(\gamma)=0$ \emph{can} lead to non-CP evolution in SQPT.

To see why this occurs consider an initial state $\gamma=\frac{1}{2}\left(\rho_H\otimes\rho_H + \rho_V\otimes\rho_D \right)$ which, by Eqn.~(\ref{eqn:qddecompb}), satisfies $\Dcal_{B:A}(\gamma)=0$. Now if we use projective measurements to prepare a set of input states $\{\rho_i=\op{i}{i}\}$ where $i=H,V,D,R$, the post-preparation states of the environment are
\begin{eqnarray*}
\tau_H &=& \rho_H \\
\tau_V &=& \rho_D \\
\tau_D &=& \frac{1}{2}(\rho_H+\rho_D) \\
\tau_R &=& \frac{1}{2}(\rho_H+\rho_D).
\end{eqnarray*}

Even though the preparation procedure puts $SE$ into a product state, since the state of the environment is not fixed, the evolution need not be CP.

The situation for ensuring that the evolution of a state $\gamma$ will \emph{always} be CP when using SQPT and preparation by projective measurements is \emph{more} restrictive than normal. Due to the nature of the preparation procedure, we not only require that the initial state be simply separable, $\gamma=\rho_S\otimes\tau_E$, we also require that the reduced state of the system, $\rho_S$, not be orthogonal to any of our required input states.
%
%

\subsection{Non-Linearities from Preparation by Measurement} \label{ssec:projbilin}

Our next issue with preparation by measurement is to do with the non-linearity of the procedure. In \cite{Kuah:2007} the authors raise the issue of non-linearity with this preparation procedure. While the process matrix determined by SQPT from this procedure will be linear, the preparation procedure itself is not.

For example, if we wished to prepare the system into a state $\rho = \frac{1}{2}(\rho_1 + \rho_2)$, where $\rho_i$ and $\rho_j$ are elements of our input set, then we have
\begin{eqnarray*}
\Pcal^{meas}(\gamma)
&=& \frac{(\Pi_1+\Pi_2)^A\gamma (\Pi_1+\Pi_2)^A}{4\tr\left[\frac{1}{2}(\Pi_1+\Pi_2)^A\gamma\right]}\\
&=& \frac{1}{2(\Gamma_1+\Gamma_2)}\left( \Pi^A_1 \gamma \Pi^A_1 + \Pi^A_1 \gamma \Pi^A_2
                    + \Pi^A_2 \gamma \Pi^A_1 + \Pi^A_2 \gamma \Pi^A_2 \right)\\
&\ne& \frac{1}{2}\left(\Pcal_1^{meas}+\Pcal_2^{meas}\right).
\end{eqnarray*}
So the preparation procedure is in fact \emph{bilinear}. A method of dealing with this situation by describing the evolution with a bilinear process matrix is proposed~\cite{Kuah:2007}. The method of using a bilinear process matrix is significantly more complicated than the linear case, and we discuss this approach in section~(\ref{sec:bqpt}). Another problem with this assumption is that using projective measurements, we can only ever prepare our system into a pure state, as mixed states do not satisfy $\rho^2=\rho$.

We propose an alternative protocol of preparation by measurement to ensure a linear procedure. Our set of preparation maps $\{\Pcal_j^{meas}\}$ correspond to a basis for density matrices. If we wish to prepare a state $\rho=\sum_i p_i\rho_i$, where $\rho_i$ are the basis states, we simply perform the stochastic preparation procedure $\Pcal^{meas}(\gamma)=\sum_i p_i \Pcal_i^{meas}(\gamma)$. This is done by randomly performing each of the basis preparation procedures with probability $p_i$. This is analogous to the preparation of mixed states by randomly preparing one of several different pure states weighted by their respective probabilities.

It should be noted that this in general will leave the joint system in a state $\Pcal^{meas}(\gamma)=\sum_i p_i\rho_i\otimes\tau_i$, which will still in general give rise to non-CP dynamics.

\section{Examples of Preparation Procedures in SQPT} \label{sec:tomoex}
We will now consider an explicit example of performing SQPT with the two preparation procedures mentioned in Sections~\ref{ssec:qptrot} and \ref{sec:sqptproj}. We will consider a system $S$ and its environment $E$, both qubits, with an interaction given by a controlled-NOT gate. This gate performs a bit flip operation $(\ket{0}\leftrightarrow\ket{1})$ of the \emph{target} qubit if the \emph{control} qubit is in a state $\ket{1}$. If the control qubit is in state $\ket{0}$ the gate does nothing. We will let $S$ be the target qubit, and $E$ the control. The unitary matrix for this gate is given by
\[
U_{CNOT}=\left(\begin{array}{ c c c c }
 						1 & 0 & 0 & 0   \\
						0 & 0 & 0 & 1 \\
						0 & 0 & 1 & 0 \\
    					0 & 1 & 0 & 0 \end{array} \right).
\]

Suppose that the initial state of joint system $SE$ is the maximally entangled state $\rho_{\phi}=\op{\phi}{\phi}$ where $\ket{\phi}=(\ket{00}+\ket{11})/\sqrt{2}$. Our tomographic input states will given by $\rho_i$ where $i=H,V,D,R$.

We now consider the resulting process matrix when we preform SQPT using the two preparation methods described in Sections~\ref{ssec:qptrot} and \ref{sec:sqptproj}:
\begin{itemize}
\item \textbf{Method A:} We use a single projective measurement $\Pi_H$ to prepare system $A$ to the state $\rho_H$, and then use unitary operations $R_i$ to rotate $\rho_H$ to the state $\rho_i$.
\item \textbf{Method B:} We use a set of projective measurement $\Pi_i$ to prepare each of the initial states $\rho_i$.
\end{itemize}

The process matrix reconstructed though SQPT are
\begin{equation}
\Lambda_A=
        \left[
            \begin{array}{cccc}
             1 & 0 & 0 & 1 \\
             0 & 0 & 0 & 0 \\
             0 & 0 & 0 & 0 \\
             1 & 0 & 0 & 1
            \end{array}
        \right],
\quad\mbox{and}\quad
\Lambda_B=\frac{1}{2}
        \left[\begin{array}{ c c c c }
 			2 & 0 &-1-i & 1   \\
			0 & 0 & 1 & 1+i \\
			-1+i & 1 & 2 & 0 \\
    		1 & 1-i & 0 & 0 \end{array}
        \right],
\end{equation}
for method $A$, and method $B$ respectively.

$\Lambda_A$ has eigenvalues $(2,0,0,0)$, and hence $\Ecal_A$ the process reconstructed from preparation by a measurement followed by rotations, is completely positive.
$\Lambda_B$ on the other hand has eigenvalues $(1+\frac{\sqrt{3}}{2},-\frac{\sqrt{3}}{2},\frac{\sqrt{3}}{2},1-\frac{\sqrt{3}}{2})$ and hence $\Ecal_B$ the process reconstructed from preparation by measurements alone, is not completely positive.

In the first case the evolution is not very interesting, it is simply the identity map $\Ecal_A=\Ical$, this is because the state of the environment after the preparation procedure is $\op{0}{0}$.

In the second case we have a genuine non-CP map. To see why this happens let us consider the state of the environment after each preparation procedure. We have that
\[
\tau_H=			\left(\begin{array}{ c c }
 							1 & 0   \\
    							0 & 0 \end{array} \right), \
\tau_V=   		\left(\begin{array}{ c c }
 							0 & 0   \\
    							0 & 1 \end{array} \right),\
\tau_D=			\frac{1}{2}\left(\begin{array}{ c c }
 							1 & 1   \\
    							1 & 1\end{array} \right),\
\tau_R=	 	\frac{1}{2}	\left(\begin{array}{ c c }
 							1 & i   \\
    							-i & 1 \end{array} \right).
\]
So in each case the state of the environment is different. For example, in the case of projecting onto $\rho_{H}$ the state of the control qubit $\tau_H$ means the evolution of qubit $A$ will be the identity, $U=\id$, while in the case of projecting onto $\rho_{V}$ our gate becomes a bit flip, $U=X$. Even though our initial state is simply separable, the state of its environment is dependent on the state of the system due to the nature of our preparation procedure.

This is an interesting result, as conceptually the only difference between the two preparation procedures was an extra rotation stage that was introduced in Method A. This means that in a linear optics tomography experiment, having a polarizer to prepare states can result in a dramatically different result than a polarizer with a wave plate placed after it. Even though these elements act only on system $S$, the initial correlations between $S$ and $E$ give rise to a CP result in one case, yet a non-CP result in the other. This is explained in more detail in Chapter~\ref{chap:linoptics} where we propose an experimental implementation to illustrate this observation.

While we used a maximally entangled input state, this result holds even for separable inputs. For example if we use the input state $\rho=\frac{1}{2}\left(  \rho_H\otimes\rho_A + \rho_D\otimes\rho_V   \right)$,
then we get a process matrix with eigenvalues $\{1.642, 0.507, -0.253, 0.105\}$, so the evolution is still non-CP. In this case the quantum discord for $\rho$ is $\Dcal_{A:B}(\rho)=\Dcal_{B:A}=0.1443$.

\section{State Preparation in AAPT} \label{sec:aaptstateprep}

We now discuss state preparation in AAPT. This situation is much simpler than in SQPT as we only need a single input state. The difference is now we are dealing with three systems, an ancilla $A$, principal system $S$ and environment $E$ as shown in Fig.~(\ref{fig:stateprepaapt}). We will assume that the joint system $ASE$ is initially in a tripartite state $\gamma_{ASE}$.

To prepare our input state for AAPT we perform a preparation procedure on the ancilla and principle system analogous to the $\Xi$ map from Section~\ref{sec:qptsto}. This procedure prepares $AS$ in the maximally entangled state $\rho^{AS}_\phi=\op{\phi}{\phi}$, where $\ket{\phi}=\sum_i\ket{i}\otimes\ket{i}/\sqrt{d}$, and is given by a map
\begin{eqnarray}
\Xi(\gamma_{AS})&=\op{\phi}{\phi} \quad\forall \gamma_{AS} \mbox{ of the joint system $AS$}.
\end{eqnarray}
Hence the preparation procedure on our whole system is given by
\begin{eqnarray*}
\Pcal(\gamma_{ASE})
    &=&(\Xi\otimes\Ical)(\gamma_{ASE}) \\
    &=& \rho_\phi^{AS}\otimes\tau_E,
\end{eqnarray*}
where $\tau_E=\tr_{AB}\left[(\Xi\otimes\Ical)\gamma_{ASE}\right]$ is the post preparation state of the environment. This preparation procedure could be implemented by a Bell-state measurement. This is a joint measurement over two systems, which detects the maximally entangled states~\cite{Mohseni:2008}.

\begin{figure}[htb]
\begin{center}
\includegraphics[width=0.9\textwidth]{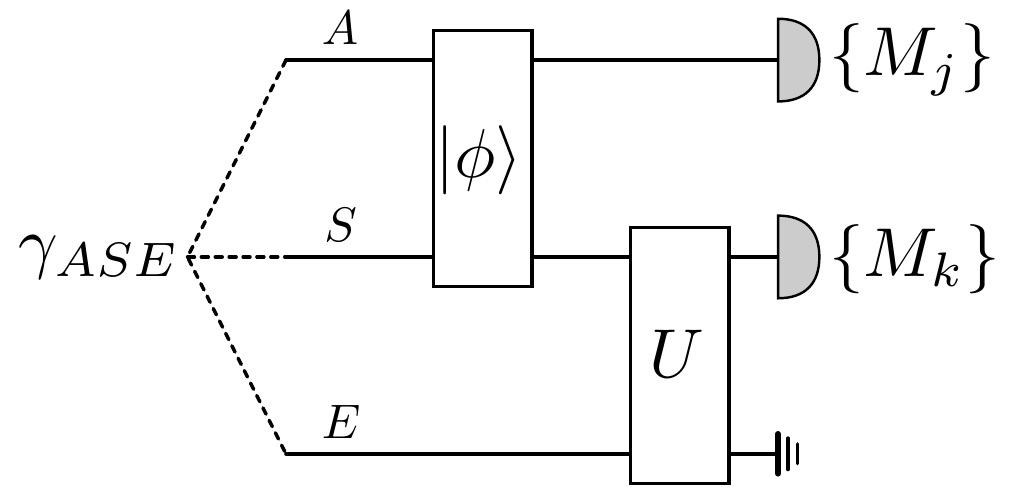}
\end{center}
\caption{State preparation in SQPT}
\label{fig:stateprepaapt}
\end{figure}

Since the interactions between our system and environment are described by a unitary operation $U$, as shown in Fig.~(\ref{fig:stateprepaapt}), the process map for this interaction, as described by the output state $\rho_\Ecal$ from AAPT is then given by
\begin{eqnarray*}
\rho_\Ecal
    &=& \tr_E \left[(\id\otimes U)\Pcal(\gamma_{ASE})(\id\otimes U^\dagger)\right] \\
    &=& \tr_E \left[(\id\otimes U)(\rho^{AS}_\phi\otimes\tau_E)(\id\otimes U^\dagger)\right]
\end{eqnarray*}

Since we only need prepare this one input state to perform AAPT (in practice we could use another initial state so long as it satisfied the requirements in Section~\ref{sec:aapt}, the state of the environment is fixed. Thus performing AAPT will be equivalent to performing SQPT with a fixed environment $\tau_E$ and so, by Thm.~\ref{thm:productcp}, the process matrix for this evolution will always be CP.

\section{Bilinear Quantum Process Tomography} \label{sec:bqpt}

Even when we have a situation where preparation by projective measurements gives a CP results (for example when either direction discord is zero), if we treat the preparation procedure as an assignment map it will still be be bilinear.

A method of performing bilinear process tomography was proposed in~\cite{Kuah:2007}. We will briefly present this method, and as done previously we will also convert their formulas into more familiar notation.

We consider the same situation as section~(\ref{sec:sqptproj}), where we have a system $A$ interacting with environment $B$, both of dimension $d$. The initial state and evolution of the joint system being given by $\gamma_0$, and $U$ respectively. If we project the initial state of the system onto $\rho^n$ for our preparation procedure, then the output is given by
\[
\Ecal(\rho^{(n)})=\frac{1}{\Gamma^{(n)}}\tr_{B}\left[U (\rho^{(n)}\otimes I )\gamma_{0} (\rho^{(n)}\otimes I )U^{\dagger}\right].
\]

The idea proposed in~\cite{Kuah:2007} is to express this in terms of matrix elements, and then rearrange the elements so that
\begin{eqnarray*}
\Ecal(\rho^{(n)})_{r,s}
	&=& \frac{1}{\Gamma^{(n)}}
		\sum_{r^{\prime},r^{\prime\prime}}
		\sum_{s^{\prime},s^{\prime\prime}}
		\sum_{\alpha,\beta,\epsilon}
			U_{r\epsilon,r^{\prime}\alpha} \, P^{(n)}_{r^{\prime}r^{\prime\prime}}\,
			\gamma_{0\, r^{\prime\prime}\alpha, s^{\prime\prime}\beta}\,
			P^{(n)}_{s^{\prime\prime}s^{\prime}}\, U^{\dagger}_{s^{\prime}\beta, s \epsilon} \\
	&=& \frac{1}{\Gamma^{(n)}}
		\sum_{r^{\prime},r^{\prime\prime}}
		\sum_{s^{\prime},s^{\prime\prime}}
		P^{(n)^{*}}_{r^{\prime\prime} r^{\prime}}
			\left( \sum_{\alpha,\beta,\epsilon} \, U_{r\epsilon,r^{\prime}\alpha} \,
			\gamma_{0\, r^{\prime\prime}\alpha, s^{\prime\prime}\beta}\,
			U^{\dagger}_{s^{\prime}\beta, s \epsilon} \right)
		P^{(n)}_{s^{\prime\prime}s^{\prime}}\,  \\
	&=& \frac{1}{\Gamma^{(n)}}
		\sum_{r^{\prime},r^{\prime\prime}}
		\sum_{s^{\prime},s^{\prime\prime}}
		P^{(n)^{*}}_{r^{\prime\prime} r^{\prime}}
			\Mcal^{(r,s)}_{r^{\prime\prime}r^{\prime}; s^{\prime\prime} s^{\prime}}\,
		P^{(n)}_{s^{\prime\prime}s^{\prime}}, \\
\end{eqnarray*}
where we have defined a new matrix $\Mcal$ by
\begin{equation}
\Mcal^{(r,s)}_{r^{\prime\prime}r^{\prime}; s^{\prime\prime} s^{\prime}}=
\sum_{\alpha,\beta,\epsilon} \, U_{r\epsilon,r^{\prime}\alpha} \,
			\gamma_{0\, r^{\prime\prime}\alpha, s^{\prime\prime}\beta}\,
			U^{\dagger}_{s^{\prime}\beta, s \epsilon}.
\label{eqn:bilinkuah}
\end{equation}

The matrix $\Mcal$ to called the \emph{bilinear process matrix}, and we can see it depends on both the initial joint state $\gamma_0$ and the evolution $U$.

This representation of $\Mcal$ in is not particularly useful so we propose a more computationally convenient one. Our idea is to break the $d^2\times d^2$ matrices $U$ and $\gamma_0$ into $d\times d$ block matrices, where each block element is a $d\times d$ matrix. I.e
\[
U=  \left(
    \begin{array}{cccc}
    [U]_{11}    & [U]_{12}  & \ldots& [U]_{1d}\\
    \left[U\right]_{21}   & \ddots    &       & \vdots   \\
    \vdots      &           &       &           \\
    \left[U\right]_{d1}   & \ldots    &       & [U]_{dd}    \\
               \end{array}
    \right),
\]
where $([U]_{ij})_{m,n}=U_{(i-1)d+m,(j-1)d+n}$. Similarly we form the block matrices for $\gamma_0$ and $U^\dagger$ noting that $[U^\dagger]_{ij}=[U]_{ji}^\dagger$.

In this notation we can calculate each entry of $\Mcal$ by
\begin{equation}
\Mcal^{(r,s)}_{r''r'; s'' s'}
= \tr\left(
        [U]_{rr'}[\gamma_0]_{r''s''}[U]_{ss'}^\dagger
    \right).
\label{eqn:bilinmatrix}
\end{equation}
$\Mcal$ itself is a block matrix, where $(r''r';s''s')$ denotes the block elements of $M$, and $(r,s)$ are the comments of a given block. Evolution is then given by
\begin{equation}
Q_n=\frac{1}{\Gamma_n}\bra{\rho_m}M\ket{\rho_n},
\label{eqn:bilinevo}
\end{equation}
where we have vectorized $\rho$, and each element of $\rho$ acts on a block matrix element of $\Mcal$. Verification of these expressions can be found in Appendix~\ref{app:bilin}

It was shown in \cite{Kuah:2007} that $\Mcal$ is hermitian. However it was mistakenly asserted that $\tr(\Mcal)=1$. We found that $\tr(\Mcal)$ is not equal to one, in fact $\tr(\Mcal)=d$, the proof of which is in Appendix~\ref{app:bqptproof}.

If we repeat the example from Section~\ref{sec:tomoex} using preparation by projective measurements with a CNOT gate and a maximally entangled input. We calculate the bilinear process matrix to be
\[\Mcal=
\frac{1}{2}\left(
  \begin{array}{cccc}
    E_{11}  & E_{12}& [0] & [0] \\
    E_{21}  & E_{22}& [0] & [0] \\
    \left[0\right]     & [0]   & E_{22} & E_{21} \\
    \left[0\right]     & [0]   & E_{12} & E_{11} \\
  \end{array}
\right),
\]
where $E_{ij}=\op{i}{j}$, and $[0]$ is the matrix with all entries zero. $\Mcal$ has two non-zero eigenvalues $\lambda_1=\lambda_2=1$, and $\tr(\Mcal)=2$.

Since $\Mcal$ is a large matrix, $d^3\times d^3$ performing tomography is a far more excruciating task, as we now require $\frac{d^2}{2}(d^2+1)$ input states to completely characterize $\Mcal$ instead of the $d^2$ required for linear process tomography. We did not get time to fully investigate the actually procedure of performing bilinear tomography, and investigating this is a possible avenue for future research. It would be interesting to see how initial correlations can effect a bilinear assignment map, and under what conditions the evolution would always be completely positive. We suspect that similarly a condition for complete positivity would be requiring that the bilinear process matrix $\Mcal$ is positive. 


\chapter{Statistical Noise in Process Tomography}           \label{chap:statsim}

In the last chapter we provided examples of idealized SQPT experiments where state preparation resulted in a non-completely positive process maps in the presence of initial correlations. The problem is that in process tomography non-CP results are frequently observed due to statistical noise. For example, in the case of photonic qubits this is due to the Poissonian count statistics from the spontaneous sources used for creation of photons.

Often when an experimenter identifies a non-CP process map in a tomography experiment it is assumed to be due to noise, and optimization techniques such as maximum likelihood estimates are employed to map it to the closest \emph{physical} process map, i.e. a completely-positive one. However, if we accept that non-CP maps can occur for reasons other than statistical noise, we need some way to distinguish between non-physical results due to the noise, and those which arise legitimately due to initial correlations. In this chapter we investigate this issue for an optical implementation of quantum computing where the statistical noise is Poissonian. In the following chapter we will outline an optical experiment to demonstrate the results of this thesis.

\section{Statistical Noise in State Tomography}

We start by considering state reconstruction of a single qubit. This will allow us to build an intuitive picture for the statistical noise present in state reconstruction before we move to the two-qubit case needed for describing a single qubit process matrix.

As mentioned in Section~\ref{ssec:mltomo}, due to the spontaneous nature of photon sources the distribution in recorded counts in an experiment will be Poissonian. If we are measuring a single qubit with density matrix $\rho$, the expected number of counts for $\Ncal$ measurements of an operator $M_m$ is
\begin{equation}
\label{eqn:nexp}
n_m=\Ncal\tr[\rho M_m].
\end{equation}
The density matrix $\rho$ is reconstructed from the expected counts using the dual basis, $\{D_m\}$, of the measurement operators, and expected probability coefficients $p_m=n_m/\Ncal$ recovered from the relative frequencies $n_m$. The reconstruction is given by,
\begin{equation}
\rho=\sum_m p_m D_m.
\label{eqn:pexp}
\end{equation}

The influence of statistical noise means that for any given tomography experiment, the measured count data will be a set of random variables obeying Poissonian distributions $\mathbb{P}(\lambda)$ with the distribution parameter given by the expected number of counts from Eqn~(\ref{eqn:nexp}). For example, the distribution of the measured counts for measurement operator $M_m$ is $\,n_m^e \sim \mathbb{P}(n_m)$.

We will assume that our experiments run for sufficiently long to approximate this with a Gaussian distribution $\mathbb{N}(\mu,\sigma^2)$ with mean and variance $(\mu_m,\sigma_m^2)=(n_{m}, n_{m})$~\cite{Feller:1966}. Hence the distributions of the observed counts are given by
\begin{equation}
\label{eqn:nmeas}
n_m^e \sim \mathbb{N}(n_m,n_m).
\end{equation}

This approximation is very accurate for large values of $n_m$, and hence is most valid when dealing with large $\Ncal$. However, its accuracy will be reduced for states $\rho$ which are orthogonal or close to orthogonal to one of our measurement operators, $\tr(\rho M_m)\approx0$), as in these cases the expected count number will be very low.
By making this approximation we can use some useful properties of the Gaussian distribution. For a normally distributed variable $X \sim \mathbb{N}(\mu,\sigma^2)$ and scalar $\alpha$, $\alpha X$ is normally distributed with mean $\mu^\prime=\alpha\mu$ and variance $\sigma^{2\prime}=\alpha^2\sigma^2$. Also, for a set of normally distributed random variables $X_i \sim \mathbb{N}(\mu_i,\sigma_i^2)$, by the \emph{Central Limit Theorem}~\cite{Feller:1966} the sum of the variables is normally distributed with mean $\mu=\sum_i \mu_i$ and variance $\sigma^2=\sum_i\sigma_i^2$.

Hence we have that the measured values of the probability coefficients from Eqn~(\ref{eqn:pexp}), $p_m^e=n_m^e/\Ncal$, are normally distributed variables
\begin{eqnarray}
p_m^e
&\sim& \mathbb{N}\left(\frac{n_m}{N},\frac{n_m}{\Ncal^2}\right) \\
&\sim& \mathbb{N}\left(p_m,\frac{p_m}{\Ncal}\right)
\end{eqnarray}
where $p_m = n_m/\Ncal$ is the expected value of $p_m^e$.

Combining these we obtain our distribution for density matrices reconstructed via state tomography. The measured density matrices $\rho^e$ will be distributed according to
\begin{equation}
\label{eqn:rhodist}
\rho^e \sim \sum_m  \mathbb{N}\left(p_m,\frac{p_m}{\Ncal}\right) D_m
\end{equation}
which is a multidimensional sum of Gaussian distributions, as the dual matrices $D_m$ define different directions on the Bloch sphere.

If we examine the distribution in a specific dimension, for example $X\equiv\sigma_1$, then
\begin{eqnarray}
\tr[X\rho^e]
&\sim& \sum_m \tr[X D_m]\mathbb{N}\left(p_m,\frac{p_m}{\Ncal}\right)\nonumber\\
&\sim& \sum_m \mathbb{N}\left(p_m\tr[X D_m],\frac{p_m}{\Ncal}\tr[X D_m]^2\right)\nonumber\\
&\sim& \mathbb{N}\left(\sum_m p_m\tr[X D_m],\sum_m \frac{p_m}{\Ncal}\tr[X D_m]^2\right)\nonumber\\
&\sim& \mathbb{N}\left(\tr[X\rho],\,\frac{1}{\Ncal}\sum_m p_m\tr[X D_m]^2\right),\label{eqn:rhodimdist}
\end{eqnarray}
which is a Gaussian distribution with mean $\mu=\tr (X\rho)$ and variance $\sigma^2=\sum_m p_m\tr[X D_m]^2/\Ncal$. So the variance of the distribution scales as $1/\Ncal$. This holds for all dimensions.

\begin{figure}[ht]
\begin{center}
\subfigure[$\Ncal=10^2$]
    {\label{fig:Drecon1}
       \includegraphics[width=0.35\textwidth]{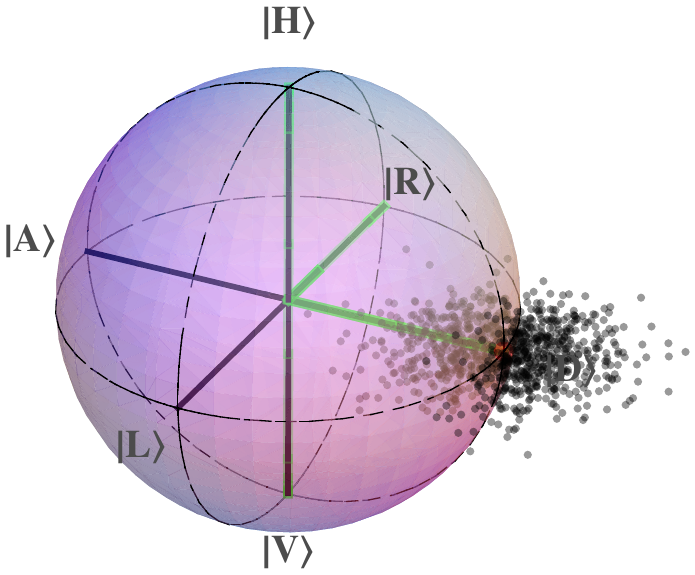}}
\subfigure[$\Ncal=10^3$]
    {\label{fig:Drecon2}
        \includegraphics[width=0.3\textwidth]{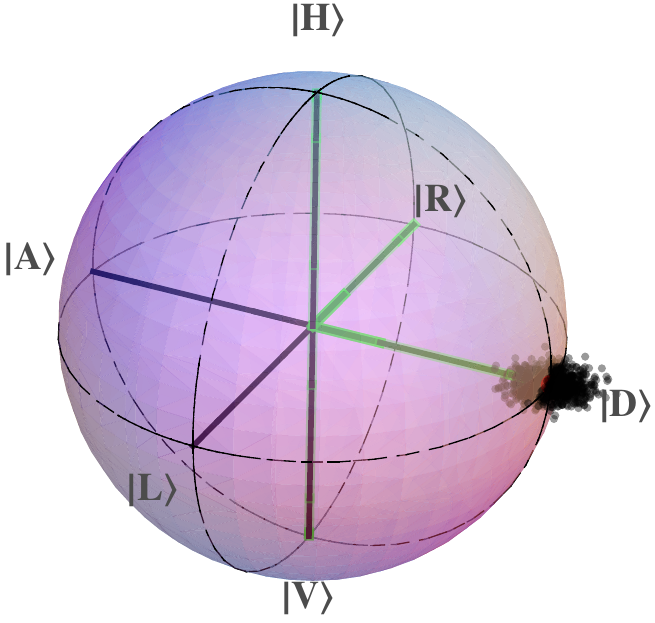}}
\subfigure[$\Ncal=10^4$]
    {\label{fig:Drecon3}
        \includegraphics[width=0.3\textwidth]{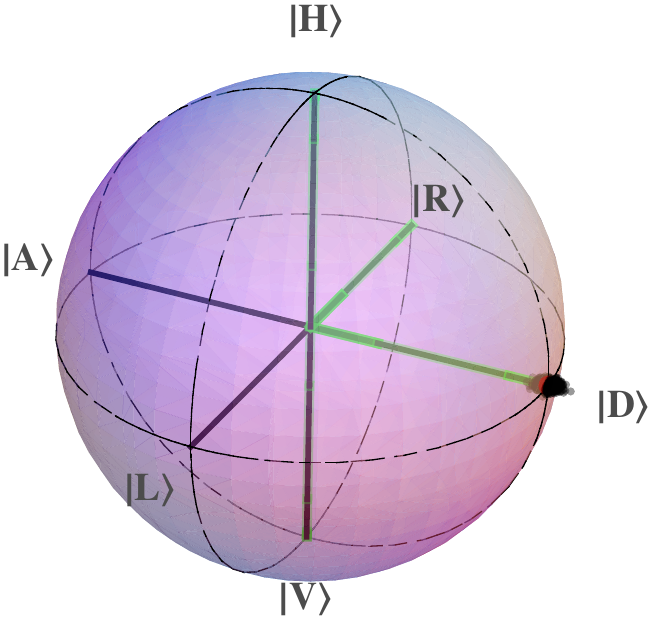}}
\caption{Distribution of reconstructed states due to Poissonian noise for $\rho=\op{D}{D}$.}
\label{fig:Drecon}
\end{center}
\end{figure}

In Fig.~(\ref{fig:Drecon}) we plot the reconstructed density matrix for a $\rho=\op{D}{D}$, for 1000 repeats of the reconstruction process. This is done for $\Ncal=10^2, 10^3,$ and $10^4$. Here we can see that the distribution is indeed centered on $\rho$, and has the appearance of a three dimensional Gaussian ball. We can see that the variance decreases with $\Ncal$ and that roughly half the data represents unphysical states, as they lie beyond the surface of the Bloch sphere.

\section{Statistical Noise in Process Tomography}

Now that we have an understanding of the single qubit case, we move to a process map for a single qubit. By the Jamiolkowski isomorphism reconstructing a process matrix for a single qubit is equivalent to reconstructing a two-qubit density matrix. A two-qubit state is characterized by 12 independent parameters, and the reconstruction procedure requires 16 independent measurements, instead of the four needed for a single qubit. We can follow the same procedure as the single qubit case and reconstruct a probability distribution of output states, though we no longer have a convenient visualization.

For the reconstruction of a process matrix $\Lambda_\Ecal$ we can generalize Eqn.~(\ref{eqn:nexp}). The expected number of counts when measuring input state $\rho_j$ with measurement operator $M_m$ is
\begin{equation}
\label{eqn:processnexp}
n_{jm} = \Ncal \tr_A\left[(\rho_j^T\otimes M_m)\Lambda_\Ecal \right] \\
\end{equation}

The excepted reconstruction for the process matrix is then
\begin{equation}
\label{eqn:processexp}
\Lambda_\Ecal  = \sum_{jm} p_{jm} \bar{D}_j^*\otimes D_m
\end{equation}
where $p_{jm}=n_{jm}/\Ncal$ is the expected probability coefficient for $\bar{D}_j^*\otimes D_m$, and $\{\bar{D}_j\}$ and $\{D_m\}$ are the dual bases to the input basis $\{\rho_j\}$ and measurement basis $\{M_m\}$ respectively. We note that if our measurement and input bases are the same, so too are the corresponding dual bases.

It is then trivial to extend the distribution of reconstructed density matrices to process matrices. The distribution of measured process matrices $\Lambda_\Ecal^e$ is
\begin{equation}
\label{eqn:processdist}
\Lambda_\Ecal^e \sim \sum_{jm}  \mathbb{N}\left(p_{jm},\frac{p_{jm}}{\Ncal}\right) \bar{D}_j^*\otimes D_m.
\end{equation}
which is a 16-dimensional sum of Gaussian distributions.

If we examine the distribution in a particular dimension, say $X\otimes Y$ for example, then analogous to Eqn.~(\ref{eqn:rhodimdist}) it is given by the Gaussian distribution
\[
\tr[(X\otimes Y) \Lambda_\Ecal^e] \sim \mathbb{N}\left( \mu, \sigma^2 \right),
\]
with mean $\mu=\tr[(X\otimes Y)\Lambda_\Ecal]$ and variance $\sigma^2=\frac{1}{\Ncal}\sum_{jm}\tr[X \bar{D}_j]^2\tr[Y D_m]^2$. This result holds for all dimensions, hence like the case of a single-qubit state, the variances in the distribution of a reconstructed the process matrix scales as $1/\Ncal$ in all dimensions.

\subsection{Non-CP Process Matrices}

The reconstruction distribution for a non-CP process matrix will be identical to the CP case, with respect to the appropriate mean and variance. The only difference now is that the distribution will be centered outside the physical subspace. Since the variance of the distribution scales as $1/\Ncal$, this gives us a possible way of distinguishing between non-CP cases due to noise, and those arising legitimately. If $\Ncal$ is sufficiently large, then an overwhelming majority of the distribution from a non-CP process matrix will also be non-CP. However in the case due to noise, a larger portion of the distribution will still correspond to CP processes. With this assumption, we should be able to employ standard statistical hypothesis testing to determine, within a confidence bound proportional to $\Ncal$, the probability of a measured non-CP result being due to noise.

We can illustrate this idea using a single qubit density matrix as an example. Consider two single-qubit density matrices $\rho_A$ and $\rho_B$, with Bloch vectors $\vec{\alpha_A}=(1/\sqrt{2},0,1/\sqrt{2})$ and $\vec{\alpha_B}=(1.2/\sqrt{2},0,1.2/\sqrt{2})$ respectively. These two vectors share a common direction on the Bloch sphere, however while $\rho_A$ is a pure state on the surface of the Bloch sphere, $\rho_B$ is not a valid density matrix. It has a greater than unit Bloch vector and thus lies outside the physical space of the Bloch sphere. We see here that $\rho_B$ is a non-positive density matrix, it has a negative eigenvalue of $-1/4$. We plot the distributions of reconstructed density matrices based on these expected states in Fig~(\ref{fig:pnonp}). We can see in Fig~(\ref{fig:pnonp100}) that for $\Ncal=100$ there is a large degree of overlap between the two distributions. However by time time $\Ncal$ reaches $1000$, as in Fig.~(\ref{fig:pnonp1000}) we can distinguish between the two distributions with a reasonable level of accuracy. The case of the process matrix is essentially the same however we don't have such a convenient visualization.

\begin{figure}[htbp]
\begin{center}
\subfigure[$\Ncal=100$]
    {\label{fig:pnonp100}
       \includegraphics[width=0.33\textwidth]{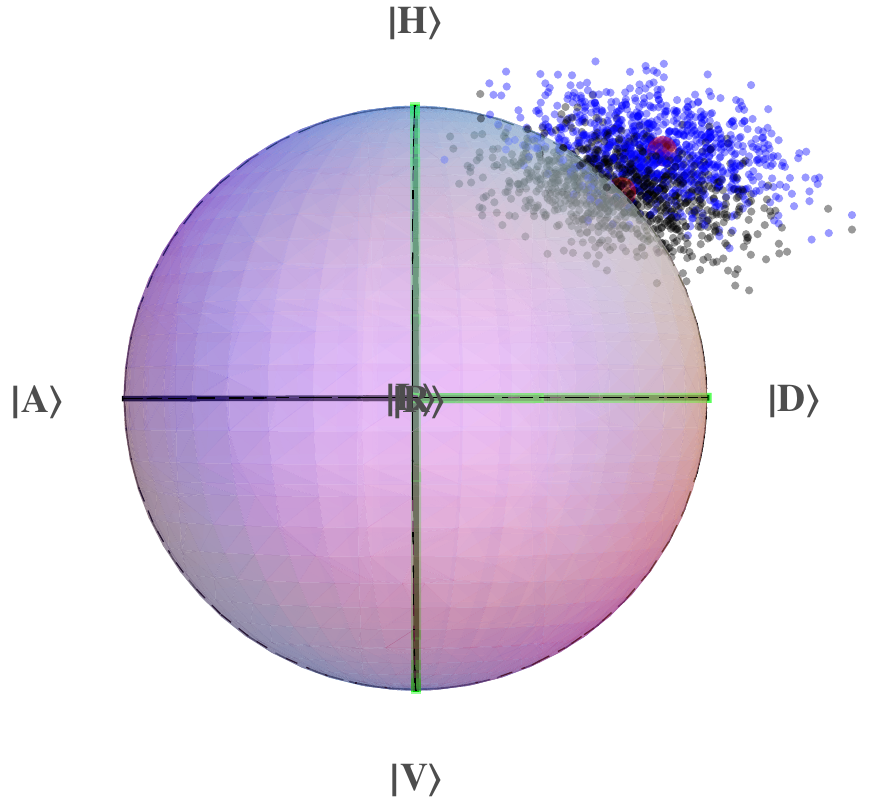}}
\subfigure[$\Ncal=500$]
    {\label{fig:pnonp500}
        \includegraphics[width=0.31\textwidth]{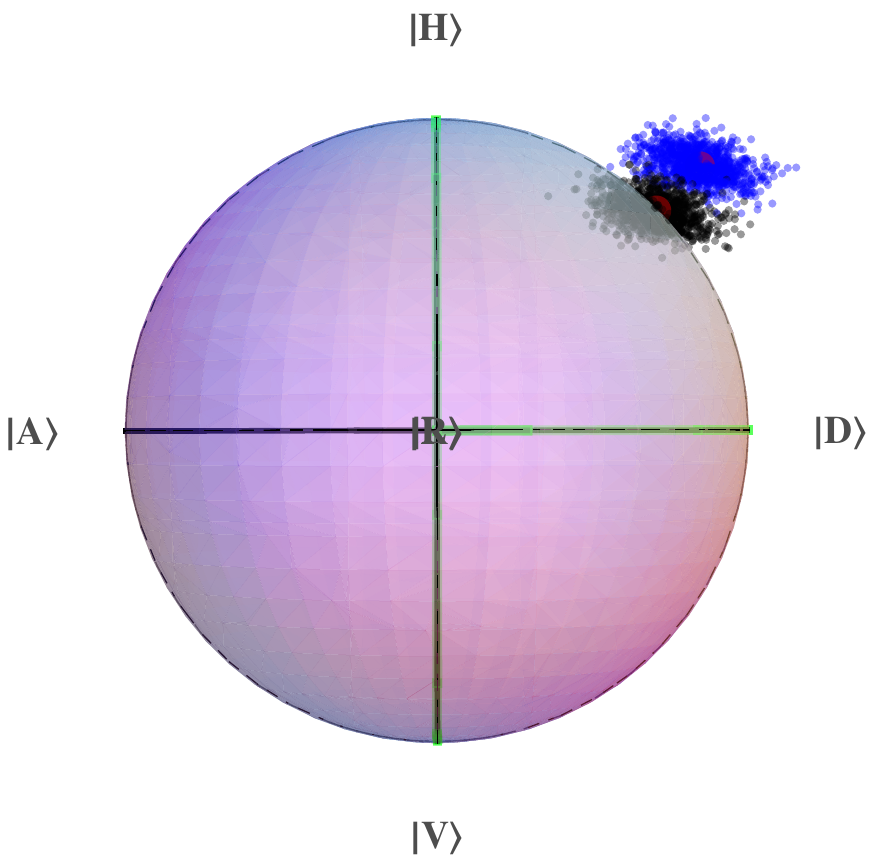}}
\subfigure[$\Ncal=1000$]
    {\label{fig:pnonp1000}
        \includegraphics[width=0.31\textwidth]{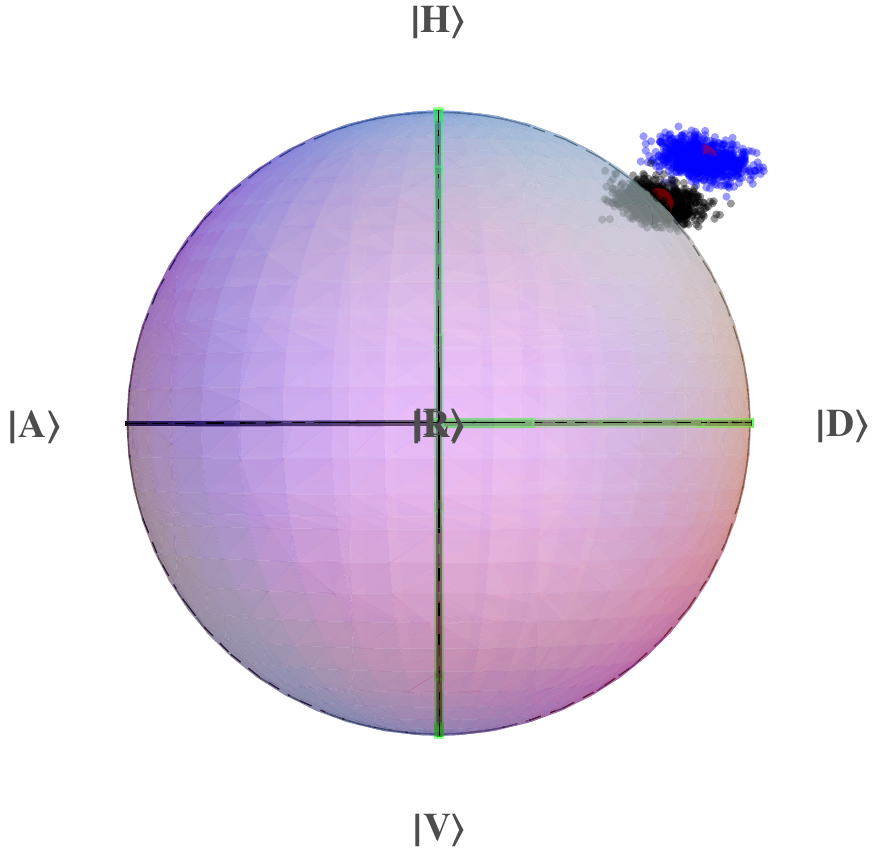}}
\caption{Distinguishing between reconstruction distributions of two density matrices with Bloch vectors in the same direction, one positive (black distribution) and one non-positive (blue distribution).}
\label{fig:pnonp}
\end{center}
\end{figure}

\section{Summary and Future Work}

We require a method of testing a measured non-CP process matrix to assign probabilities for the deviation from complete positivity being due to statistical noise and being due to initial correlations. As we previously mentioned, the distribution of process matrices reconstructed by process tomography (both CP and non-CP) is a multi-dimensional Gaussian distribution. In each dimension however the variance is inversely proportional to $\Ncal$, so by greatly increasing $\Ncal$ we can increase the distinction between the two causes for a non-CP result. However, in certain situations, such as a non-CP process map lying very close to the CP boundary, increasing $\Ncal$ to the required level may not be realistically achievable.

A more rigorous formulation of a statistical model to allow for non-CP process matrices in tomography optimization would be an entire study in itself. This would be an important direction for future work in the area of dealing with non-CP results in process tomography. A possible direction this research could take is to consider more advanced sampling schemes for our reconstruction data, such as sequential analysis or adaptive schemes.

In sequential analysis the sample size $\Ncal$ is not fixed in advance. Instead data is evaluated as it is accumulated. We start with a certain sample size $\Ncal_1$, and perform tomography measurements, then following reconstruction we repeat with process with another sample $\Ncal_2$. This count data is added to $\Ncal_1$ and reconstruction is repeated with the cumulative data. This process repeats until some predefined termination rule when we can reconstruct a significant result us satisfied. A possible termination rule is reaching a high confidence level in identifying the measured process matrix as a true non-CP result, or as an artifact of statistical noise.

In an adaptive scheme we would change our measurement procedure over the course of the experiment. For example, if we combined an adaptive measurement scheme with sequential analysis, each time we collect a new set of count data we could change our measurement basis. This would make the reconstruction algorithm more complicated, however it has the potential to increase the accuracy of our results. 


\chapter{Experimental Implementation With Linear Optics}           \label{chap:linoptics}

The major part of this thesis has been focused on the mathematical description of open quantum systems, and their evolution in the presence of initial correlations. In this chapter we move into a physical picture where we impart to the reader a sense of how such abstract concepts might be demonstrated in an \emph{actual} experiment. We will focus on an implementation using \emph{linear optical quantum computing} (LOQC) with photonic qubits.

LOQC is one of several major architectures currently used to implement quantum information techniques~\cite{Kok:2007}. It is sufficiently advanced to perform quantum gates, it has well characterized sources of noise, and well developed measurement and tomography techniques. However, a thorough description of LOQC is far beyond the scope of this thesis. Instead we will give a \emph{very} brief introduction to some of the key concepts involved in translating abstract mathematical entities into actual physical devices. Our goal is to give an outline for an experiment designed to implement some ideas proposed in this thesis. Specifically, we are concerned with implementing the example shown in Section~\ref{sec:tomoex} which highlighted how different state preparation techniques could lead to to CP or non-CP results when used in SQPT.

\section{Photonic Qubits}
In LOQC there are two main methods of encoding a photon as a qubit. The first, which we mentioned in Section~\ref{ssec:bloch}, is to use the polarization state of the photon. Here the logical states $\ket{0}$ and $\ket{1}$ are taken to correspond to the horizontal and vertical polarization states respectively. The second method, known as the \emph{dual rail} representation, is to represent a qubit by a photon in one of two separate optical modes. The logical states $\ket{0}$ and $\ket{1}$ are taken to correspond the photon being in the first mode $\ket{1,0}$, or the second mode $\ket{0,1}$ respectively. We should note that polarization encoding can be considered as a specific type of dual rail encoding. The implementation for performing one and two qubit gates that we will examine is based principally on polarization encoding.

To generate photonic qubits, single photon sources are required. Currently this is done by a process known as \emph{spontaneous parametric down conversion} (SPDC)~\cite{Nielsen:2000}. This process involves pumping a nonlinear optical medium with light of frequency $\omega_0$. With some small probability a pump photon will spontaneously down convert into two \emph{daughter} photons, each of frequency $\omega_0/2$. The careful implementation of filtering allows us to use these daughter photons in an experiment. However, since the down conversion process is spontaneous and photon detection is destructive, to use both the daughter photons an experimenter will usually employ post-selection techniques. This means that only instances when two photons are detected at the end at the end of the experiment will be considered valid.

\section{Performing Single Qubit Unitary Operations}

LOQC has everything we need to implement arbitrary single qubit gates. This is done through the use of wave plates, phase delays, beam splitters, and polarizing beam splitters.

For polarization-encoded photons, single qubit operations can be implemented using combinations of wave plates. These are optical devices made from birefringent crystal which induce a phase shift between orthogonal polarization components of a light beam. In practice we only need half-wave plates (HWP) and quarter-wave plates (QWP), which induce a phase difference of $\varphi=\pi$ and $\varphi=\pi/2$ respectively~\cite{Langford:2007}. The unitary operators corresponding to HWP and QWP with the axis of the induced phases shift rotated an angle $\theta$ from horizontal are given by
\[
U_{hwp}(\theta_h)=e^{i\frac{\pi}{2}}\left[
  \begin{array}{cc}
    \cos2\theta_h & \sin2\theta_h \\
    \sin2\theta_h & -\cos2\theta_h \\
  \end{array}
\right],
\quad\quad U_{qwp}(\theta_q)=\frac{1}{\sqrt{2}}\left[
  \begin{array}{cc}
    1+i\cos2\theta_q & i\sin2\theta_q \\
    i\sin2\theta_q & 1-i\cos2\theta_q \\
  \end{array}
\right].
\]
From here \emph{any} single qubit unitary gate can be implemented on a polarization encoded photon using a combination of appropriately aligned HWPs and QWPs.

\section{Two-Qubit Entangling Gates}\label{sec:2qubitgate}

To actually perform \emph{quantum} computation we required a two-qubit gate which can induce (or remove) entanglement between a pair of photons. Such a gate is called an \emph{entangling gate}. Implementing a two-qubit entangling gate is more challenging than a single-qubit gate, as inducing interactions between photons is quite difficult. This issue has been circumvented by using a process called \emph{measurement induced non-linearity}~\cite{KLM:2001}. In effect, this simulates a nonlinear interaction with a measurement. The trade off here is that the process is non-deterministic, which means there is only a certain probability of the gate being successful, and success cannot be determined without some form of measurement.

Conceptually, the simplest such gate is the \emph{controlled-Z} (CZ) gate. This gate induces a phase shift of $\pi$ on the $\ket{11}$ term of a two-qubit state. Its action is given by the unitary matrix
\[
U_{CZ}=\left[
  \begin{array}{cccc}
    1 & 0 & 0 & 0 \\
    0 & 1 & 0 & 0 \\
    0 & 0 & 1 & 0 \\
    0 & 0 & 0 & -1 \\
  \end{array}
\right]
\]
Recently three groups~\cite{Langford:2005, Kiesel:2005, Okamoto:2005} independently arrived at similar designs for a particularly simple implementation of a CZ gate in LOQC. The implementation by Langford \emph{et al.}~\cite{Langford:2005} requires only three partially polarizing beam splitters and two half-wave plates. The design of their gate is shown in Fig.~\ref{fig:langcz}.

\begin{figure}[htb]
\begin{center}
\includegraphics[width=0.8\textwidth]{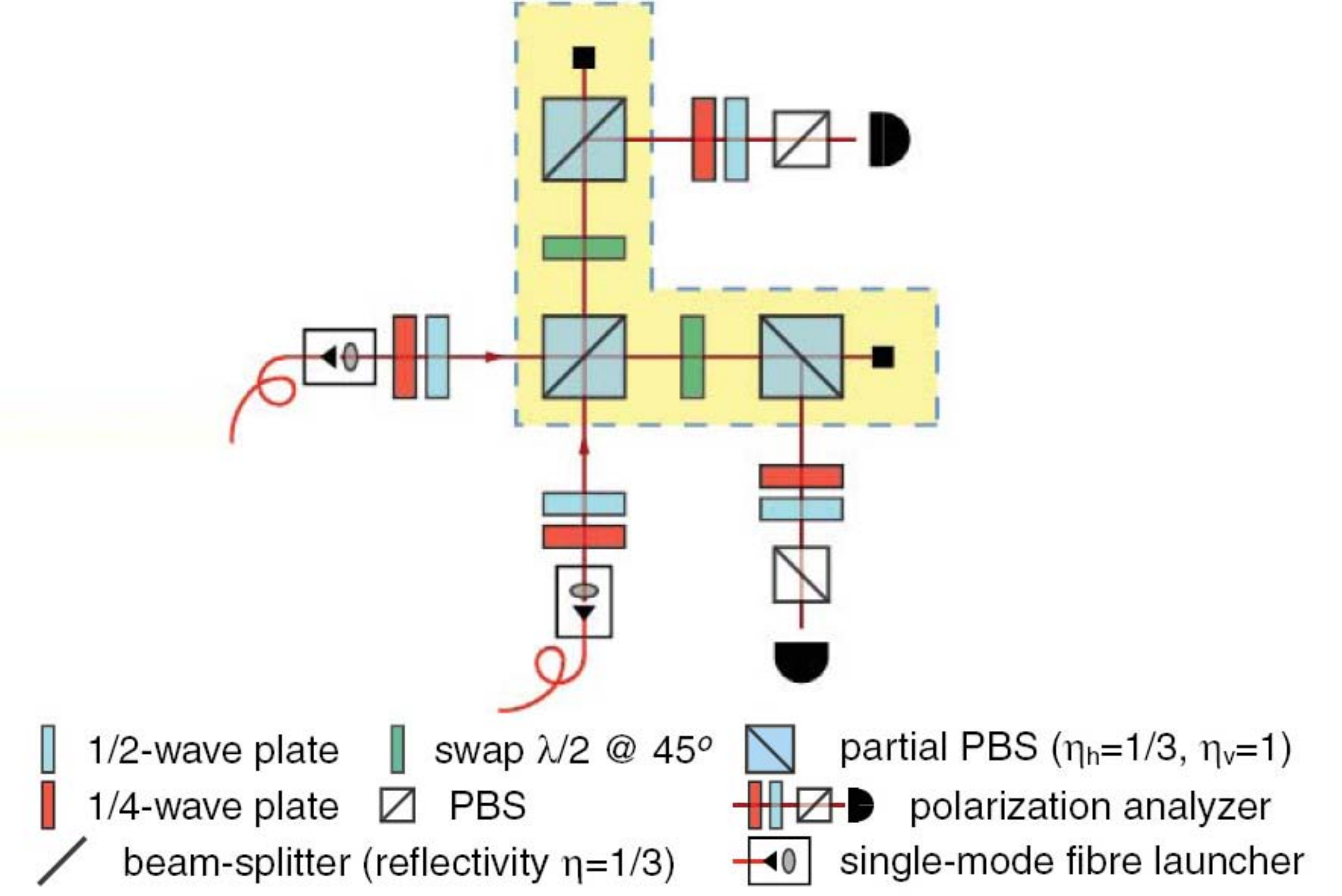}
\end{center}
\caption{Linear optical implementation of a non-deterministic Controlled-Z gate gate by Langford \emph{et al.}~\cite{Langford:2005}. Figure reproduced with permission.}
\label{fig:langcz}
\end{figure}

CZ gates in combination with single-qubit gates can reproduce \emph{any} two-qubit gate~\cite{Nielsen:2000}. To implement a CNOT gate we require the application of two \emph{Hadamard} gates, one applied to the target qubit before the CZ gate, and one after. The unitary matrix for the Hadamard gate is given by
\[
H=\frac{1}{\sqrt{2}}\left[
  \begin{array}{cc}
    1 & 1 \\
    1 & -1 \\
  \end{array}
\right],
\]
which up to a global phase shift of $\pi$ is simply a HWP with $\theta$ set to $22.5^\circ$ from horizontal. With the necessary tools for producing qubits, performing a CNOT gate, and determining the output states with SQPT, all that is left is the initial state preparation.

\section{State Preparation in LOQC}

In comparison to the theory presented in Chapter~\ref{chap:tomography}, the only novel factor introduced in an experimental implementation of SQPT is the method of state preparation for the tomography input states. In Section~\ref{ssec:qptrot}, we proposed two different methods of preparation which involved projective measurements, one using only measurements, and the other using a single measurement followed by unitary rotations. We will now recast these procedures in terms of optical components. We are assuming that our experiment consists of two photonic qubits which can be prepared in a correlated initial state $\gamma_0$. We will call the first qubit the \emph{system}, denoted $S$, and the second qubit the \emph{environment}, denoted $E$. Our preparation procedures are only applied to the system qubit $S$ while the environment qubit is left alone.

\subsection{Method I: State Preparation by Measurement and Rotations}\label{ssec:method1}

First we will consider the preparation map described in Section~\ref{ssec:qptrot}. Recall that it consisted of performing a single projective measurement to a fixed pure state, and then using unitary rotations to transform the post-measurement state to the required inputs states for performing SQPT. We wish to prepare qubit $S$ into the input states $\ket{H}, \ket{V}, \ket{D}$, and $\ket{R}$.

Our first problem with implementing this procedure in LOQC is that we cannot truly perform projective measurements. This is because measurements in optics rely on photon detection, which destroys the photon being measured. One way to simulate the effect of a projective measurement is to use a polarization filter and to post-select photons that get transmitted as verified by a final detection. This filter can be rotated to only transmit light of any chosen plane-polarized state, the unitary matrix for a polarization filter with its polarization axis set at an $\theta$ from horizontal is given by
\[
P(\theta_p)=\frac{1}{2}
\left[
  \begin{array}{cc}
    1+\cos2\theta_p & \sin2\theta_p \\
    \sin2\theta_p & 1-\cos2\theta_p \\
  \end{array}
\right].
\]
If we align a polarization filter to transmit only $\ket{H} (\theta=0)$, then we can use a HWP and QWP to give our desired input states (relative to a total phase shift experienced by the qubit). The preparation apparatus for this procedure is depicted in Fig.~(\ref{fig:loqcsprep}). For Method I the polarizer is fixed at $P(0)$. The alignment of the HWP and QWP required to give our required input states, and the induced phase shift of the wave plates, are shown in Table~\ref{tab:prepsettings}. We note that when using this preparation procedure for SQPT we \emph{always} expect to reconstruct a CP process matrix.

\begin{figure}[htb]
\begin{center}
\includegraphics[width=0.6\textwidth]{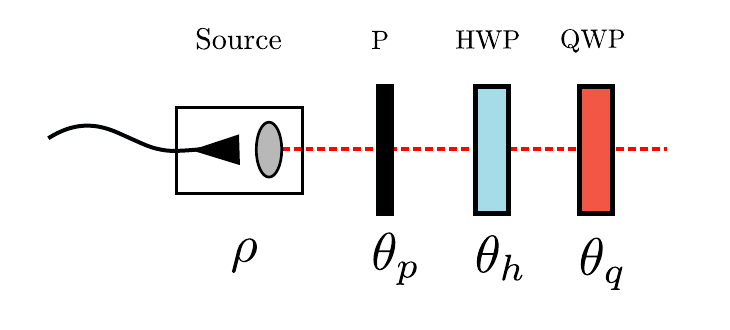}
\end{center}
\caption{Apparatus for performing state preparation in linear optics for Methods I and II. Settings for the angles of optical elements to prepare the input states H,V,D,R are shown in Table~(\ref{tab:prepsettings}).}
\label{fig:loqcsprep}
\end{figure}

\begin{table}[htbp]
\begin{center}
\begin{tabular}{|c| c c c c | c c c c |}
  \hline
           & \multicolumn{4}{|l|}{Method I}          & \multicolumn{4}{|l|}{Method II}           \\
           &$\theta_p$& $\theta_h$ & $\theta_q$ & $z$-shift & $\theta_p$ &$\theta_h$ & $\theta_q$ & $z$-shift \\
  \hline
  $\ket{H}$&$0^\circ$ & $0^\circ$   & $0^\circ$  &$135^\circ$& $0^\circ$  &$0^\circ$  & $0^\circ$  & $0^\circ$ \\
  $\ket{V}$&$0^\circ$ & $45^\circ$  & $0^\circ$  & $45^\circ$& $90^\circ$ &$0^\circ$  & $0^\circ$  & $225^\circ$ \\
  $\ket{D}$&$0^\circ$ & $22.5^\circ$& $45^\circ$ &$135^\circ$& $45^\circ$&$22.5^\circ$& $22.5^\circ$& $135^\circ$ \\
  $\ket{R}$&$0^\circ$ & $45^\circ$  & $-45^\circ$& $0^\circ$ & $0^\circ$  &$45^\circ$ & $-45^\circ$& $0^\circ$ \\
  \hline
\end{tabular}
  \label{tab:prepsettings}
  \caption{Settings for optical implements to prepare required input states for SQPT for methods I and II. Angles are rotations of optical axes from horizontal for a linear polarizer ($\theta_p$), HWP $\theta_h$, and QWP $\theta_q$. $z$-shift is the relative phase shift experienced by qubit $S$ compared to qubit $E$.}
\end{center}
\end{table}

\subsection{Method II: State Preparation by Measurement}\label{ssec:method2}

Now we consider the second preparation procedure, which we outlined in Section~\ref{sec:sqptproj}. This procedure described preparing our four input states $\{\ket{H},\ket{V},\ket{D},\ket{R}\}$ by only using projective measurements. As with Method I, since we cannot perform a true projective measurement, we simulate the effect of one using a linear polarizer. However, we now have an additional problem as we cannot directly produce elliptically polarized light in such a manner. We propose then that we overlook this problem and prepare the state $\ket{R}$ in the same manner as in Method I. For the vertical and diagonal states however, we can change the angle of the linear polarizer instead of using the HWP and QWP as in method 1. As we will show below, this revised preparation procedure can still result in a non-CP process matrix.

We will use the same preparation apparatus as Method I as depicted in Fig.~(\ref{fig:loqcsprep}). However, now the angles on the optical implements will be different. The settings for the polarizer, HWP, QWP and the total induced phase shift are also shown in Table~\ref{tab:prepsettings}. We make note that in the case of producing diagonally polarized light, the angle settings for the HWP and QWP were not zero, this is because for such an alignment, the combination of the wave plates would rotated the diagonally polarized light transmitted by the polarizer to anti-diagonally polarized light. We could just remove the HWP and QWP from the setup. However they are still needed to produce our right-circularly polarized state. In addition using this setup gives us a constant state preparation apparatus consisting of a polarizer, HWP, and QWP.

As previously mentioned we cannot perform a projective measurement for the state $\ket{R}$, and so we produce $\ket{R}$ by rotating $\ket{H}$. We will now revise our SQPT example from Section~\ref{sec:tomoex} to accommodate this.

We suppose our system and environment, both photonic qubits, are initially in the maximally entangled state $\ket{\phi}=(\ket{00}+\ket{11})/\sqrt{2}$. The interaction between the two photons is given by a CNOT gate with the system photon the target qubit. If we perform SQPT with this hybrid preparation procedure the resultant process map is given by
\[
\Lambda_\Ecal=\frac{1}{2}
\left[
\begin{array}{cccc}
 2 & 0 & -1-i & 2 \\
 0 & 0 & 0 & 1+i \\
 -1+i & 0 & 2 & 0 \\
 2 & 1-i & 0 & 0
\end{array}
\right]
\]
which has eigenvalues $\{2.039, -1.039, 0.863, 0.137\}$. Hence we still obtain a non-CP result with this method.

\section{Outline of an Optical Demonstration of Non-CP Maps}

The previous sections illustrate all the components needed to perform an experiment to demonstrate non-CP maps, which could be implemented with currently available techniques in LOQC. A simplistic schematic for the experiment is shown in Fig.~(\ref{fig:loqcexp}). The details on preparing the initially correlated state, and performing state tomography at the output are to be determined according to the planned experiment set-up.


\begin{figure}[htb]
\begin{center}
\includegraphics[width=\textwidth]{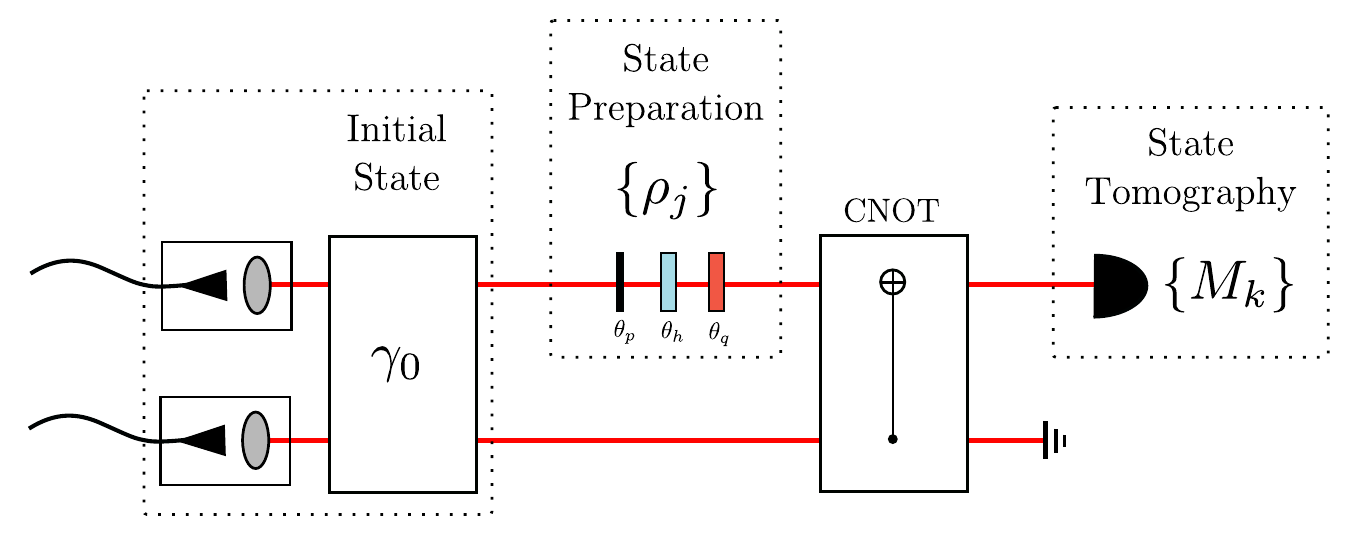}
\end{center}
\caption{Design of an experiment to demonstrate non-CP maps with currently available technology in linear optics.}
\label{fig:loqcexp}
\end{figure}

Following the preparation of an initially correlated joint system state $\gamma_0$, the experimenter will perform SQPT of the CNOT gate, which is constructed using the CZ and Hadamard gates. The input states for SQPT should be prepared using both Methods I and II as perviously outlined, and the reconstructed process matrices should be compared.
For the initially correlated states we suggest the experimenter first prepare the maximally entangled Bell state $\gamma_0=\op{\phi}{\phi}$. Following this the experimenter should prepare separable states, for example $\gamma_0=\frac{1}{2}\left(  \rho_H\otimes\rho_A + \rho_D\otimes\rho_V   \right)$, and $\gamma_0=\frac{1}{2}\left(\rho_H\otimes\rho_H + \rho_V\otimes\rho_D \right)$.


\subsection{Analysis of Results}
 Using the mentioned initial states $\gamma_0$, our expected results are a CP process matrix for Method I, and a non-CP process matrix for Method II. Problems may arise in SQPT because we can no longer use maximum likelihood tomography since we are explicitly looking for the possibility of a non-CP process map. To reduce the influence of statistical noise in the count statistics we suggest running the experiment for much longer than usually done so as  to minimize the effect of the Poissonian noise on the count data. Based on the statistical methods introduced in Chapter~\ref{chap:statsim} we should increase $\Ncal$, the number of copies of a state used in the tomographic reconstruction, to as large as feasibly possible. This will increase the ability to determine the probability that a measured process matrix arose from our expected process matrix due to statistical noise. 

\chapter{Conclusion} \label{chap:conclusion}

\section{Summary}
In this thesis we examined the evolution of open quantum systems in the presence of initial correlations in relation to non-completely positive maps. The study of open quantum systems is a crucial subject in quantum information science, as all real world quantum devices will experience some degree of interaction with their surroundings. In the case where we are not interested in the explicit time evolution of such systems we describe their evolution by quantum operations. These are CPTP maps on density matrices. In the early chapters we examined the theory for CPTP maps.

In Chapter~\ref{chap:correlations} we introduced non-CP maps. These can occur in quantum systems when the state of the system and its environment are initially correlated. In this chapter we investigated a new scheme of classifying the correlations of quantum systems by using a quantity known as quantum discord. We found that the proposed new definition of a classically correlated state being one with zero quantum discord is inconsistent. We showed, by a counter example, that not only is the quantity of quantum discord not symmetric, but if the state of a joint system has zero quantum discord, then the discord of the state with the roles of the two systems interchanged in generally need not be zero.

If we define correlations as quantum when a state has quantum discord greater than zero, this can lead to a situation where a system is \emph{both} quantum and classically correlated --- two situations which intuitively \emph{should} be mutually exclusive. This result implies that when using the proposed definition of classifying correlations based on quantum discord, correlations in quantum mechanics are \emph{directional}. We also investigated a result which stated that the evolution of initial states with only classical correlations would always lead be CP. In light of the found asymmetry we a led to conjecture that this theorem would not hold if we reversed the roles of the systems in our definition of quantum discord.

In Chapter~\ref{chap:tomography} and Chapter~\ref{chap:stateprep} our research moved into the direction of quantum process tomography. We introduced the two main schemes of implementing this important technique for completely characterizing an unknown physical process: SQPT and AAPT.  We investigated the effect of state preparation on these procedures, examining methods proposed in the literature, and also proposed our own, which has number of advantages. We investigated how these procedures influenced the nature of the reconstructed process matrix, either CP or non-CP, an question which has not been asked before.

We found that when using AAPT to characterize an unknown quantum operation, the reconstructed map would always be CP regardless of initial correlations. This is because only a single input state was required, effectively truncating the influence of the environment on the system by fixing it in a single state in accordance with the standard theory of CP maps. In the case of SQPT however, the nature of the preparation method was found to have a large influence on the nature of the reconstructed map. We investigated to specific examples which could be implemented in linear optics, a preparation procedure from the literature which used only projective measurement, and a procedure of our own design which used a single projective measurement followed by unitary rotations. We found that our proposed method would always lead to the reconstruction of a CP map, however the procedure using only measurements was found to lead to non-CP evolution. It is known that the preparation by measurement is a bilinear quantum operation, as apposed to linear operation. We proposed a novel method of implementing the bilinear preparation procedure as a linear process, which does not affect the issue of non-CP dynamics.

At the end of Chapter~\ref{chap:stateprep} we investigated a proposed procedure for describing bilinear process matrices. We noticed that one of the properties of the bilinear process matrix asserted in the literature was incorrect. Namely, the author stated that a bilinear process matrix has a trace equal to one, when in fact the trace is equal to the dimension of the system undergoing evolution.

An important issue which arose in our investigation of process tomography was the effect of statistical noise on the reconstructed process matrices. We examined the case of quantum optics where the spontaneous nature of photon emission leads to Poissonian distributed count statistics. This statistical noise could lead to the reconstruction of non-CP process matrices when we expect a CP result. This presents a challenge of having to distinguish between non-CP results due to noise, and true non-CP reconstructions in any actual tomography experiment.

In Chapter~\ref{chap:statsim} we formulated a statistical distribution for reconstructed process matrices. This was given by a multidimensional sum of Gaussian distributions, each with variance inversely proportional to the number of copies of our states used to perform tomography. We proposed that by increasing the number of copies one could distinguish between true non-CP results and ones due to noise, as the distribution for true non-CP results would largely lie outside CP process space, when in the noise case the majority of the distribution would correspond to CP processes.

Finally, in Chapter~\ref{chap:linoptics} we brought together many of the ideas proposed in the earlier chapters to design an experiment which could demonstrate non-CP maps. This was done using currently available equipment from linear optics, and demonstrated how different state preparation procedures in SQPT could greatly influence the identified quantum process. Our two proposed preparation procedures both consisted of three optical elements, a linear polarizer followed by a HWP and QWP. The only difference between the schemes was the orientation of the optic axes of each device. The successful implementation of this experiment would illustrate how non-CP makes can arise in real world situations, and also show how a seemingly minor change in preparation can have a large impact on our final tomographic output

\section{Directions for Future Investigation}

Our research left several avenues for further study into non-CP quantum processes. An obvious step would be to actually implement our proposed experiment from Chapter~\ref{chap:linoptics}. In addition a more rigorous study of the statistical distribution of reconstructed density matrices would be beneficial in helping use distinguish between true non-CP results, and ones due to statistical noise. Some possible directions for this investigation include developing a tomography protocol which incorporates sequential analysis and or adaptive schemes. One possible example for this would be an implementation of tomography where the number of copies prepared is not fixed, but we can resample as required to increase accuracy, changing the direction of our measurements to minimize the variance of the resulting statistical distribution. Finally, our study of quantum discord in Chapter~\ref{chap:correlations} illustrated inconsistencies with the proposed method for classifying correlations of quantum systems. While the approach appears to have merit, the lack of symmetry poses an interesting question about the symmetry of system correlations. 

\appendix

\chapter{Mathematical Proofs}
\section{Proofs for Chapter \ref{chap:quantumops}}
\subsection{Process Matrix Equations} \label{app:processmatrix}
We wish to prove that for $\Lambda_\Ecal$ defined by $\Lambda_\Ecal=\sum_{ij} E_{ij}\otimes \Ecal(E_{ij})$, the quantum operation $\Ecal(\rho)$ is given by Equation~(\ref{eqn:pmapevo}).
Note that we can expand any density matrix in terms of $E_{ij}$ as $\rho=\sum_{ij} \rho_{ij}E_{ij}$. Hence $\rho^T=\sum_{ij} \rho_{ij}E_{ji}$. Then the right hand side of Equation~(\ref{eqn:pmapevo}) is,
\begin{eqnarray*}
\tr_A\left(\Lambda(\rho^T\otimes\id)\right)
&=& \tr_A\left[ \left(\sum_{ij} E_{ij}\otimes \Ecal(E_{ij})\right)\left( \sum_{mn} \rho_{mn}E_{nm}\otimes\id\right)\right]\\
&=& \sum_{ij}\sum_{mn} \rho_{mn} \tr_A\left[ (E_{ij}E_{nm})\otimes\Ecal(E_{ij})\right]\\
&=& \sum_{ij}\sum_{mn} \rho_{mn} \Ecal(E_{ij}) \,\tr\left[E_{ij}E_{nm}\right]\\
&=& \sum_{ij}\sum_{mn} \rho_{mn} \Ecal(E_{ij}) \,\tr\left[\ket{i}\ip{j}{n}\bra{m}\right]\\
&=& \sum_{ij}\sum_{mn}  \Ecal(\rho_{mn}E_{ij}) \delta_{jn}\delta_{im}\\
&=& \sum_{ij}  \Ecal(\rho_{ij}E_{ij})\\
&=& \Ecal(\rho)        \quad\qed.
\end{eqnarray*}

\subsection{Process Matrix to Kraus Representation} \label{app:pmaptokraus}

We wish to construct a Kraus representation for a quantum operation $\Ecal$ described by a process matrix $\Lambda_\Ecal$. To do this we first prove that for a basis $E_{ij}=\op{i}{j}$, we can express evolution by the quantum operation $\Ecal$ as,
\begin{equation}
\Ecal(\rho)=\sum_{r r'}\sum_{s s'} (\Lambda_\Ecal)_{rr',ss'} E^T_{rr'}\rho E_{ss'}.
\label{eqn:canonicalkraus}
\end{equation}
This is called a \emph{canonical Kraus representation}.

Our first step is to expand $\rho$ in terms of $E_{ij}$, as $\rho=\sum_{ij}\rho_{ij}E_{ij}$. Hence
\[
E_{rr'}^T\rho E_{ss'}
= \sum_{ij} \rho_{ij} E_{r'r}E_{ij}E_{ss'}
= \rho_{rs} E_{r's'}
\]
Next, from the definition of $\Lambda$ we have
\begin{eqnarray*}
(\Lambda_\Ecal)_{rr',ss'}
&=&\sum_{ij} \left[E_{ij}\otimes \Ecal(E_{ij})\right]_{rr',ss'} \\
&=& \sum_{ij} (E_{ij})_{rs} \Ecal(E_{ij})_{r's'} \\
&=& \sum_{ij} \Ecal(E_{ij})_{r's'} \delta_{ir}\delta_{js}\\
&=& \Ecal(E_{rs})_{r's'}
\end{eqnarray*}
In addition, we notice that we can expand out $\Ecal(\rho)$ as
\[
\Ecal(\rho)=\sum_{rs}\Ecal(\rho_{rs}E_{rs})
\quad\Longrightarrow\quad \Ecal(\rho)_{r's'}=\sum_{rs}\Ecal(\rho_{rs}E_{rs})_{r's'}
\]
Finally we put these three facts together to get
\begin{eqnarray*}
\sum_{r r'}\sum_{s s'} (\Lambda_\Ecal)_{rr',ss'} E^T_{rr'}\rho E_{ss'}
&=& \sum_{r r'}\sum_{s s'} \rho_{rs} \Ecal(E_{rs})_{r's'} E_{r's'}\\
&=& \sum_{r r'}\sum_{s s'} \Ecal(\rho_{rs}E_{rs})_{r's'} E_{r's'}\\
&=& \Ecal(\rho),
\end{eqnarray*}
so equation~(\ref{eqn:canonicalkraus}) does indeed describe evolution by $\Ecal$.

Now we construct the Kraus operators $K_n$. First we collect the double indices in Equation~(\ref{eqn:canonicalkraus}) to single indices, so $\Ecal(\rho)=\sum_{mn} (\Lambda_\Ecal)_{mn}E_m\rho E_n^\dagger$. Next, since the process matrix is positive, we can find its spectral decomposition $\Lambda_\Ecal= \sum_i \lambda_i\op{e_i}{e_i}$. Hence,
\begin{eqnarray*}
\sum_{mn} (\Lambda_\Ecal)_{mn}E^T_m\rho E_n
&=& \sum_i\sum_{mn} \lambda_i (e_i)_m(e_i)^*_n E^T_m\rho E_n\\
&=& \sum_i \left(\sum_{m}\sqrt{\lambda_i}(e_i)_m E_m^T \right) \rho
\left(\sum_n \sqrt{\lambda_i}(e_i)^*_n E_n\right)\\
&=& \sum_i K_i \rho K_i^\dagger,
\end{eqnarray*}
where $K_i=\sum_{m}\sqrt{\lambda_i}(e_i)_m E_m^T$, and $(e_i)_m$ is the $m^{th}$ element of the $i^{th}$ eigenvector of $\Lambda_\Ecal$.
Finally, we notice that $\sum_{m}\sqrt{\lambda_i}(e_i)_m E^T_m$ can be rewritten as $\sqrt{\lambda_i}\,mat\ket{e_i}$.

Hence we have shown that for $\Ecal$ defined by a process map $\Lambda_\Ecal$, we can construct a Kraus representation by forming the Kraus operators $K_i = \sqrt{\lambda_i}\,mat\ket{e_i}$, where $\lambda_i$ and $\ket{e_i}$ are the eigenvalues, and eigenvectors of $\Lambda_\Ecal$ respectively.

\subsection{Superoperator Definition} \label{app:superop}

We will show that our definition of the superoperator in Eqn~(\ref{eqn:sodef}) is consistent with the evolution described by Eqn.~(\ref{eqn:sopevo}). First we note that the set $\{E_{ij}\}$ is an orthonormal basis with respect to the Hilbert-Schmidt inner product. To check
\begin{eqnarray*}
\ip{E_{ij}}{E_{kl}}
&=& \tr(E_{ij}^\dagger E_{kl})\\
&=&\tr\left( \ket{j}\ip{i}{k}\bra{l} \right) \\
&=&\delta_{ik}\delta_{jl}
\end{eqnarray*}
Hence, any density matrix can be written $\rho=\sum_{ij}\rho_{ij}E_{ij}$, or in vectorized form $\ket{\rho}=\sum_i \rho_{ij}\ket{E_{ij}}$. So the evolution of a vectorized matrix $\ket{\rho}$ is given by
\begin{eqnarray*}
\Phi_\Ecal\ket{\rho}
&=& \Phi_\Ecal \left(\sum_{ij}\rho_{ij}\ket{E_{ij}}\right)\\
&=& \sum_{ij}\rho_{ij} \Phi_\Ecal\ket{E_{ij}}\\
&=& \sum_{ij}\rho_{ij}\sum_{kl}\ket{\Ecal(E_{kl})}\ip{E_{kl}}{E_{ij}}\\
&=& \sum_{ij}\rho_{ij}\sum_{kl}\ket{\Ecal(E_{kl})}\delta_{ki}\delta_{lj}\\
&=& \sum_{ij}\rho_{ij}\ket{\Ecal(E_{ij})}\\
&=& \sum_{ij}\ket{\Ecal(\rho_{ij}E_{ij})}\\
&=& \ket{\Ecal(\rho)}.
\end{eqnarray*}

\subsection{Product states with fixed environment always give completely positive dynamics} \label{app:pstatecp}

Assume our joint system is in a simply separable state $\gamma=\rho\otimes\tau$, where $\rho$ is an arbitrary state of system $A$ and $\tau$ is a fixed state of system $B$, independent of the state of system $A$. We will assume system $AB$ evolves via an arbitrary unitary operation $U$. To prove complete positivity we will find a Kraus representation for the quantum operation $\Ecal$ given by $\Ecal(\rho)=\tr_B\left[U(\rho\otimes\tau)U^\dagger\right]$.

Since $\tau\ge0$ we can take its square root, so the initial state is
\[
\gamma=\rho\otimes\sqrt{\tau}\sqrt{\tau} = (\id\otimes\sqrt{\tau})(\rho\otimes\id)(\id\otimes\sqrt{\tau}).
\]
We can then expand out the middle $\id$ as $\sum_{i} \op{i}{i}$ where $\{\ket{i}\}$ is an orthonormal basis for $B$.
Hence $\gamma=\sum_i (\id\otimes\sqrt{\tau})(\rho\otimes\op{i}{i})(\id\otimes\sqrt{\tau})$.

Our evolution is then given by
\begin{eqnarray}
\Ecal(\rho)
	&=&\tr_B\left[U \gamma U^\dagger \right] \\
	&=& \sum_i \tr_B\left[U(\id\otimes\sqrt{\tau})(\rho\otimes\op{i}{i})(\id\otimes\sqrt{\tau})U^\dagger\right]\\
	&=&\sum_{ij} \bra{j}^B U (\sqrt{\tau}\ket{i})^B \rho (\bra{i}\sqrt{\tau})^B U^\dagger\ket{j}^B\\
	&=& \sum_{ij} K_{ij} \rho K_{ij}^\dagger,
\end{eqnarray}
where we are using the vector description of the partial trace from section~(\ref{ssec:partialtrace}), and the superscript $B$ denotes an operator acting on system $B$.

So we have a Kraus representation given by operators $K_{ij}=\bra{j}^B U (\sqrt{\tau}\ket{i})^B$.
Now we just need to check these operators satisfy the completeness relation.
\begin{eqnarray}
\sum_{ij} K_{ij}^\dagger K_{ij}
&=& \sum_{ij} (\bra{i}\sqrt{\tau})^B U^\dagger \op{j}{j}^B U (\sqrt{\tau}\ket{i})^B\\
&=& \sum_{i} (\bra{i}\sqrt{\tau})^B U^\dagger U (\sqrt{\tau}\ket{i})^B \\
&=& \sum_{i} \id_A \bra{i}\tau\ket{i}\\
&=& \id_A \tr(\tau)\\
&=& \id_A.
\end{eqnarray}
Hence $\Ecal$ is CP $\qed$.

\section{Proofs for Chapter \ref{chap:correlations}}

\subsection{$\Dcal_{B:A}(\rho)=0$ implies completely positive evolution }\label{app:discordbacp}

Consider a system $A$ with its environment given by the system $B$. We wish to prove that if the initial state, $\rho_{AB}$, of the joint system $AB$ is classically correlated as defined by $\Dcal_{B:A}(\rho)=0$, then evolution of the system $A$ will always be CP.

Let $\rho_{AB}$ satisfy $\Dcal_{B:A}(\rho_{AB})=0$. According to equation~(\ref{eqn:qddecomp}) this means there exists a set of orthogonal projectors $\{\Pi^A_j\}$ such that
\begin{eqnarray*}
\rho_{AB}   &=& \sum_j \Pi_j^A \rho_{AB} \Pi_j^A \\
            &=& \sum_j p_j \Pi_j\otimes\tau_j,
\end{eqnarray*}
where $\tau_j=\tr_{A}(\Pi_j^A \rho_{AB} \Pi_j^A)$ and $p_j=\tr(\Pi_j^A\rho_{AB})$.
The initial state of system $A$ is
\[
\eta=\tr_B(\rho_AB)=\sum_j p_j \Pi_j.
\]
Hence evolution of $\eta$ is given by a unitary operation $U$ on the joint system as follows:
\begin{eqnarray}
\eta\prime
&=& \tr_B\left(U(\sum_j p_j \Pi_j\otimes\tau_j)U^\dagger\right)\\
&=& \sum_j p_j \sum_n \bra{n}^B\left(U \Pi_j\otimes\tau_j U^\dagger\right)\ket{n}^B,
\end{eqnarray}
where the partial trace is expressed in terms of $\{\ket{n}\}$, an orthonormal basis for state vector space of $B$, and $\ket{n}^B=\id\otimes\ket{n}$.
Since $\tau_j$ is a positive matrix we can take its square root $\sqrt{\tau_j}$, and write it as $\tau_j=\sqrt{\tau_j}\,\id\,\sqrt{\tau_j}=\sum_m \sqrt{\tau_j}\op{m}{m}\sqrt{\tau_j}$.
Hence we have
\begin{eqnarray}
\eta\prime&=&
\sum_{j}\sum_{n,m} p_j \bra{n}^B U (\Pi_j\otimes\sqrt{\tau_j}\op{m}{m}\sqrt{\tau_j}) U^\dagger\ket{n}^B \nonumber\\
&=& \sum_{j}\sum_{n,m} p_j \bra{n}^B U (\sqrt{\tau_j}\ket{m})^B \,\Pi_j\, (\bra{m}\sqrt{\tau_j})^B U^\dagger\ket{n}^B \nonumber\\
&=& \sum_j \sum_{n,m} p_j D^j_{nm}\,\Pi_j\,D^{j\dagger}_{nm},\label{eqn:discordproof2}
\end{eqnarray}
where $D^j_{nm}=\bra{n}^B U (\sqrt{\tau_j}\ket{m})^B$ is an operator acting on system $A$ only.

Now we wish to remove the $j$ dependence from $D^j_{nm}$. We do this noting that $D^j_{nm}=\sum_l D^l_{nm}\delta_{lj}$, and so we can rewrite equation~(\ref{eqn:discordproof2}) as
\begin{equation}
\eta\prime=\sum_j \sum_{n,m} p_j \left(\sum_l D^l_{nm} \delta_lj \Pi_j\right)\Pi_j \left(\sum_l \Pi_j\delta_lj\, D^{l\dagger}_{nm}\right),\label{eqn:discordproof3}
\end{equation}
where $\Pi_j\Pi_j\Pi_j=\Pi_j$ as $\Pi_j$ are projectors.

Now since $\{\Pi_i\}$ are orthogonal, $\delta_lj\Pi_j=\Pi_l\Pi_j$ Hence we have
\begin{eqnarray}
\eta\prime&=&
\sum_j \sum_{n,m}p_j \left(\sum_l D^l_{nm} \Pi_l \right)\Pi_j\Pi_j\Pi_j\left(\sum_l \Pi_l D^{l\dagger}_{nm}\right)\\
&=& \sum_j \sum_{n,m} K_{nm} p_j\Pi_i K^\dagger_{nm}\\
&=& \sum_{\alpha} K_\alpha \eta K^\dagger_\alpha,
\end{eqnarray}
where we collapse $(n,m)$ to a single index $\alpha$, and the Kraus operators are defined
\[
K_{\alpha}=K_{nm}=\sum_l D^l_{nm} \Pi_l
=\sum_l \bra{n}^B U (\sqrt{\tau_l}\ket{m})^B \Pi_l.
\]
Now all that remains is to show the Kraus operators satisfy the completeness relation.

\begin{eqnarray*}
\sum_\alpha K^\dagger_\alpha K_\alpha
&=& \sum_{n,m}\sum_{l,k} \Pi_l D^l_{nm}  D^k_{nm} \Pi_k\\
&=& \sum_{n,m}\sum_{l,k} \Pi_l   (\bra{m}\sqrt{\tau_l})^B U^\dagger \op{n}{n}^B U (\sqrt{\tau_k}\ket{m})^B \Pi_k\\
&=& \sum_{m}\sum_{l,k} \Pi_l   (\bra{m}\sqrt{\tau_l})^B U^\dagger U (\sqrt{\tau_k}\ket{m})^B \Pi_k
\quad\mbox{as } \sum_n \op{n}{n}=\id_B\\
&=&  \sum_{m}\sum_{l,k} \Pi_l\Pi_k   \bra{m}\sqrt{\tau_l}\sqrt{\tau_k}\ket{m}
\quad\mbox{as $U$ is unitary}\\
&=& \sum_m\sum_k \Pi_k \bra{m}\tau_k\ket{m}
\quad\mbox{as }\, \Pi_l\Pi_k=\delta_{lk}\Pi_k\\
&=& \sum_k \Pi_k \tr(\tau_k)\\
&=& \id_A \quad\mbox{as $\tr(\tau_k)=1$, and}\sum_k \Pi_k=\id_A.
\end{eqnarray*}
So the operators $\{K_\alpha\}$ are indeed Kraus operators, hence the evolution is completely positive. $\qed$

\section{Proofs for Chapter \ref{chap:tomography}}

\section{Calculation of the Dual Basis for (H,V,D,R)}\label{app:duals}

We wish to calculate the dual basis $\{D_i\}$for the tomographically complete measurement set $\{M_i\}$, where $i=H,V,D,R$. Starting with Equation~\ref{eqn:dualdef}, if we vectorize our matrices we have
\begin{eqnarray*}
\ip{M_i}{D_j}&=&\delta_{ij}\\ \Rightarrow
\ket{M_i}\ip{M_i}{D_j}&=&\delta_{ij}\ket{M_i} \\ \Rightarrow
\left(\sum_i \op{M_i}{M_i}\right)\ket{D_j} &=& \ket{M_j} \\ \Rightarrow
\ket{D_j} = \Mcal^{-1}\ket{M_j},
\end{eqnarray*}
where $\Mcal=\sum_i \op{M_i}{M_i}$, and $\Mcal^{-1}$ denotes the Moore-Penrose generalized inverse of $\Mcal$~\footnote{The generalized inverse for a matrix $\Mcal$ is a unique matrix $\Mcal^{-1}$ satisfying $\Mcal\Mcal^{-1}\Mcal=\Mcal$, $\Mcal^{-1}\Mcal\Mcal^{-1}=\Mcal^{-1}$, $(\Mcal\Mcal^{-1})^\dagger=\Mcal\Mcal^{-1}$, $(\Mcal^{-1}\Mcal)^\dagger=\Mcal^{-1}\Mcal$. In the case where $\Mcal$ is an invertible matrix, its generalized inverse is the usual inverse matrix.}.

For our chosen measurement set,
\begin{eqnarray*}
\Mcal&=& \op{M_H}{M_H} +\op{M_V}{M_V} +\op{M_D}{M_D} +\op{M_R}{M_R} \\
    &=&
    \left[
      \begin{array}{cccc}
        1 & 0 & 0 & 0 \\
        0 & 0 & 0 & 0 \\
        0 & 0 & 0 & 0 \\
        0 & 0 & 0 & 0 \\
      \end{array}
    \right]+
     \left[
      \begin{array}{cccc}
        0 & 0 & 0 & 0 \\
        0 & 0 & 0 & 0 \\
        0 & 0 & 0 & 0 \\
        0 & 0 & 0 & 1 \\
      \end{array}
    \right]+
     \left[
        \begin{array}{cccc}
        \frac{1}{4} & \frac{1}{4} & \frac{1}{4} & \frac{1}{4} \\
        \frac{1}{4} & \frac{1}{4} & \frac{1}{4} & \frac{1}{4} \\
        \frac{1}{4} & \frac{1}{4} & \frac{1}{4} & \frac{1}{4} \\
        \frac{1}{4} & \frac{1}{4} & \frac{1}{4} & \frac{1}{4}
        \end{array}
     \right]+
      \left[
        \begin{array}{cccc}
        \frac{1}{4} & -\frac{i}{4} & \frac{i}{4} & \frac{1}{4} \\
        \frac{i}{4} & \frac{1}{4} & -\frac{1}{4} & \frac{i}{4} \\
        -\frac{i}{4} & -\frac{1}{4} & \frac{1}{4} & -\frac{i}{4} \\
        \frac{1}{4} & -\frac{i}{4} & \frac{i}{4} & \frac{1}{4}
        \end{array}
      \right]\\
    &=& \frac{1}{4} \left[
                      \begin{array}{cccc}
                        6 & 1-i & 1+i & 2 \\
                        1+i & 2 & 0 & 1+i \\
                        1-i & 0 & 2 & 1-i \\
                        2 & 1-i & 1+i & 6 \\
                      \end{array}
                    \right]
\end{eqnarray*}
Hence taking the inverse gives
\[
\Mcal^{-1}=\left[
    \begin{array}{cccc}
    2 & -1+i & -1-i & 0 \\
    -1-i & 6 & 2 i & -1-i \\
    -1+i & -2 i & 6 & -1+i \\
    0 & -1+i & -1-i & 2
    \end{array}
    \right]
\]
Now finally, we use this in $D_j=mat(\Mcal^{-1}\ket{M_j})$ to get,
\begin{eqnarray*}
D_1=\frac{1}{2}\left[
        \begin{array}{cc}
        2    & -1+i \\
        -1-i & 0
        \end{array}
    \right] \quad&&\quad
D_3=\left[
        \begin{array}{cc}
        0 & 1 \\
        1 & 0
        \end{array}
        \right]\\
D_2=\frac{1}{2}\left[
        \begin{array}{cc}
        0    & -1+i \\
        -1-i & 2
        \end{array}
    \right] \quad&&\quad
D_4=\left[
        \begin{array}{cc}
        0 & -i \\
        i & 0
        \end{array}
        \right].
\end{eqnarray*}

\section{SQPT Process Matrix Definition}\label{app:sqptpmap}

We wish to show that the definition of the process matrix in SQPT
\[
\Lambda_\Ecal=\sum_j D_j^*\otimes\Ecal(\rho_j),
\]
is equivalent to the definition in Section~\ref{ssec:processmap}. We do this by verifying Equation~(\ref{eqn:pmapevo}) with this new definition.

First we express $\rho$ in terms of our input states, ie $\rho=\sum_{k}p_{k}\rho_{k}$. Then the RHS of Equation~(\ref{eqn:pmapevo}) becomes
\begin{eqnarray*}
\tr_{B}\left[ \Lambda_{\Ecal}(\id\otimes\rho^{T}) \right]
		&=&\tr_{A}\left[  \left(\sum_{j} D_{j}^{*}\otimes\Ecal(\rho_j)\right)
		\left(\sum_{k} p_{k}\rho_{k}^{T}\otimes\id \right) \right] \\
		&=& \sum_j\sum_k p_{k}\tr_A\left[(D_{j}^{*}\rho_k^T)\otimes\Ecal(\rho_j)\right] 		\\
		&=& \sum_j\sum_k p_{k}\Ecal(\rho_j)\tr\left[D_{j}^{*}\rho_{k}^{T} \right]\\
		&=& \sum_j\sum_k p_{k}\Ecal(\rho_j)\tr\left[\rho_{k}D_{j}^{\dagger} \right]
		\quad\mbox{$\rho_{k}$ and $D_{j}$ are hermitian}\\
		&=& \sum_j\sum_k p_{k}\Ecal(\rho_j)\delta_{jk} \quad\mbox{definition of the $D_j$}\\
		&=&  \Ecal\left(\sum_{k}p_{k}\rho_{k}\right) \\
		&=& \Ecal(\rho).
\end{eqnarray*}
Since this is true for all $\rho$ we must have $\Lambda_\Ecal=\sum_j D_j^*\otimes\Ecal(\rho_j) \, \qed$.

\section{Proofs for Chapter \ref{chap:stateprep}}

\subsection{A stochastic preparation map does not effect the state of the environment}\label{app:ximap}

If we are preparing our system by a map $(\Xi\otimes\Ical)(\gamma)=\op{\psi}{\psi}\otimes\tau$. Then by combining the spectral and Schmidt decompositions, we can express the initial state $\gamma$ as
\begin{eqnarray*}
\gamma
    &=& \sum_k \lambda_k\op{e_k}{e_k}\\
    &=& \sum_{k,i,j} \lambda_k\mu_{ki}\mu_{kj} \op{i_k^A}{j_k^A}\otimes\op{i_k^B}{j_k^B}
\end{eqnarray*}
Hence,
\begin{eqnarray*}
(\Xi\otimes\Ical)\gamma
    &=& \sum_{k,i,j} \lambda_k\mu_{ki}\mu_{kj} \Xi(\op{i_k^A}{j_k^A})\otimes\op{i_k^B}{j_k^B}\\
    &=& \op{\psi}{\psi}\otimes\tau.\\
\Rightarrow \tau
    &=& \tr_A \left[ (\Xi\otimes\Ical)\gamma\right]\\
    &=& \sum_{k,i,j} \lambda_k\mu_{ki}\mu_{kj}\op{i_k^B}{j_k^B}\tr[\Xi(\op{i_k^A}{j_k^A})]
\end{eqnarray*}

If $\Xi$ is trace preserving then
\[
\tr[\Xi(\op{i_k^A}{j_k^A})]=\tr[\op{i_k^A}{j_k^A}]=\delta_{ij}
\]
Hence,
\[
\tau = \sum_{k,i} \lambda_k\mu_{ki}^2\op{i_k^B}{i_k^B} = \tr_A(\gamma)
\]
So state of the environment post preparation is independent of the map $\Xi$, instead it \emph{only} depends on the initial state $\gamma$.

\subsection{Bilinear process matrix equations} \label{app:bilin}

We wish to show that our definition of the process matrix in Equation~(\ref{eqn:bilinmatrix}) is equivalent to the original definition in Equation~(\ref{eqn:bilinkuah}). We do this by expanding out our expression into the index notation of the original equation. So,
\begin{eqnarray*}
\Mcal^{(r,s)}_{r''r'; s'' s'}
    &=& \tr\left([U]_{rr'}[\gamma_0]_{r''s''}[U]_{ss'}^\dagger\right) \\
    &=& \sum_\epsilon \left([U]_{rr'}[\gamma_0]_{r''s''}[U^\dagger]_{s's}\right)_{\epsilon\epsilon}\\
    &=& \sum_\epsilon \sum_{\alpha, \beta}
        \left([U]_{rr'}\right)_{\epsilon \alpha}
        \left([\gamma_0]_{r''s''}\right)_{\alpha \beta}
        \left([U^\dagger]_{s's}\right)_{\beta \epsilon}\\
    &=& \sum_{\alpha, \beta, \epsilon}
        U_{r\epsilon, r'\alpha}(\gamma_0)_{r''\beta,s''\epsilon}(U^\dagger)_{s'\beta,s\epsilon},
\end{eqnarray*}
which is the original definition of $\Mcal$ $\qed$.

\subsection{Trace of bilinear process matrix is $d$} \label{app:bqptproof}

Let $M$ be a bi-linear process map given by an initial state $\gamma_0$ and unitary $U$ for a $d^2$-dimensional joint system $AB$. Hence $M$ is given by
\[
M^{(r,s)}_{r''r'; s'' s'}= \tr\left(
        [U]_{rr'}[\gamma_0]_{r''s''}[U]_{ss'}^\dagger
    \right).
\]
We now compute the trace of $M$.
\begin{eqnarray*}
\tr(m)
    &=& \sum_{r,r',r''} M^{(r,r)}_{r''r'; r'' r'}\\
    &=& \sum_{r,r',r''} \tr\left([U]_{rr'}[\gamma_0]_{r''r''}[U]_{rr'}^\dagger\right) \\
    &=& \tr\left[ \left( \sum_{r,r'}[U]_{rr'}^\dagger[U]_{rr'}\right)\left(\sum_{r''}[\gamma_0]_{r''r''}\right)\right].
\end{eqnarray*}
But since $U$ is unitary
\[
\id_{d^2}=U^\dagger U
= \left(
    \begin{array}{cccc}
      \sum_{i}[U]_{i1}^\dagger[U]_{i1} & \times & \ldots & \times \\
      \times       &  \sum_{i}[U]_{i2}^\dagger[U]_{i2} &  & \vdots \\
      \vdots &  & \ddots & \times \\
      \times & \ldots & \times &  \sum_{i}[U]_{id}^\dagger[U]_{id} \\
    \end{array}
  \right),
\]
where $\times$ are cross terms $[U]^\dagger_{mn}[U]_{kl}$. Hence we have
\[
\sum_i [U]_{ij}^\dagger[U]_{ij} = \id_d \Rightarrow \sum_{ij} [U]_{ij}^\dagger[U]_{ij} = d\id_d.
\]
Hence
\begin{eqnarray*}
\tr(m)
    &=& \tr\left[ d\id_d\left(\sum_{r''}[\gamma_0]_{r''r''}\right)\right]\\
    &=& d\,\sum_{r''}\tr\left([\gamma_0]_{r''r''}\right)\\
    &=& d \tr\left(\gamma_0 \right) \\
    &=& d.              \quad\quad\qed
\end{eqnarray*} 

\backmatter


\bibliography{references}

\end{document}